%% file: main.tex
\documentclass[%
reprint,
superscriptaddress,
nofootinbib,
longbibliography,
noeprint,
amsmath,amssymb,
aps
]{revtex4-2}
\usepackage[utf8]{inputenc}
\usepackage{bbm}
\usepackage{bm}
\usepackage[left=2cm,right=2cm,top=2cm,bottom=2cm]{geometry}
\usepackage{enumitem}
\usepackage{amsmath,amsfonts,amssymb,amsthm,graphics,graphicx}

\usepackage{hyperref}
\usepackage{tabularx}

\theoremstyle{definition}
\usepackage [autostyle, english = american]{csquotes}
\usepackage[caption=false]{subfig}
\graphicspath{{figures/}}
\MakeOuterQuote{"}
\newcommand{\ket}[1]{|{#1}\rangle}
\newcommand{\bra}[1]{\langle {#1} |}

\def\equationautorefname~#1\null{Eq. (#1)\null}

\newcommand{\appref}[1]{\hyperref[#1]{App.~\ref*{#1}}}
\renewcommand{\vec}[1]{\mathbf{#1}}

\newcommand{\gs}[1]{#1}
\newcommand{\david}[1]{#1}
\newcommand{\comment}[1]{}
\newcommand{\CNOT}{\mathrm{CNOT}}
\newcommand{\iSWAP}{\mathrm{iSWAP}}

\usepackage{tikzit}
\usepackage{subfiles}

\usepackage{hyperref}
\newcommand{\pauliplus}{\oplus}
\usepackage{colortbl}
\newcommand{\cosetZ}{B_t^{(s)}}
\newcommand{\coset}[1]{\mathbf{A}_{t}^{(L,{{{#1}}})}}
\makeatletter 
    
\renewcommand\onecolumngrid{
\do@columngrid{one}{\@ne}%
\def\set@footnotewidth{\onecolumngrid}
\def\footnoterule{\kern-6pt\hrule width 1.5in\kern6pt}%
}

\renewcommand\twocolumngrid{
        \def\footnoterule{
        \dimen@\skip\footins\divide\dimen@\thr@@
        \kern-\dimen@\hrule width.5in\kern\dimen@}
        \do@columngrid{mlt}{\tw@}
}%

\makeatother    
\begin{document}
\input{tree-styles.tikzstyles}

\subfile{main-singletree2}

\input{bibliography.bbl}

\clearpage
\onecolumngrid
\setcounter{figure}{0}
\setcounter{section}{0}
\setcounter{page}{1}
\let\oldthefigure\thefigure
\renewcommand{\thefigure}{S\oldthefigure}
\setcounter{equation}{0}
\renewcommand{\theequation}{S\arabic{equation}}
\newpage
\subfile{supplement}

\end{document}

%% file: tree-styles.tikzstyles

\tikzstyle{stabilizer}=[fill={rgb,255: red,191; green,191; blue,191}, draw={rgb,255: red,191; green,191; blue,191}, shape=rectangle]
\tikzstyle{error}=[fill={rgb,255: red,255; green,242; blue,0}, draw=red, shape=circle]
\tikzstyle{physical}=[fill={rgb,255: red,128; green,128; blue,128}, draw={rgb,255: red,128; green,128; blue,128}, shape=circle]
\tikzstyle{U}=[fill=white, draw=black, shape=circle, minimum height=0.6 cm, minimum width=0.6 cm, line width=1 pt, tikzit fill=magenta]
\tikzstyle{cnot}=[fill=white, draw=black, shape=rectangle, minimum height=0.7 cm, minimum width=0.7 cm, line width=1 pt]
\tikzstyle{X-input}=[fill=red, draw=red, shape=rectangle]
\tikzstyle{Z-input}=[fill=blue, draw=blue, shape=rectangle]
\tikzstyle{CNOT}=[fill=white, draw=blue, shape=circle, minimum width=0.7 cm, minimum height=0.7 cm, line width=1 pt, tikzit fill=cyan]
\tikzstyle{NOTC}=[fill=white, draw=red, shape=circle, minimum height=0.7 cm, minimum width=0.7 cm, line width=1 pt, tikzit fill={rgb,255: red,255; green,191; blue,191}]
\tikzstyle{x spider}=[fill={rgb,255: red,232; green,165; blue,165}, draw=black, shape=circle]
\tikzstyle{z spider}=[fill={rgb,255: red,216; green,248; blue,216}, draw=black, shape=circle]
\tikzstyle{bigU}=[fill=white, draw=black, shape=circle, minimum width=0.7 cm, minimum height=0.7 cm, line width=1 pt]
\tikzstyle{trace}=[fill=white, draw=black, shape=rectangle]
\tikzstyle{bulkerror}=[fill={rgb,255: red,0; green,255; blue,255}, draw=red, shape=circle]

\tikzstyle{gray}=[-, draw={rgb,255: red,128; green,128; blue,128}, line width=1 pt]
\tikzstyle{stack}=[-, draw={rgb,255: red,140; green,175; blue,255}]
\tikzstyle{grayout}=[-, fill={rgb,255: red,255; green,191; blue,191}, opacity=0.6, draw={rgb,255: red,128; green,128; blue,128}, dashed]
\tikzstyle{thick}=[-, line width=1.5 pt, draw=magenta]

%% file: main-singletree2.tex
\title{ 
Dynamically generated concatenated codes and their phase diagrams}
\author{Grace M. Sommers}
\address{Physics Department, Princeton University, Princeton, NJ 08544}
\author{David A. Huse}
\address{Physics Department, Princeton University, Princeton, NJ 08544}
\author{Michael J. Gullans}
\address{Joint Center for Quantum Information and Computer Science,
NIST/University of Maryland, College Park, Maryland 20742, USA}
\date{\today}

\begin{abstract} 
 We formulate code concatenation as the action of a unitary quantum circuit on an expanding tree geometry and find that for certain classes of gates, applied identically at each node, a binary tree circuit encodes a single logical qubit with code distance that grows exponentially in the depth of the tree. When there is noise in the bulk or at the end of this encoding circuit, the system undergoes a phase transition between a coding phase, where an optimal decoder can successfully recover logical information, and a non-coding phase. Leveraging the tree structure, we combine the formalism of ``tensor enumerators'' from quantum coding theory with standard recursive techniques for classical spin models on the Bethe lattice to explore these phases. In the presence of bulk errors, the coding phase is a type of spin glass, characterized by a distribution of failure probabilities. When the errors are heralded, the recursion relation is exactly solvable, giving us an analytic handle on the phase diagram.
\end{abstract}
\maketitle

In statistical physics, models defined on trees are often analytically tractable owing to the absence of loops, serving as a kind of infinite-dimensional limit~\cite{Gujrati1995,Baxter2007}. A classic example is the Ising/Potts model on the Bethe lattice, which, under fully polarized boundary conditions or free boundary conditions, exhibits a transition to a ferromagnetic phase~\cite{Chayes1986,Haggstrom1996,Wagner2000,Baxter2007} or spin glass phase~\cite{Higuchi1977,Chayes1986,Bleher1995,Ioffe1996,Mezard2001,Mezard2006}, respectively. The absence of loops has also enabled the construction of exactly solvable models of measurement-induced phase transitions in quantum circuits defined on an (expanding or contracting) tree geometry~\cite{Nahum2021,Feng2023,Ferte2024,Ferte2024haar,Feng2024,Ravindranath2025}. But beyond being a setting for analytic tractability, the tree geometry also arises naturally in classical and quantum error correction, in the context of concatenated codes~\cite{Poulin2006,Yadavalli2023}. Code concatenation is a well-worn strategy for constructing high-distance codes out of small building blocks and underlies classic proofs of the threshold theorem~\cite{shor1996fault,aharonov1997,Knill1998,aharonov2008,Nielsen2010,Gottesman2024}. In this work we formulate code concatenation as the action of a unitary quantum circuit on an expanding tree geometry and study "coding transitions" under various error models.

For certain classes of gates, applied identically at each node, the tree circuit encodes zero-rate codes with code distance that grows exponentially in the depth of the tree. When noise is applied to the leaves (``surface'') and/or branches (``bulk'') of the tree, these codes exhibit transitions between a ``coding phase'' at low error rates and ``noncoding phase'' at high error rates. The task of a maximum likelihood decoder is to determine, given a syndrome, the most likely logical class of a correction operator. Using the language of tensor enumerators~\cite{Cao2024,Cao2024expansion}, we relate this problem to that of classical broadcasting/reconstruction on trees~\cite{Kesten1966,Kesten1967,Evans2000,Mossel2003,Mossel2004,Mezard2006}, with the bulk error rate playing the role of temperature, and the coding phase exhibiting the rich configurational landscape of a spin glass. When the code is a CSS code~\cite{Calderbank1996,Steane1996} subject to independent bit and phase flips, the connection to classical spin models is especially transparent, as the distribution of logical class probabilities across syndromes can be interpreted as a distribution of magnetizations at the root of the tree. We numerically simulate the recursion relation of this distribution to map out the phase diagram as a function of the bulk and surface error rates. When the errors are heralded, the recursion becomes exactly solvable, turning into a set of ``flow equations'' for the survival probability of logical information~\cite{Ferte2024}. 

\gs{In the most idealized setting of quantum error correction, quantum information is (1) perfectly encoded into a stabilizer code, which is then (2) fed through an (uncorrelated) Pauli channel, followed by (3) a perfect syndrome measurement. For concatenated codes, this setup corresponds to placing noise only on the leaves. As in Ref.~\cite{Yadavalli2023}, which independently introduced similar models of dynamically generated concatenated codes, we generalize beyond this simplistic scenario by allowing step (1) to also be noisy, i.e. introducing errors into the bulk of the tree.} The present work differs from Ref.~\cite{Yadavalli2023} in several ways. First, we focus on the optimal decoder, formulated as a tensor network. Second, we emphasize the statistical mechanics interpretation of these error models, probing the spin glass structure in the coding phase and examining the scaling behavior using both numerical and analytical methods. 
\gs{In a forthcoming work, we also go beyond the ``singletree'' geometry to develop a fault-tolerant state preparation protocol~\cite{Sommers2025}, which incorporates mid-circuit measurements with noisy ancillas and thus lifts constraint (3).}

\gs{An additional motivation for studying noisy trees comes from the realm of low-density parity-check LDPC codes. With a slight change in perspective---interpreting the tree graph as a \textit{finite-rate}, $\textit{classical}$ Tanner code rather than the encoding circuit of a single logical qubit---the recursive methods detailed here serve as probes of the finite-temperature phases of classical LDPC codes on locally tree-like expander graphs~\cite{Tibor}. The energy barriers of these classical LDPC codes are, in turn, closely related to those of quantum LDPC codes obtained via product constructions~\cite{Breuckmann2021,Placke2024}.}

The paper proceeds as follows. In~\autoref{sect:setup} we define and classify the encoding circuits using tensor enumerators.~\autoref{sect:methods} discusses optimal decoding of Pauli errors, deriving a recursion relation for the distribution of logical class probabilities.~\autoref{sect:unheralded} applies these methods to two models: depolarizing noise applied to the surface of a code with optimal distance scaling, and independent bit and phase flips applied to the ``Bell tree''~\cite{Yadavalli2023}, which generates a CSS code. Turning to heralded noise models,~\autoref{sect:herald} presents the theory of heralded surface errors, while~\autoref{sect:bell} offers a case study of the Bell tree with bulk and surface errors. We conclude in~\autoref{sect:conclude}, reserving several details for the Supplement~\cite{supp-ref}.\nocite{VandeWetering2020,oeis,maclagan2021,Brown2012,Brown13,Bravyi2009,Kalachev2022}

\section{Binary tree encoding circuits}\label{sect:setup}

\begin{figure}[t]
\includegraphics[width=\linewidth]{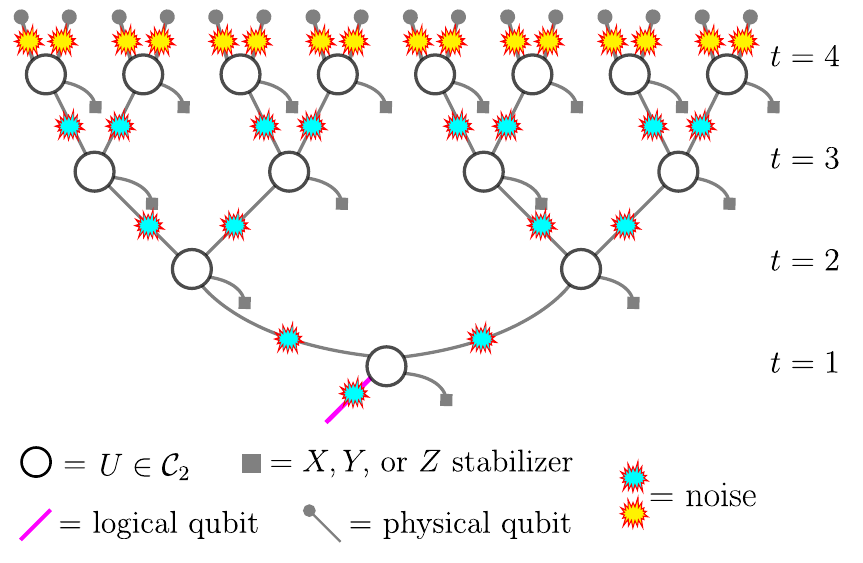}
\caption{Binary tree encoding circuit. This tree has depth 4 and produces a $[[16,1,d(4)]]$ stabilizer code. Blue and yellow ``starbursts'' indicate single-qubit errors on the branches and leaves, respectively, of the tree; the rates of these errors parameterize the phase diagrams studied in this work. \label{fig:tree}}
\end{figure}

Consider a $b$-ary tree, directed upwards as in~\autoref{fig:tree}. To each node, we associate an $b$-qubit gate, feeding in $b-1$ fresh qubits (by convention, on the right side). Thus this expanding circuit geometry generates an $b^t$-qubit state, where $t$ is the tree depth, starting from one qubit (shown in purple) at the root of the tree. If we further impose that each gate is a Clifford gate, and all fresh qubits are initialized in a given stabilizer state (henceforth referred to as ``stabilizer inputs''), this circuit encodes the root qubit (``logical input'') into an $[[n,k,d]]=[[b^t,1,d(t)]]$ stabilizer code, where $n$ is the number of physical qubits, $k=1$ the number of logical qubits, and $d$ the code distance, the minimal weight of a logical operator, which depends on the depth. This encoding circuit contains all self-concatenated qubit codes as special cases, by taking each gate and stabilizer input to be identical; then each node defines the $[[b,1,d(1)]]$ base code. 
While code concatenation is usually applied to a small code with distance $d>1$, we find that even binary trees ($b=2$), for which the local code necessarily has $d(1)=1$, can generate codes with distance $d(t)$ that grows exponentially in the depth $t$ of the tree. These binary trees are the focus of this paper. \gs{Different classes of binary tree codes discussed in this work are summarized in~\autoref{tab:table}.}

\begin{table*}[t]
\begin{tabularx}{\linewidth}{|>{\hsize=0.75\hsize}X|>{\hsize=0.65\hsize}X|>{\hsize=1.62\hsize}X|>{\hsize=0.98\hsize}X|}
\hline
    \textit{Name/description} & \textit{Distance scaling} & \textit{Unheralded error performance} & \textit{Heralded error performance} \\
    \cline{1-4}
    \textbf{Bell tree:} CSS code, alternating repetition in the X and Z basis [\autoref{eq:copy}, \autoref{eq:deloc}]. Different forms summarized in~\autoref{app:bell} of the Supplement~\cite{supp-ref}. & $d(t) = 2^{\lfloor t/2 \rfloor}$ [\autoref{eq:d2} &  
    \textbf{Independent bit/phase flips:} $p_{cx}(q) < p_{cz}(q)$ at fixed $q < q_c$. First-order transition at bulk threshold $q_c = 0.0066 \pm 0.0002$. [\autoref{sect:bell},\autoref{fig:phase-unheralded}] \vspace{4pt} \newline
      \textbf{Surface depolarizing:} Quantum info preserved up to $p_{c1} \approx 0.158$, classical bit preserved up to $p_{c2} \approx 0.22$ [\autoref{app:unheralded-bell} of Supplement] \vspace{4pt} \newline
      \textbf{Postselected on trivial syndrome:} $q_{FM} = 0.0417...$, $\nu = 1/2$ [\autoref{app:ferro} of Supplement]
   & \textbf{Independent X/Z heralding:} $p_c(q) < p_c(q)$, with $p_c(0) = 1-p_c(0) = (1+\sqrt{5})/2$, $\xi_c(0) = 3.2705...$. Bulk threshold: $q_c = 0.05505...$ with $\xi_c(q) \sim (q_c-q)^{-1/2}$. [\autoref{fig:phase}] \vspace{4pt} \newline
   \textbf{Surface errors, full phase diagram:} See \autoref{app:erasure-bell} of Supplement. \\
    \hline
    \textbf{Optimal-distance self-concatenated code:} Optimal Clifford isometry from 1 to 2 qubits, same on each node. Representative: \autoref{eq:best-tensor}. & $d(t) \propto 1.521^t$ [\autoref{eq:best-distance}]. Varying node types only improves the distance by a constant factor [\autoref{app:non-identical} of Supplement]. & \textbf{Surface depolarizing:} $p_c \approx 0.188, \xi_c \approx 2.9$  [\autoref{fig:tree-opt}] & \textbf{Surface erasures:} $p_c = 1/2, \xi_c = 2.603...$ [\autoref{eq:best-xi}, \autoref{fig:surface-e}] \\ \hline
    \textbf{Random binary tree:} Random Clifford isometry from 1 to 2 qubits at each node. & $d(t) \propto 1.153^t$ [\autoref{app:non-identical} of Supplement] & \textbf{Surface depolarizing:} $p_c \approx 0.188$, $\xi_c \approx 7.1$ [\autoref{app:subthreshold} of Supplement]  & \textbf{Surface erasures:} $p_c=1/2, \xi_c = 6.1126...$ [\autoref{app:herald-random} of Supplement] \\ \hline
        \rowcolor{red!10}\textbf{Repetition code encoded using a binary tree:} Each node copies information in one basis [\autoref{eq:copy}], providing no defense against the opposite type of errors.  & $d_X(t) = 2^t$, $d_Z(t) = 1$ & \textbf{Phase flips:} No threshold \vspace{4pt} \newline
    \textbf{Bit flips:} Ising model on Bethe lattice, spin glass phase for $q<q_{c} = \frac{1}{2}(1-\frac{1}{\sqrt{2}})$ \cite{Higuchi1977,Chayes1986,Bleher1995,Ioffe1996,Evans2000,Mezard2001,Mezard2006}. Continuous transition, with paramagnet unstable to spin glass~\cite{Chayes1986,Ioffe1996} ($p_{cx}(q) = 1/2$) up to $q_c$. \vspace{4pt} \newline
    \textbf{Bit flips with postselection:}
    Ising model with polarized BCs, ferromagnetic phase for $q < q_{FM} = 1/4$~\cite{Chayes1986,Haggstrom1996,Wagner2000,Baxter2007}. Continuous transition, with paramagnet unstable to ferromagnet ($p_{cx}(q) = 1/2$) up to $q_{FM}$~\cite{Chayes1986}. &
    \textbf{Heralded phase flips:} No threshold \vspace{4pt} \newline
    \textbf{Heralded bit flips:} Percolation on the Bethe lattice $\Rightarrow$ continuous transition at $q_{cx} = 1/2$~\cite{Grimmett1999}. \\ \hline
    \end{tabularx}
\caption{\gs{Properties of various classes of codes defined on binary trees. The shaded row is the classical repetition code, which is not a good quantum code but serves as a useful point of comparison throughout this work.}\label{tab:table}}
\end{table*}
\subsection{Tensor network formulation}
Recursion relations for the code distance are obtained using the method of tensor weight enumerators introduced in Ref.~\cite{Cao2024}, which we review briefly here and cast in a slightly different language.\footnote{Similar tensor network methods are presented in Refs.~\cite{Farrelly2021,Farrelly2022}. These tensors can be stitched together into more complex tensor networks~\cite{Ferris2014,Farrelly2021,Farrelly2022,Cao2022lego}, making the construction far more general than what we present here.}

The basic object for a binary tree is the rank-4 tensor describing a single node,
\begin{equation}\label{eq:gate-enum}
\input{R-tensor.tikz} = R^{\alpha\beta}_{i j} = \begin{cases}
1 & P_\alpha P_\beta P_i P_j \in \mathcal{S} \\
0 & \mathrm{otherwise}
\end{cases}
\end{equation}
where $\mathcal{S}$ is the stabilizer group of the ``encoding state''~\cite{Cao2024} constructed by purifying both input legs with a reference, i.e., it is generated by
\begin{align}
\mathcal{S} &= \langle U(ZI)U^\dag \otimes (ZI), U(XI)U^\dag \otimes (XI), \notag \\
& U(IZ)U^\dag \otimes (IZ), U(IX) U^\dag \otimes (IX) \rangle
\end{align}
where $U$ is the two-qubit gate on that node. We use the convention that $P_0 = I, P_1 = X, P_2 = Z, P_3 = Y.$\footnote{The ordering of $Y$ and $Z$ is nonstandard, but is chosen for consistency with the binary symplectic representation of stabilizer circuits~\cite{Gottesman2024}, where $X = \begin{pmatrix} 1 & 0 \end{pmatrix}, Z = \begin{pmatrix} 0  & 1 \end{pmatrix}, Y = \begin{pmatrix} 1 & 1\end{pmatrix}$.} In other words, $R^{\alpha \beta}_{ij} = 1$ if the logical $P_i P_j$ is encoded as $P_{\alpha}P_\beta$. This is an unsigned version of the circuit tensor recently introduced in \cite{Kukliansky2024}.

From this tensor, we have the flexibility to leave any of the ``stabilizer legs'' open, to construct a code with $k>1$. Fixing one of these legs to be a $P_{\alpha}$ stabilizer -- i.e., feeding in a fresh qubit in the +1 eigenstate of $P_\alpha$ -- means contracting the above tensor with the vector $e_I + e_{P_\alpha}$, which we will denote by:
\begin{equation}\label{eq:stabilizer}
\input{stabilizer.tikz} = e_I + e_{P_\alpha},
\end{equation}
the circuit tensor for state preparation~\cite{Kukliansky2024}. {Here we have followed the notation of Refs.~\cite{Cao2024,Kukliansky2024} in symbolically denoting the four-component basis vectors:
\begin{equation}\label{eq:symbol}
e_I = \begin{pmatrix} 1 \\ 0 \\ 0 \\ 0 \end{pmatrix}, \, e_X = \begin{pmatrix} 0 \\ 1 \\ 0 \\ 0 \end{pmatrix}, \,e_Z = \begin{pmatrix} 0 \\ 0 \\ 1 \\ 0 \end{pmatrix}, \, e_Y = \begin{pmatrix} 0 \\ 0 \\ 0 \\ 1 \end{pmatrix}.
\end{equation}}

For example, if we fix all the right legs to be $Z$ stabilizers, then we obtain the rank-3 tensor
\begin{equation}\label{eq:Rtilde}
\tilde{R}^{\alpha\beta}_i = R^{\alpha\beta}_{i 0} + R^{\alpha \beta}_{i 2}.
\end{equation}
In words, $\tilde{R}_i^{\alpha\beta}=1$ if the logical coset $\overline{P}_i$ of the associated $[[2,1,1]]$ code contains the operator $P_\alpha P_\beta$, {or in the symbolic notation of~\autoref{eq:symbol}, $\tilde{R}$ contains the term $e_{P_i}^{P_\alpha P_\beta}$.}

A concrete example that figures prominently in the analysis below is obtained by taking the gate $U=(H\otimes H) \CNOT$ with $Z$ stabilizer inputs:
\begin{equation}\label{eq:tildeR-bell}
\tilde{R} = e^{II}_{I} + e^{XX}_I + e^{ZZ}_X + e^{YY}_X + e^{IX}_Z + e^{XI}_Z + e^{ZY}_Y + e^{YZ}_Y.
\end{equation}
Placing this tensor on every node of the tree encodes a generalized Shor code~\cite{ECZoo}, and is referred to in Ref.~\cite{Yadavalli2023} as the \textit{Bell tree}. Without the Hadamard gates, a binary tree composed of only CNOT gates would generate a concatenated repetition code: \gs{each node copies the state on the input onto each of the outputs,} so a depth $t$ tree encodes a \textit{classical} bit into the codewords $\ket{0}^{\otimes 2^t}$ and $\ket{1}^{\otimes 2^t}$. The repetition code has an optimal threshold against bit flip errors, but it is a poor quantum code, providing no protection against phase flips. In contrast, with the added Hadamard gates, the Bell tree alternatively concatenates a ``bit flip'' repetition code and a ``phase flip'' repetition code, thus defending against both types of errors. \gs{We will revisit the similarities and differences between the phase diagrams of the concatenated repetition code and Bell tree, summarized in~\autoref{tab:table}, throughout this paper.}

\subsection{Vector enumerators}
Viewing the tree circuit as the encoding circuit of a logical fed into the root, we can associate to this code a four-component vector, the complete balanced vector enumerator~\cite{Cao2024}, whose $e_P$ component enumerates the elements of the logical $P$ class:
\begin{equation}\label{eq:complete-enum}
\vec{A}^{(L)}(\vec{u}) = \sum_{P} e_P \sum_{E\otimes P  \in \tilde{\mathcal{S}}}  \vec{u}^{\mathrm{wt}(E)}  \end{equation}
where $\tilde{S}$ is the stabilizer group of the $n+1$-qubit encoding state of the $[[n,1,d]]$ code, and $E$ is an element of the $n$-qubit Pauli group modulo phases. Here we have adopted the notation of Ref.~\cite{Cao2024} in defining
\begin{equation}
\vec{u}^{\mathrm{wt}(E)} = x^{\mathrm{wt}_X(E)} z^{\mathrm{wt}_Z(E)} y^{\mathrm{wt}_Y(E)} w^{\mathrm{wt}_I(E)}
\end{equation}
where $wt_F(E)$ is the number of Pauli $F$'s in the Pauli string $E$.

The complete vector enumerator can be obtained from a network of $\tilde{R}$ tensors [\autoref{eq:Rtilde}] as follows. Place the vector 
\begin{equation}
\input{physical.tikz} = e_I
\end{equation}
on each leaf of the tree: these are the ``physical qubits.'' Then let $\pauliplus$ denote the operation:
\begin{equation}\label{eq:pauliplus}
i \pauliplus j = k\quad \mathrm{if} \quad P_i P_j \sim P_k
\end{equation}
where $\sim$ means ``equal up to a phase''\footnote{{Writing $i\in \{0,1,2,3\}$ as a two-component bitvector, $\pauliplus$ is just vector addition modulo 2.}}.
Finally, let 
\begin{equation}
F^{\alpha}_{ij} = \substack{i \\ \input{error.tikz} \\ j} = f_{\alpha \pauliplus i \pauliplus j}.
\end{equation}
{Here $\alpha \in \{0,1,2,3\}$ labels the insertion of a Pauli error $P_\alpha$, and $\vec{f} = (f_0, f_1, f_2, f_3)$ parameterizes the probability of applying each Pauli}, in a sense that will be made clearer below.

In this notation, the vector enumerator of a depth $t$ tree is
\begin{align}\label{eq:tree-enum}
&\vec{A}^{(L)}_t(w,x,z,y) \equiv (I_t, X_t, Z_t, Y_t) \notag \\
&=\scalebox{0.8}{\input{tree-enumerator.tikz}}
\end{align}
where each $F^{0}$ tensor is evaluated at $\vec{f} = w, x, z, y$.\footnote{The top two layers are just a roundabout way of contracting with the vector $(w,x,z,y)$ on each leaf, since $\delta_{i0} F^{0}_{ij} = f_j$, but this decomposition will prove useful when we add noise.}

The tree structure naturally induces a recursion relation on the vector enumerator. To obtain the layer $t+1$ enumerator from layer $t$, we contract the logical legs of two generation-$t$ vector enumerators with the two outgoing legs of the $\tilde{R}$ tensor. If the generation-$t$ vector enumerators are identical, this amounts to the substitution
\begin{equation}\label{eq:recursion-A}
\mathbf{A}^{(L)}_{t+1}(w,x,z,y) = \mathbf{A}^{(L)}_1(I_t, X_t,Z_t,Y_t).
\end{equation}

\subsection{Code distance classes}
While the complete vector enumerator distinguishes between different types of non-identity Paulis ($X,Y,Z$)~\cite{Hu2020}, the more traditional weight enumerator just keeps track of the total non-identity weight (Hamming weight) of each element. In general, an $[[n,k,d]]$ stabilizer code is described by a pair of scalar enumerators, the Shor-Laflamme weight enumerators~\cite{Shor1997}. The $A$ enumerator is a polynomial in $u$ where the coefficient of $u^m$ is the number of elements of the stabilizer group with weight $m$. The $B$ enumerator is a polynomial where the coefficient of $u^m$ is the number of logical operators with weight $m$. These enumerators can be recovered from~\autoref{eq:tree-enum} evaluated at $w=1,x=z=y=u$, through the relation:
\begin{equation}
A_t(u)=I_t(u), \quad B_t(u) = I_t(u) + X_t(u) + Z_t(u) + Y_t(u).
\end{equation}

The code distance is the minimal weight of a nontrivial logical, hence the smallest power of $u$ for which $B_t(u) - A_t(u) = X_t(u) + Z_t(u) + Y_t(u)$ has a nonzero coefficient. Letting $d_P(t)$ be the smallest power of $u$ with nonzero coefficient in $P_t(u)$, this implies
\begin{equation}
d(t) = \min(d_X(t), d_Y(t), d_Z(t)).
\end{equation}

Across all two-qubit Cliffords, identically applied at each node, there are ten distinct classes of $\vec{A}^{(L)}$, three of which produce codes with exponentially growing code distance.

The class with the fastest-growing code distance obeys the relation
\begin{equation}\label{eq:best-distance}
d(t) = 2 d(t-3) + d(t-2), d(0)=d(1)=1, d(2) = 2,
\end{equation}
which asymptotically grows as $d(t)\sim a^t$ where $a=1.5213...$ . A representative circuit in this class has the gate $U = (\mathbbm{1} \otimes R_X[\pi/2]) \iSWAP$, with $X$ stabilizer inputs, yielding
\begin{equation}\label{eq:best-tensor}
\tilde{R} = e_I^{II} + e_I^{YY} + e_X^{XX} + e_X^{ZZ} + e_Z^{IY} + e_Z^{YI} + e_Y^{ZX} + e_Y^{XZ}.
\end{equation}
We will refer to this class as the ``optimal-distance self-concatenated code.'' Remarkably, even when we allow each gate to be different, the maximum possible distance still only grows as $1.521^t$~\cite{supp-ref}.

The remaining two ``good'' classes, one of which contains the Bell tree introduced above, have code distance
\begin{equation}\label{eq:d2}
d(t) = 2 d(t-2) = 2^{\lfloor t/2\rfloor}.
\end{equation}
\subsection{CSS Codes}
While suboptimal in terms of distance, the Bell tree is appealing because it makes a CSS code, in which the logical operators and stabilizer generators can all be chosen to be all $Z$'s or all $X$'s~\cite{Calderbank1996,Steane1996}. If bit flip and phase flip errors occur independently, then we can decode them separately, essentially reducing the problem to decoding two classical codes.

In~\autoref{eq:tildeR-bell}, we included Hadamard gates to give a representation of the Bell tree in which every node is identical. An equivalent representation that will prove more convenient in the ensuing analysis is obtained by pushing the Hadamards through to the beginning of the circuit. Then, on odd levels, we feed $Z$ stabilizers into a CNOT gate with the fresh stabilizers as the target, and on even levels, we feed $X$ stabilizers into a CNOT gate with the fresh stabilizers as the control (see~\autoref{app:bell} of the Supplement for more details~\cite{supp-ref}). In this form, odd layers of the tree ``copy'' information in the $Z$ basis:
\begin{equation}\label{eq:copy}
    \ket{0}\rightarrow \ket{00},\quad \ket{1} \rightarrow \ket{11}.
\end{equation}
and ``delocalize'' information in the $X$ basis\footnote{The language of ``copy'' and ``delocalizer'' nodes is due to Shiv Akshar Yadavalli and Iman Marvian.}:
\begin{equation}\label{eq:deloc}
\ket{+} \rightarrow (\ket{++} + \ket{--})/\sqrt{2}, \quad \ket{-} \rightarrow (\ket{+-}+\ket{-+})/\sqrt{2}.
\end{equation}
The role of the two bases is exchanged in even layers. Correspondingly, the minimal weight of an X logical operator doubles in odd layers, while the minimal weight of a Z logical operator doubles in even layers, hence the distance scaling in~\autoref{eq:d2}.

\section{Decoding Pauli Channels}\label{sect:methods}
\gs{In this section, we discuss the optimal decoder for i.i.d. Pauli errors, first on the leaves of the tree, and then on the links. We relate this task to that of classical broadcasting on trees and interpret the coding phase as a spin glass.}

\subsection{Surface errors}
Suppose each qubit of the code generated by the \david{depth $t$} tree circuit is subject to the single-qubit Pauli channel
\begin{equation}\label{eq:pauli-channel}
\mathcal{N}_{\vec{r}}(\rho) = (1-p) \rho + p_x X \rho X + p_y Y \rho Y + p_z Z \rho Z,
\end{equation}
where $\vec{p}=(p_x,p_y,p_z)$ and $p = |\vec{p}|_1$.

Given a syndrome $s$, a maximum likelihood decoder works as follows: find some "canonical error" $E_s$ with syndrome $s$. Then calculate the probability of each logical class $L$ relative to $E_s$---that is, the total probability of an error that is stabilizer-equivalent to $E_s \overline{L}$. If $L^*$ is the most likely logical class, then applying the correction operator $E_s \overline{L^*}$ is the optimal decoding strategy. A logical error occurs if the actual physical error belongs to a different logical class from $E_s \overline{L^*}$.

Our $[[2^t, 1, d(t)]]$ codes encode only one logical, indicated by the purple logical leg at the root of the tree. Thus, our decoder only needs to compute the relative weights of four logical cosets of the error subspace with syndrome $s$. These weights are encapsulated in the \textit{coset enumerator}~\cite{Cao2024expansion}: If $\Pi$ is a projector onto the error-free subspace, then for a given Pauli error $E_s$ applied to the physical qubits, we define
\begin{equation}\label{eq:coset-enumerator}
\coset{E_s} = (I_t(E_s \Pi), X_t(E_s\Pi), Z_t(E_s\Pi), Y_t(E_s\Pi))
\end{equation}
where $L(E_s\Pi)$ is the weight enumerator of operators that are stabilizer-equivalent to $E_s \bar{L}$. Given $E_s = P_{\alpha_1} \otimes P_{\alpha_2} \otimes \cdots \otimes P_{\alpha_n}$, we can compute~\autoref{eq:coset-enumerator} by replacing the $F^{0}$ tensor on the $i$th leaf in~\autoref{eq:tree-enum} with $F^{\alpha_i}$, and evaluating each $F^{\alpha}$ at  $(f_0=w=1-p, f_1 = x = p_x, f_2 = z = p_z, f_3 = y = p_y)$~\cite{Poulin2006,Cao2024expansion}. {This is the sense in which $F^{\alpha}$ ``inserts'' a Pauli operator $P_{\alpha}$.} 

The failure probability of the optimal decoder applied to this syndrome class is then
\begin{equation}\label{eq:pfail-s}
P(s) = 1 - \frac{\max(\coset{E_s})}{\cosetZ} \equiv 1 - \max_j(\eta^{(E_s)}_j).
\end{equation}
where $\cosetZ = |\coset{E_s}|_1$ and $\bm{\eta}^{(E_s)}$ denotes the vector of probabilities, conditioned on observing syndrome $s$, of the four logical classes relative to $E_s$\footnote{Note that for a fixed syndrome, the particular permutation of the components of $\bm{\eta}$ depends on whether we take the ``canonical error'' to be $E_s, E_s \overline{X}, E_s \overline{Z},$ or $E_s \overline{Y}$. This freedom manifests as a four-fold symmetry in the distribution $Q$ [\autoref{eq:distr}].} 

Let $Q^{(t)}$ denote the distribution of conditional probability vectors, weighted according to the probabilities of the associated syndrome classes, in a depth $t$ tree. Since a given syndrome appears with total probability proportional to $\cosetZ$, this distribution takes the form:
\begin{equation}\label{eq:distr}
Q^{(t)}(\bm{\eta}) \propto \sum_{s} \cosetZ \sum_L \delta\left(\bm{\eta} - \bm{\eta}^{(E_s \overline{L})}\right). 
\end{equation}
The average failure probability of the optimal decoder is then
\begin{equation}\label{eq:pfail-eta}
P_F(t) = 1 - \langle \max_j (\eta_j) \rangle_{t}
\end{equation}
where $\langle \cdot \rangle_t$ indicates an average taken over the distribution $Q^{(t)}$.

The parameters of the Pauli channel [\autoref{eq:pauli-channel}] set the initial condition of this distribution:
\begin{equation}\label{eq:init-cond}
Q^{(0)}(\bm{\eta}) = \frac{1}{4} \sum_{\alpha=0}^3 \delta(\bm{\eta} - \scalebox{0.8}{\input{error-init.tikz}}).
\end{equation}
evaluated at $f_0, f_1, f_2, f_3 = 1-|\vec{p}|_1, p_x, p_z, p_y$.

\subsection{Bulk errors}
Now suppose there are also errors in the bulk of the tree, on the links between gates. In the tensor network implementation of the optimal decoder, this is a simple modification: consider a pattern of errors (``fault'') with Pauli $P_{\alpha_i}$ at the $i$th possible error location. Then, inserting $F^{\alpha_i}$ at each location $i$, evaluated at the parameters of the channel at that site, yields a tensor network whose output is a vector of probabilities of the four logical classes relative to that fault.\footnote{In technical terms, these logical classes are classes of faults that are equivalent under stabilizers and gauge operators of the \textit{spacetime code}~\cite{Bacon2017,Gottesman2022,Delfosse2023} associated with the faulty circuit; see~\autoref{app:spacetime} of the Supplement for details~\cite{supp-ref}.}

\gs{Taking inspiration from the classical setting discussed below~\cite{Mezard2006}, recast in the tensor network language, we obtain a recursion relation for $Q^{(t)}$:
\begin{equation}\label{eq:recursion}
Q^{(t+1)}(\bm{\eta}) \propto \frac{1}{4} \sum_{\alpha = 0}^3 \int \delta\left(\bm{\eta} - f_\alpha(\{\bm{\eta}\})\right) z(\{\bm{\eta}\})\prod_{i=1}^b dQ^{(t)}(\bm{\eta}_i)
\end{equation}
where, setting $b=2$,
\begin{subequations}\label{eq:f-eta}
\begin{align}
z(\{\bm{\eta}\}) &=  \sum_i \scalebox{0.8}{\input{recursion-bulk.tikz}} = \sum_i \scalebox{0.8}{\input{recursion.tikz}}, \\
     f_{\alpha}(\{\bm{\eta}\})_i &= \scalebox{0.8}{\input{recursion-bulk.tikz}}/z(\{\bm{\eta}\}),
\end{align}
\end{subequations}}
and the cyan coloring indicates that $F^\alpha$ is evaluated at the parameters of the bulk error channel, $(1-|\vec{q}|_1, q_x, q_z, q_y)$.

In the ensuing analysis, we will separately tune the noise rates $\vec{p}$ and $\vec{q}$ on the surface and in the bulk, respectively. 

\subsection{Spin glass interpretation}
~\autoref{eq:distr} and the associated recursion relation,~\autoref{eq:recursion}, can be interpreted through the lens of classical broadcasting on trees~\cite{Evans2000,Mossel2003,Mossel2004,Mezard2006}. In that setting, a classical state is fed into the root of a tree, which the receiver tries to infer based on the states of the spins on the leaves. For example, if the classical encoding circuit is a concatenated repetition code, then applying bit flips at rate $q$ on each branch maps onto the Ising model at temperature $\exp(-2\beta)=q/(1-q)$ on this tree. The broadcasting process answers the question of how best to define ``free boundary conditions'' for the Ising model on a finite tree: generating the distribution of boundary spins by broadcasting from a root spin properly samples from the Gibbs measure of the infinite graph and captures the ``point-to-set'' correlations associated with replica symmetry breaking~\cite{Mezard2009}. At temperatures/error rates above the ``reconstruction threshold'', the conditional distribution for the root spin given the leaves is trivial: the disordered free-boundary Gibbs state is pure~\cite{Bleher1995,Ioffe1996,Evans2000,Mezard2006}, and no information is successfully transmitted. Below the threshold, the leaves remain correlated with the root even at infinite depth, signaling a spin glass phase.

The quantum analog to a distribution of boundary spins is a distribution over syndrome classes, weighted by the factor of $\cosetZ$ in~\autoref{eq:distr}. The components of the vector $\bm{\eta}^{(s)}$ are analogous to the conditional probabilities of the root spin given the set of spins on the leaves. At high rates of bulk and/or surface errors, the associated model is in a paramagnetic/noncoding phase, with the fixed point distribution
\begin{equation}\label{eq:noncoding}
    Q^*(\bm{\eta}) = \delta(\bm{\eta} - (1/4,1/4,1/4,1/4)).
\end{equation}
The nature of the coding phase will depend on the noise model.

More generally, the coset enumerator is just another name for the partition function $Z_E$ along the Nishimori line in the optimal decoder - stat mech mapping~\cite{Cao2024expansion,Dennis2002,Chubb2021}. In the general construction, stabilizer generators are mapped onto spin degrees of freedom, and different syndrome configurations correspond to different realizations of quenched disorder in the interactions between the spins. Whereas applying this mapping directly to the tree codes would yield a model with long-range interactions owing to the high weight of some stabilizers, a recursive treatment of the problem reveals that the quenched disorder can be pushed onto the boundary conditions, with the more frustrated boundary conditions corresponding to less likely syndromes. 

In our models, bulk errors---insertion of $F^\alpha$---play the role of introducing a finite temperature. Meanwhile, surface errors set the initial condition [\autoref{eq:init-cond}], akin to applying random entropic fields with magnitude controlled by $\vec{p}$ on the leaves.

\subsection{Independent bit and phase flips}
For CSS codes, such as those produced by the Bell tree, the analysis and interpretation simplifies when the noise factorizes into a bit flip and phase flip channel, i.e.
\begin{equation}
\mathcal{N}_{(r_x(1-r_z),r_x r_z, r_z(1-r_x))} = \mathcal{N}_{(r_x,0,0)} \circ \mathcal{N}_{(0,0,r_z)}.
\end{equation}
Then, since the X and Z decoding problems can be solved separately, we have essentially returned to the classical setting. The (coset) enumerators now only have two components, and the distribution over probability vectors $\bm{\eta}$ can be reinterpreted as a distribution of magnetizations $m_{x,z}$ at the root of a tree:
\begin{equation}\label{eq:mag}
\bm{\eta}_{x,z} = \left(\frac{1+m_{x,z}}{2},\frac{1-m_{x,z}}{2} \right).
\end{equation}
The logical failure probability of an optimal decoder (\autoref{eq:pfail-eta}) then becomes
\begin{equation}\label{eq:fail-m}
P_{X,Z} = \frac{1 - \langle |m_{x,z}|_t \rangle}{2}.
\end{equation}
Following the arguments of Ref.~\cite{Yadavalli2023}, we can interpret~\autoref{eq:fail-m} in terms of a state preparation task as follows. Consider the X decoding problem, sensitive to phase flip errors. Successful decoding is tantamount to being able to distinguish the logical basis states $\ket{\overline{\pm}}$. By the Holevo-Helstrom theorem~\cite{Helstrom1969,Holevo1973}, the failure probability for depth-$t$ state preparation is lower-bounded by
\begin{equation}\label{eq:hh}
\frac{1 -  ||\mathcal{E}_t(\ket{+}\bra{+}) - \mathcal{E}_t(\ket{-}\bra{-})||_1/2}{2}
\end{equation}
where $\mathcal{E}_t$ is the channel which maps input logical states to noisy physical states on $2^t$ qubits. The trace distance between \textit{quantum} states in~\autoref{eq:hh} is twice the total variation distance between the two distributions of \textit{classical} bit strings, obtained by measuring every qubit in the $\mathcal{E}_t(\ket{\pm}\bra{\pm})$ in the X basis (fully decohering in the Z basis)~\cite{Yadavalli2023}. Reinterpreting these classical bit strings as classical spin configurations, the total variation distance is precisely $\langle |m_z| \rangle_t$, recovering~\autoref{eq:fail-m}.  

\section{Phase diagram under unheralded errors}\label{sect:unheralded}
In this section, we study phase diagrams under unheralded noise, leveraging a ``population dynamics'' method~\cite{Mezard2006} to numerically sample the recursion relation. We focus on two models: the phase diagram of a representative optimal-distance self-concatenated code [\autoref{eq:best-tensor}] tuned by the rate of depolarizing noise on the leaves, and the two-parameter phase diagram of the Bell tree under independent bit and phase flips on the surface and in the bulk. {Further details on the method and comparisons to other codes are presented in~\autoref{tab:table} and in~\autoref{app:unheralded-methods} - \ref{app:unheralded-bell-bulk} of the Supplemental Material~\cite{supp-ref}.}

\subsection{Population dynamics}\label{sect:population}
To handle the factor of $z(\{\bm{\eta}\})$ in ~\autoref{eq:recursion}, we follow Ref.~\cite{Mezard2006} in defining a set of distributions $\{\tilde{Q}^{i}\}_{i=0}^3$, where
\begin{equation}\label{eq:distr-i}
\tilde{Q}_i (\bm{\eta}) = 4 \eta_i Q(\bm{\eta}) \Rightarrow Q(\bm{\eta}) = \frac{1}{4} \sum_{i=0}^3 \tilde{Q}_i (\bm{\eta}).
\end{equation}

In terms of $\tilde{Q}$, the recursion relation~\autoref{eq:recursion} becomes~\cite{supp-ref}
\begin{align}\label{eq:recursion-i}
    \tilde{Q}^{(t+1)}_i(\bm{\eta}) \propto \int \delta(\bm{\eta}-f_0(\{\bm{\eta}\})) &\sum_{\{\alpha\}}\left(\sum_k \tilde{R}^{\alpha_1 ... \alpha_b}_k q_{k\pauliplus i}\right) \notag \\ \prod_{j=1}^b &d\tilde{Q}^{(t)}_{\alpha_j}(\bm{\eta}_j)
\end{align}
with the initial condition
\begin{equation}\label{eq:init-i}
\tilde{Q}^{(0)}_i(\bm{\eta}) = \eta_i \sum_{\alpha=0}^3 \delta(\bm{\eta} - \scalebox{0.8}{\input{error-init.tikz}}) = \sum_{\alpha=0}^3 p_{\alpha \pauliplus i} \prod_j \delta(\eta_j - p_{\alpha \pauliplus j}).
\end{equation}
In the classical setting, $\tilde{Q}_i$ is the distribution of $\bm{\eta}$, conditioned on starting the broadcast from state $i$. In the quantum setting, we can think of the index $i$ as fixing the logical class to $\overline{P_i}$, in the sense that, in the absence of errors, $\tilde{Q}_i$ is a delta function peaked at $\hat{e}_i$.

Now we can simulate~\autoref{eq:recursion-i} as follows. Initialize four large populations, $\{S_i\}_{i=0}^3$, each of size $M$, according to $\tilde{Q}^{(0)}_i$. Then for $t=0,...,t_{max}-1$, for each $i$,
\begin{enumerate}
\item \label{step1}Draw $\alpha_1,...,\alpha_b$ (where $b=2$ for binary trees) with probability proportional to $\sum_k \tilde{R}_k^{\alpha_1,..,\alpha_b} q_{k \pauliplus i}$ (the term in parentheses in~\autoref{eq:recursion-i}).
\item \label{step2}For each $\alpha_j$, draw a vector $\bm{\eta}_j$ from $S^{(t)}_j$.
\item \label{step3}Add the vector $f_0(\{\bm{\eta}\})$ to the population $S_i^{(t+1)}$.
\item \label{step4}Repeat steps 1-3 $M$ times.
\end{enumerate}
$M$ must be chosen large enough so that the result is insensitive to the population size, and independent runs are consistent~\cite{supp-ref}.

\subsection{Depolarizing surface errors}
For trees that generate codes with growing code distance, subject to errors only on the leaves, we expect at least two phases: the noncoding phase where the fixed point distribution is trivial (\autoref{eq:noncoding}), and a coding phase characterized by a \gs{zero-temperature fixed point:}
\begin{equation}\label{eq:coding-surface}
\tilde{Q}^s_i(\bm{\eta}) = \delta(\bm{\eta} - \hat{e}_i).
\end{equation}
\gs{That is, in the absence of bulk errors, the coding phase has asymptotically perfect recovery as $t\rightarrow\infty$.} 
Depending on the gate and the bias in the error channel, there may also exist intermediate phases where only classical information survives. 

\begin{figure}[t]
\centering
\includegraphics[width=0.8\linewidth]{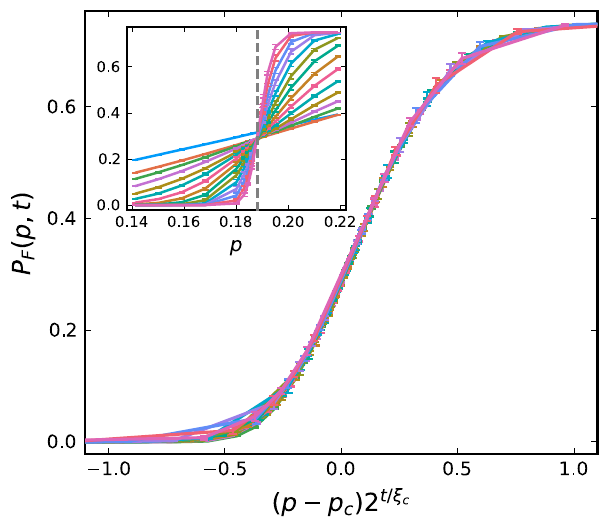}
\caption{Average failure probability of the optimal decoder for a $d(t)\propto 1.521^t$ tree code [\autoref{eq:best-tensor}] subject to depolarizing surface errors ($p_x=p_y=p_z=p/3$ in \autoref{eq:pauli-channel}). A scaling collapse with $p_c=0.188,\xi_c=2.9$ is shown for depths $t=6$ to $t=18$. Unscaled data are shown in the inset, for $t=4$ to $t=18$.\label{fig:tree-opt}}
\end{figure}

To identify the threshold between these phases, we compute the average failure probability [\autoref{eq:pfail-eta}], which in the population dynamics is
\begin{equation}\label{eq:pfail-sample}
P_F(\vec{p},t) = 1 - \frac{1}{4M} \sum_{i=0}^3 \sum_{\bm{\eta} \in S_i^{(t)}} \max_j(\eta_j).
\end{equation}

Let's consider depolarizing noise, $p_x=p_y=p_z=p/3$. In the vicinity of a critical point at $p_c$, $P_F(p,t)$ obeys the scaling form
\begin{equation}\label{eq:scaling}
P_F(p,t) = f((p-p_c)b^{t/\xi_c})~,
\end{equation}
where $\xi_c$ is an effective correlation length along the time direction, dependent on the choice of gate.

The threshold of random quantum codes subject to depolarizing noise saturates the so-called ``hashing bound'', which is $p_H=0.1893$ for zero-rate codes~\cite{Bennett1996}. The estimated threshold of the optimal-distance self-concatenated code [\autoref{eq:best-tensor}] is remarkably close to this bound, with $p_c \approx 0.188$ and no intermediate phase. A scaling collapse with $\xi_c = 2.9$ is shown in~\autoref{fig:tree-opt}. The subthreshold scaling of this and other codes is presented in~\autoref{app:subthreshold} of the Supplement.

Under the same noise model, the Bell tree has an intermediate phase, where only a classical bit survives. This feature is tied to the CSS nature of the code~\cite{Yadavalli2023}. In particular, alternating layers of the tree improve the robustness against bit flips (adding Z-type stabilizers) and phase flips (adding X-type stabilizers); see~\autoref{app:unheralded-bell} of the Supplement.

\subsection{Bell tree}\label{sect:bell-unheralded}
\begin{figure*}[t]
\subfloat[]{
\centering 
\includegraphics[width=0.35\linewidth]{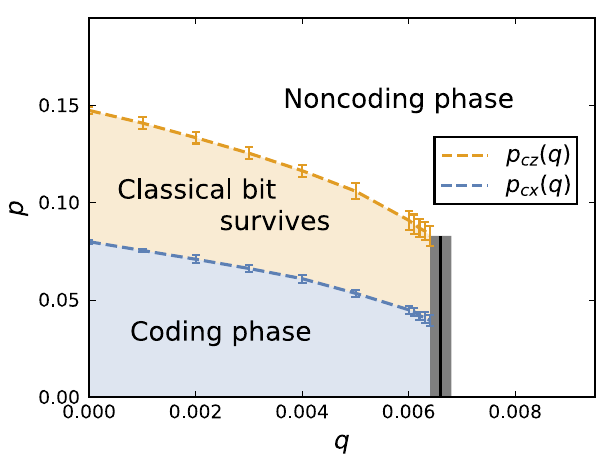}\label{fig:phase-unheralded}}
\subfloat[]{
\includegraphics[width=0.3\linewidth]{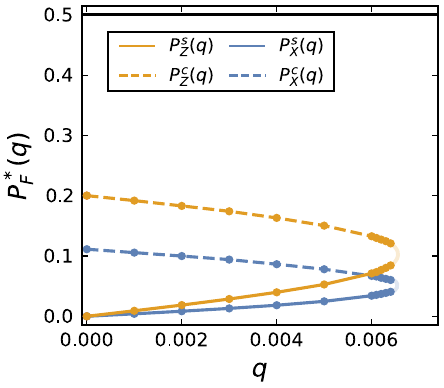}
\label{fig:fixed-points-unheralded}}
\subfloat[]{
\includegraphics[width=0.3\linewidth]{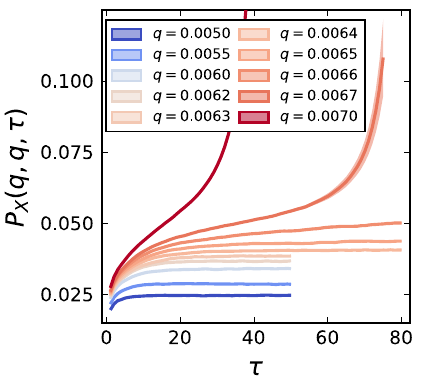}\label{fig:q-flows}
}
\caption{(a) Phase diagram of the Bell tree under unheralded bit and phase flips, at rate $p$ on the leaves and $q$ on the branches. (b) Logical failure probabilities at the stable (solid surves) and unstable (dashed curves) fixed points. Uncertainties are smaller than the markers. Lighter curves beyond $q=0.0064$ indicate the conjectured merging of the coding and critical fixed points. (c) Logical X failure probability on trees of depth $t=2\tau$, along the line $p=q$. Solid curves and light ribbons show the average and standard error, respectively, across 10-20 independent runs of population dynamics, with population size $M = 8 \times 10^5$ at $q=0.0067, 0.007$ and $M = 4\times 10^5$ at the remaining error rates.}
\end{figure*}

Turning to bulk errors, we focus on the Bell tree with independent bit and phase flips. The phase diagram shown in~\autoref{fig:phase-unheralded} exhibits three phases: a coding phase robust to both types of errors, a noncoding phase where no logical information survives, and an intermediate phase which preserves only classical information.\footnote{The basis of surviving logical information in the intermediate phase depends on the parity of the tree depth: in odd (even) depth trees the threshold for preparing states in the $Z$ ($X$) basis is higher. We consider even depths $t=2\tau$ for concreteness.}

\subsubsection{Phase diagram}

Consider state preparation in the $Z$ basis. The update rule~\autoref{eq:f-eta} associated with the ``copy node'', expressed in terms of the magnetization $m$, is
\begin{subequations}\label{eq:copy-f-m}
\begin{align}
f_0(m_1, m_2) &= \frac{(m_1 + m_2)(1-2q)}{1+m_1 m_2}, \\
z(m_1,m_2) &= 1 + m_1 m_2.
\end{align}
\end{subequations}
The ``delocalizer node'' induces the update
\begin{equation}\label{eq:deloc-f-m}
f_0(m_1, m_2) = m_1 m_2 (1-2q),\quad z(m_1,m_2)=1.
\end{equation}

For $q < q_c = 0.0066 \pm 0.0002$ in either state preparation basis, there are three fixed point distributions: a stable coding fixed point $Q^s(m)$, a noncoding fixed point  $Q^f(m)=\delta(m)$, and a critical point $Q^c(m)$. As in the previous section, the phase boundaries are identified by computing the failure probability as a function of depth, or equivalently, the first moment of the distribution $Q(|m|)$. The estimated failure probabilities at these fixed points are plotted in~\autoref{fig:fixed-points-unheralded}.

Which fixed point the system flows to depends upon the surface error rate $p$. Increasing $p$ tunes the system through a pair of 
phase transitions bounding the intermediate phase. For example, preparing a state in the $Z$ basis, which is vulnerable to logical $X$ errors, the logical failure probability $P_X$ converges to a finite value (solid blue curve) if $p<p_{cx}(q)$, and to $1/2$ if $p>p_{cx}(q)$, where $p_{cx}(q)$ is the blue phase boundary in~\autoref{fig:phase-unheralded}.

The critical point is more subtle. Note that the threshold $p_{cx}(q)$ is well below the critical failure probability $P_X^c(q)$, and likewise for $p_{cz}(q), P_Z^c(q)$. This is because the initial distribution at $p_{cx}(q)$, which is a sum of a small number of delta functions, is not a fixed point. To identify the location of the putative unstable fixed point, we initialize the population dynamics close to $p_{cx}(q)$. As a function of tree depth, the distribution first approaches the unstable fixed point, then diverges towards one of the stable fixed points. This manifests as a temporary plateau in $P_X(p,q,\tau)$, at $P_X^c(q)$.

Now consider the line $p=q$, which cuts through the coding phase and crosses a first-order phase boundary (black vertical line in~\autoref{fig:phase-unheralded}). \gs{Equal bulk and surface error rates arise naturally from the perspective of running a circuit forward in time, where our goal is to encode one logical qubit into $2^{2\tau}$ physical qubits, and each layer of concatenation is subject to errors at the same rate $q$.} As shown in~\autoref{fig:q-flows}, up to $q\approx 0.0066$ the logical failure probability converges toward the coding fixed point, approaching this fixed point more slowly as $q\nearrow q_c$. At $q>q_c$, $P_X(q,q,\tau)$, the long-time limit of $P_X(q,q,\tau)$ jumps to $1/2$.

Based on the estimated $q_c$ and the trends in the fixed points (determined up to $q=0.0064$) we conjecture that at $q=q_c$, the coding fixed point and unstable fixed point merge, forming a marginal fixed point which then disappears at larger $q$. In~\autoref{sect:bell}, we will confirm this picture for a simpler model of heralded errors.

\subsubsection{Comparison to Ising model}\label{sect:ising}
As noted above, the definition of the Bell tree is similar to the construction of a classical concatenated repetition code, except that on alternating layers, the ``copy nodes'' are swapped out for ``delocalizer nodes''. The model of bit flip errors on the concatenated repetition code tree is well understood, mapping to the ferromagnetic Ising model on a tree with free boundary conditions~\cite{Evans2000,Mossel2003,Mossel2004,Mezard2006}. A natural question, then, is how the Bell tree phase diagram compares to that of the Bethe lattice Ising model.\footnote{The statements here about the Ising model extend to the $Q$-state Potts model, corresponding to the concatenated repetition code on $Q$-ary digits subject to a $Q$-ary symmetric channel. If $Q\geq 5$, a first-order transition at a higher error rate preempts the continuous transition, but there is still a threshold below which the paramagnet is unstable~\cite{Mezard2006,Sly2011}. In the ensuing discussion, we stick with bits ($Q=2$) for simplicity.}

The essential difference is that concatenated repetition codes have a threshold below which the noncoding phase (paramagnet) is \textit{unstable}: an infinitesimal perturbation away from $m=0$ on the leaves sends the system toward a fixed point with $\langle |m| \rangle > 0$~\cite{Chayes1986,Ioffe1996}. Since the ``initial condition'' set by the error channel on each leaf is $m = \pm(1-2p)$, the instability of the paramagnet at $q=0$ has a simple coding interpretation: a repetition code can be reliably decoded all the way up to $p_c=1/2$, just by counting the number of up spins vs. down spins. In fact, the paramagnetic fixed point remains unstable up to a critical bulk error rate~\cite{Carlson1988,Carlson1990distr,Carlson1990a}, the aforementioned ``reconstruction threshold''. The transition is continuous, with $\langle |m|\rangle$ continuously going to zero at $q_c$~\cite{Chayes1986}. 

In contrast, in the Bell tree---and good quantum concatenated codes more generally---the noncoding fixed point is stable even in the absence of bulk noise. This is because $p_c=1/2$ is an upper bound on the threshold of any zero-rate classical code, which cannot be saturated by any CSS code that is simultaneously robust to phase flips~\cite{Steane1996,Calderbank1996}. The same reasoning applies to non-CSS quantum codes, and was implicit in the scaling form~\autoref{eq:scaling} positing a stable coding and noncoding phase separated by a critical point at $q=0$.
At nonzero $q$ in the Bell tree, while the $\langle |m|\rangle = 0$ trivial fixed point is unstable for $(1-2q)^2 > 1/2$ in the ``copy node'' recursion~\autoref{eq:copy-f-m} (corresponding to $q_c = \frac{1}{2}(1-\frac{1}{\sqrt{2}})$ in the classical repetition code~\cite{Chayes1986,Ioffe1996,Evans2000,Mezard2006}), introducing delocalizer nodes on alternating layers stabilizes the noncoding phase. As a consequence, the transition at $q=q_c$ is first-order, signaled by the jump in $P_F^*(q)$ at $q_c$. \gs{Nevertheless, our conjecture that the coding and critical fixed points become marginal at $q=q_c$ indicates that we can still associate a diverging correlation length, i.e. $\xi_c(q)\rightarrow \infty$, to this first-order transition.}
\subsubsection{Magnetization distribution}
\begin{figure}[t]
\includegraphics[width=\linewidth]{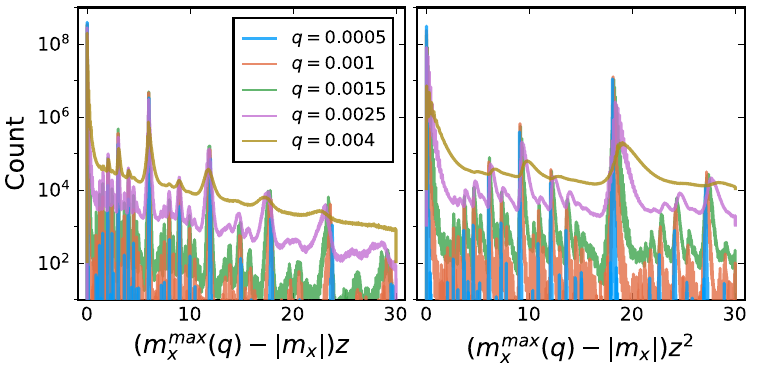}
\caption{Fixed point distribution of conditional magnetizations $|m_x|$, within the coding phase of the Bell tree subject to unheralded bit and phase flips at rate $q$ in the bulk of the tree. In each panel, the bin width is 0.01 in units rescaled by $1/z$ and $1/z^2$ respectively, where $z=q/(1-q)$. For each $q$, the total number of data points is $4\times 10^{8}$, of which $86.3\%$ - $99.7\%$ fall within the plotted interval.  \label{fig:mag-distr}}
\end{figure}

In the presence of bulk errors, the fixed point distribution of magnetizations in the coding phase no longer takes the simple form of~\autoref{eq:coding-surface}. Instead, the distribution consists of a continuum of delta functions, all the way down to $|m|=0$. Each delta function is associated with one or more syndromes and weighted according to the probability of observing that syndrome. The trivial syndrome, which plays the role of fully polarized, ``unfrustrated'' boundary conditions, has the largest $|m|=m_{max}$. In classical terms, the distributions of leaf spins resulting from preparing the logical states $\ket{\pm}$ (or $\ket{0}, \ket{1}$) are easiest to distinguish when the all-zero syndrome is observed, since the error-free states are true ground states with the largest energy barrier between them.  
We discuss the relationship between the threshold obtained by postselecting on trivial syndrome and a ``ferromagnetic'' transition in~\autoref{app:ferro} of the Supplement~\cite{supp-ref}. 

Flipping bits in the syndrome frustrates the boundary conditions and gives rise to more local minima, separated by smaller barriers, and thus with smaller $|m|$. These local minima appear as a hierarchy of peaks in~\autoref{fig:mag-distr}, which shows histograms of the fixed point distribution of $|m_x|$ at small $q=p$ deep within the coding phase. 
To  highlight the rich ``spin glass'' landscape across a range of $q$, we consider the distribution close to $m_{max}$ and plot the difference from this value, rescaled by a power of the ``Boltzmann factor'' $z=q/(1-q)$. The strongest first-order peaks are at $m_{max}-|m| = 6nz$, and the strongest second-order peaks are at $m_{max}-|m| = 9nz^2$, where $n$ is an integer. These features can be identified in a ``low-temperature'' (small $q$) expansion of the recursion relation~\cite{supp-ref}. As $q$ increases, the peaks become broader and less pronounced, while the continuous background becomes stronger.  {Prior studies of the distribution of magnetizations in the Ising Bethe lattice spin glass have revealed a similar complexity of features~\cite{Carlson1988,Carlson1990distr}, though not from the error correction perspective.}

\section{Heralded surface errors}\label{sect:herald}
In the previous section, we resorted to simulating the recursion relation~\autoref{eq:recursion} because the number of delta functions contained in the distribution proliferates. A situation in which we can do exact calculations is that of \textit{heralded} noise. Suppose each leaf of the tree is subject to the noise channel: 
\begin{widetext}
\begin{align}\label{eq:heralded-channel}
\mathcal{E}_{\vec{p}}(\rho) &= (1-p) \rho \otimes \ket{n}\bra{n} + \frac{p_a}{2}\mathbbm{1} \otimes \ket{a}\bra{a} + p_x \mathcal{N}_{(\frac{1}{2},0,0)}(\rho) \otimes \ket{x}\bra{x} + p_y \mathcal{N}_{(0,\frac{1}{2},0)}(\rho) \otimes \ket{y}\bra{y} + p_z \mathcal{N}_{(0,0,\frac{1}{2})}(\rho) \otimes \ket{z}\bra{z}
\end{align}
\end{widetext}
where now $\vec{p} = (p_x,p_y,p_z,p_a)$, $p=|\vec{p}|_1$, and $\mathcal{N}$ is the Pauli channel defined in~\autoref{eq:pauli-channel}.

In~\autoref{eq:heralded-channel}, the state $\sigma$ of the classical heralding register indicates whether a given qubit is left untouched ($\sigma=n$, occurs with probability $1-p$), replaced by a maximally mixed state ($\sigma=a$, occurs with probability $p_a$), or subject to a random $X$, $Y$, or $Z$ error ($\sigma=x,y,z$ respectively). The standard case of 
erasures corresponds to $\vec{p}=(0,0,0,p)$.

The heralding means that an efficient optimal decoder exists (even absent a tree structure)~\cite{Delfosse2020,Gullans21}, and for an optimal decoder for the given pattern of heralded errors, a given logical operator, say $X_L$, is either (1) completely lost---i.e., a logical $X$ error is undetectable, and it is equally likely that, upon returning to the code space, we will have implemented $I$ or $X_L$---or (2) completely preserved---i.e., we can perfectly decode without implementing a logical $X$ error. 

Thus, the distribution of probability vectors $Q(\bm{\eta})$ collapses onto a small number of delta functions: $(1,0,0,0)$ and permutations thereof correspond to completely preserving the quantum information; $(1/2,1/2,0,0)$ and permutations thereof correspond to preserving only a classical bit; and $(1/4,1/4,1/4,1/4)$ corresponds to all information being lost. The full distribution is then described by the coefficients $(n,x,z,y,a)$ of these delta function peaks, where $n$ ($a$) is the probability that no (all) logical information is lost, and $x, y, z$ are the probabilities that a logical $X$, $Y$ or $Z$ error is undetectable, but that a classical bit remains.

~\autoref{eq:recursion} then reduces to a set of ``flow equations'' for the probability distribution 
\begin{equation}\label{eq:pivec}
\vec{\pi}(t) = (n(t),x(t),z(t),y(t),a(t))
\end{equation}
where $t$ is the depth measured from the leaves. The parameters $\vec{p}$ of the channel~\autoref{eq:heralded-channel} applied to the leaves set the initial conditions of the flow:
\begin{equation}
\vec{\pi}(0) = (1-p, p_x, p_z, p_y, p_a).
\end{equation}
The probability vector~\autoref{eq:pivec} was introduced in Ref.~\cite{Ferte2024}, where the flow equations were derived for a model of CNOT gates and random single-qubit gates. Here, we generalize the method and demonstrate how the scaling form (\autoref{eq:scaling}) follows from the flow equations.



\subsection{Flow equations}

The flow equations can be obtained directly from the tensor~\autoref{eq:gate-enum} as follows. Let 
\begin{align}
\bm{\alpha} &= ({\alpha}_n,{\alpha}_x,{\alpha}_z,{\alpha}_y,{\alpha}_a) \notag \\
&= (\{0\}, \{0,1\}, \{0, 2\}, \{0, 3\}, \{0,1,2,3\}).
\end{align}
Then let $s(j,k)$ denote the set of logical operators obtained by "pulling back" the operators $\{P_a P_b\}_{a\in \bm{\alpha}_j, b\in\bm{\alpha}_k}$ through the gate~\cite{Ferte2024}:
\begin{equation}
s(j,k) = \{i: \exists \, \alpha \in \bm{\alpha}_j, \beta \in \bm{\alpha}_k \, : \, \tilde{R}^{\alpha\beta}_i = 1\}.
\end{equation}
Then
\begin{equation}\label{eq:flow-pi}
\pi_i' = \sum_{j, k} \pi_j \pi_k [\delta(s(j,k) = \vec{\alpha}_i]
\end{equation}
Since $\sum \pi_i = 1$, the flow is within a four-dimensional subspace, which we choose to parameterize by 
\begin{equation}
\tilde{\pi} = (n, x, z, a).
\end{equation}
When the errors occur only on the surface, every choice of gate preserves the $\mathbb{Z}_2$ symmetry $n \leftrightarrow a$, which one can interpret as the absence of a bulk ``field''~\cite{Ferte2024}. 

\subsection{Critical behavior}
In the vicinity of a fixed point $\tilde{\pi}^*$ of~\autoref{eq:flow-pi}, the linearized flow matrix is:
\begin{equation}
M_{ij}(\tilde{\pi}^*) = \frac{\partial \tilde{\pi}_i'}{\partial \tilde{\pi}_j}|_{\tilde{\pi}=\tilde{\pi}^*}.
\end{equation}
The eigenvalues of this matrix govern the flow toward or away from this fixed point. 

For any gate, two of the fixed points of~\autoref{eq:flow-pi} are immediate: the "coding phase" fixed point at $n=1$ and the "noncoding phase" fixed point at $a=1$. For the good quantum codes of interest, both of these fixed points are stable, with all zero eigenvalues.\footnote{This statement holds for codes that are robust against a finite density of both bit flips and phase flips. In the concatenated repetition code, and its perturbations in the ``quantum Darwinism phase'', both $n=1$ and $a=1$ are unstable toward a classical coding fixed point~\cite{Ferte2024}.} Thus, there must be at least one other fixed point separating these two phases, whose relevant eigenvalue(s) determine the effective correlation ``time'' $\xi_c$ in~\autoref{eq:scaling}. To wit, the failure probability of the optimal decoder applied to a depth $t$ code is a linear function of the flow variables:
\begin{equation}\label{eq:pfail-erasure}
P_F(\vec{p},t) = \frac{x(\vec{p},t)+y(\vec{p},t)+z(\vec{p},t)}{2} + \frac{3a(\vec{p},t)}{4}. 
\end{equation}

\begin{figure}
    \centering
    \includegraphics[width=\linewidth]{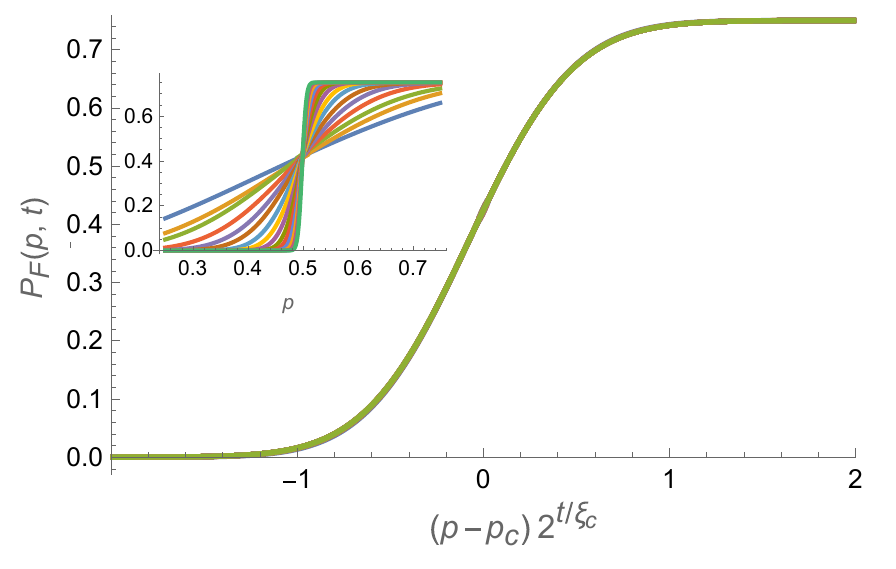}
    \caption{Failure probability under surface erasures in the optimal-distance self-concatenated code, for which $p_c=1/2$ and $\xi_c=2.602...$(\autoref{eq:best-xi}). Scaling collapse (main panel) shows depths $t=8$ to $t=40$, while the inset shows $t=2$ to $t=16$.}
    \label{fig:surface-e}
\end{figure}

As an example, consider the representative optimal-distance self-concatenated code [\autoref{eq:best-tensor}]. If each leaf is erased at rate $p$, $\pi$ flows to $n=1$ if $p<p_c=1/2$, $a=1$ if $p>p_c$, and the nontrivial fixed point
\begin{equation}\label{eq:pistar-152}
    \tilde{\pi}^* = (0.305193, 0.0784792, 0.268924, 0.305193)
\end{equation}
if $p=p_c$. The latter fixed point has one relevant eigenvector $(1, 0, 0, -1)/\sqrt{2}$, associated with an odd perturbation of the initial condition $\tilde{\pi}=(1/2-\delta,0,0,1/2+\delta)$, where $\delta = p-p_c$. Thus from the eigenvalue $\lambda =1.30519...$, we obtain
\begin{equation}\label{eq:best-xi}
    \xi_c = \frac{1}{\log_2(\lambda)}=2.60328...
\end{equation}
A scaling collapse is shown in~\autoref{fig:surface-e}.

As in the unheralded models, the correlation ``time'' $\xi_c$ depends on the gate, but for all gates considered, under i.i.d. surface errors of the form~\autoref{eq:heralded-channel}, $\xi_c > 2$.  Therefore, in some sense the tree structure amplifies the randomness near the critical point, producing more finite-size broadening of the transition than would naively be expected due to the typical variations $\sim N^{1/2}$ of the number of erasures on the $N$ leaves of the tree. To probe this broadening, we also consider a 
model in which the erasures are placed on the leaves in a nonrandom, balanced pattern. This sharpens the transition to a width of $1$ erasure ($\xi_c=1$), and 
shifts it to an erasure fraction different from that for randomly placed erasures~\cite{supp-ref}.  Thus, in the absence of bulk errors, the finite-size scaling of the rounding of the transition is set by the quenched randomness of the surface errors, and by removing or reducing that quenched randomness we can make the transition sharper, in this case all the way down to of order a single error.

\section{Heralded errors in the Bell tree}\label{sect:bell}
Next we consider the phase diagram  as a function of two error rates: $p$ applied to each qubit on a leaf of the tree, and $q$ applied to the qubit on each link between two gates within the tree. 

Whereas introducing unheralded errors on the links is like turning on finite temperature, we can view the heralded model as remaining at zero temperature but removing some of the interactions (at heralded locations). When each node of the tree is just a copy node---i.e., when the tree encodes a classical repetition code---heralded bit flips on the branches therefore map directly onto percolation, \gs{which has a continuous transition at $q_c=1/2$ on a binary tree~\cite{Grimmett1999}}. Here we consider the next simplest model, that of the Bell tree, and compare the resulting phase diagram to the one obtained under unheralded errors (\autoref{fig:phase-unheralded}). Exploiting the relative simplicity of the model, we analytically examine the critical behavior and introduce a quantity, the conditional code distance, to characterize the evolving code conditioned on the heralded pattern.
\subsection{Phase diagram}
\begin{figure}[t]
\includegraphics[width=\linewidth]{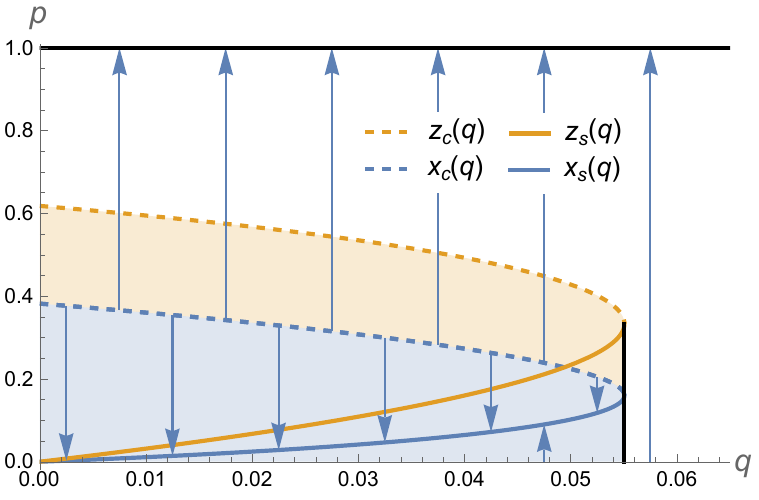}
\caption{Phase diagram and fixed points of the Bell tree with heralded bit and phase flips at rate $p$ on the surface and rate $q$ in the bulk. Blue shaded region is the coding phase, orange shaded region is where only a classical bit survives, and unshaded region is the noncoding phase. Vertical black line is the boundary between the coding phases and noncoding phase at $q=q_c$.
Blue arrows indicate the direction of the flow of $x_0$ under~\autoref{eq:x0-flow}.\label{fig:phase}}
\end{figure}

Consider a noise model of heralded bit flips and phase flips, i.e. the channels $\mathcal{E}_{(p,0,0,0)} \circ \mathcal{E}_{(0,0,p,0)}$ and $\mathcal{E}_{(q,0,0,0)}\circ \mathcal{E}_{(0,0,q,0)}$.  This simplifies the flow equations to two single-parameter flows $x_0, z_0$, where $x_0$ is the probability that an $X$ logical error is undetectable, 
irrespective of whether or not a $Z$ logical error has occurred, and {\it vice versa}. The phase diagram shown in~\autoref{fig:phase} resembles the phase diagram under (unheralded) bit and phase flips, but now the fixed point distribution is fully characterized by its first moment, i.e. the average logical failure probability of the decoder ($x_0/2$, $z_0/2$) uniquely specifies the fixed point:
\begin{equation}\label{eq:Qstar-herald}
Q^*(|m_x|) = x_0 \delta(|m_x|) + (1-x_0) \delta(|m_x|-1)
\end{equation}
and likewise for $m_z$.

If the leaves of the tree are at $t=T$, then let $\tau=(T-t)/2$ be the depth measured from the leaves in units of two layers of the tree.  Let
\begin{equation}
f_{d,q}(x) = (1-q) x + q \delta_{d 0}.
\end{equation}
Then $x_0$ and $z_0$ obey the complementary
recursion relations\footnote{As in our analysis of unheralded errors, we focus on even layers: the roles of $x_0$ and $z_0$ would be swapped if we instead looked at odd layers.}:
\begin{subequations}\label{eq:flow}
\begin{align}
x_0(\tau+1) &= f_{0,q}(f_{0,q}(2 x_0(\tau) - x_0(\tau)^2)^2) \equiv g_q(x_0(\tau))\label{eq:x0-flow}\\
z_0(\tau+1) &= f_{0,q}\left(2f_{0,q}(x_0(\tau)^2)-f_{0,q}(x_0(\tau)^2)^2\right)\label{eq:z0-flow}.
\end{align}
\end{subequations}

For $q \leq q_c = 0.05505...$, ~\autoref{eq:x0-flow} has three fixed points: a coding phase stable fixed point at $x_0^* = x_s(q)$, a noncoding phase stable fixed point at $x_0^*=x_f=1$, and an unstable fixed point governing the phase transition at $x_0^* = x_c(q)$, shown in in~\autoref{fig:phase}.
Thus, for surface error rate $p<x_c(q)$, $x_0$ flows to the coding fixed point, where a logical X error remains detectable 
to $\tau\rightarrow \infty$ with probability $1-x_s(q)$.~\autoref{eq:z0-flow} has analogous fixed points, at $z_s(q)$, $z_f=1$, and $z_c(q)$ respectively. Comparing the two flows, the fact that $x_c(q) < z_c(q)$ means that, as in the unheralded model, if bit flips and phase flips occur with the same probability, there is an intermediate phase in which only a classical bit survives. At $q=q_c$, the coding fixed point and critical point for each flow merge into one marginal fixed point. This is the threshold of bulk error rate $q$ beyond which only the noncoding fixed points remain: no logical information, classical or quantum, can survive to infinite depth for $q>q_c$. 

\subsection{Comparison to unheralded errors}
A key qualitative difference between the heralded and unheralded phase diagrams (\autoref{fig:phase} and~\autoref{fig:phase-unheralded}) is the nature of the fixed point distribution. In the coding phase of the unheralded model, we found a rich hierarchy of magnetizations, conditioned on the syndrome (\autoref{fig:mag-distr}). In contrast, any heralded fixed point~\autoref{eq:Qstar-herald} contains at most two delta functions; physically, it is just a convex sum of the noncoding (``infinite temperature'') fixed point and the \textit{zero-temperature} coding fixed point (\autoref{eq:coding-surface}). The flow variable $x_0$ uniquely specifies the distribution by quantifying the relative contributions of the infinite- and zero-temperature fixed points.

One consequence of this is that the surface error thresholds $p_{cx}(q), p_{cz}(q)$ (boundaries of the intermediate phase) coincide precisely with the critical fixed points $x_c(q)$, $z_c(q)$ (dashed curves), again in contrast to the unheralded phase diagram where $p_{cx}(q) < P_X^c(q)$. 

Another point of comparison is the scaling behavior in the vicinity of the phase boundaries. Under unheralded errors, we expect the transition at $p_c(q)$ to broaden as $q$ increases, but \david{numerical methods for estimating} $\xi_c(q)$ can be unstable near $p_c$. {Similarly, the base of the exponential decay toward the coding fixed point roughly approaches 1 as $q\rightarrow q_c$ (see ~\autoref{app:lambda} of the Supplement), but this is only a numerical result.} The analytic tractability of the heralded noise model allows us to pin down these trends exactly, a topic to which we now turn.

\subsection{Critical behavior}
Since the flows of $x_0$ and $z_0$ are independent and have identical critical behavior, in the following we focus on the behavior of $x_0(\tau)$. At $p=q<q_c$, $x_0(\tau)$ converges exponentially toward $x_s(q)$, as
\begin{equation}\label{eq:w}
w(q,\tau) \equiv x_s(q) - x_0(\tau) \sim \lambda_s(q)^\tau=2^{-2\tau/\xi_s(q)}~,
\end{equation}
where 
\begin{equation}\label{eq:lambda}
\lambda_s(q) = g_q'(x)|_{x=x_s(q)}~,
\end{equation}
and $\xi_s(q)$ is the bulk correlation length along the time direction in this coding phase.

As $q\rightarrow q_c$, $\lambda_s(q)\rightarrow 1$, and at $q=q_c$, the exponential convergence is replaced with an algebraic decay:
\begin{equation}\label{eq:wc}
w(q,\tau) \sim 1/\tau.
\end{equation}

Thus, as $q\rightarrow q_c$, the correlation length $\xi_s(q)$ in the coding phase diverges, as does correlation length $\xi_c(q)$ in the vicinity of $x_c(q)$. The critical exponent $\nu$ describing this divergence is determined by expanding~\autoref{eq:lambda} to leading order in $(q_c-q)$:
\begin{equation}\label{eq:nu}
\xi_s(q)\sim \frac{1}{\log \lambda_s(q)} \sim (q_c - q)^{-1/2} \Rightarrow \nu=1/2.
\end{equation}

One way to interpret these results is as a zero-temperature interface pinning transition~\cite{Gullans21,Li2023,Lovas2023,Sommers2024}. The logical information is put in at the root of the tree, and the place where it is lost is a type of interface.\footnote{This interpretation is closely related to statistical mechanics mappings for the entanglement entropy and mutual information~\cite{Nahum2017op,Nahum2017entanglement,Jonay2018,Zhou2019,Zhou2020,Bao2020,Jian2020,Li2021,Li2023,Sang2023}; see~\autoref{app:mutual} of the Supplemental Material~\cite{supp-ref}.}

Let $L(\tau)$ denote the probability that the logical is lost by time $t=2\tau$, given that it survives to $t=2(\tau - 1)$:
\begin{equation}
L(\tau) = \frac{x_0(\tau)-x_0(\tau-1)}{1-x_0(\tau - 1)}.
\end{equation}
This is, roughly, the probability that an interface can be placed at $\tau$ at no ``cost''.

In the noncoding phase, $L(\tau)$ asymptotes to 1, since at late enough times, the interface can be placed anywhere. In the coding phase, $L(\tau)$ decays exponentially with $\tau$: either the logical survives (no interface) or it is lost early on, when it is most vulnerable (the interface in pinned near the root of the tree).  
Finally, at $q=q_c$, we infer from~\autoref{eq:w} that $L(\tau) \sim 1/\tau^2$, so the probability that the interface can be put in decays as a power law. 
Noting that for $p=q=q_c$ the logical has a nonzero probability of surviving to infinite time, this power-law probability of the interface being there sums to less than one.

\subsection{Conditional code distance}
\begin{figure}[t]
\includegraphics[width=0.8\linewidth]{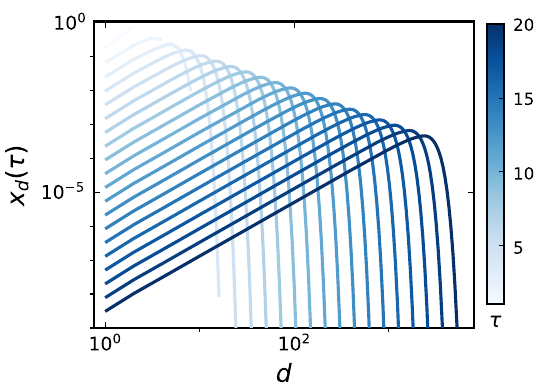}
\caption{Distribution of conditional distances in the Bell tree at $q=0.03$, $p=x_s(q)=0.0418...$, up to $\tau=20$ (darkest curve).
\label{fig:distr-dist}}
\end{figure}
Now we consider the "conditional code distance" of the instantaneous code after $t$ noisy encoding layers. This is the minimal number of additional qubits an adversary would need access to, at time $t$, in order to uncover the encoded information. If there are no errors in the bulk ($q=0$), then the conditional distance matches the (exponentially growing) distance $d(t)$. When we introduce bulk errors at locations $\vec{\sigma}$, each heralded error has the potential to turn some stabilizers into gauge qubits, which cannot be used to detect errors~\cite{Poulin2005,Kribs2006,Bacon2006}. Then $d(\vec{\sigma},t)$ is the distance of a $[[n,1,r,d(\vec{\sigma},t)]]$ \textit{subsystem} code where $r$ is the number of gauge qubits. Once the logical information is lost, we say $d(\vec{\sigma})=0$. 

Again, we consider heralded bit flips and just keep track of the conditional $X$ distance, $d_X(\vec{\sigma})$.

Let $x_d(p,q,\tau)$ denote the total probability, after $2\tau$ layers at bulk error rate $q$ and surface errors at rate $p$, of patterns $\sigma$ resulting in distance $d$, that is:
\begin{equation}
x_d(p,q,\tau) = \sum_{\vec{\sigma}} \mathbbm{P}_{p,q}(\vec{\sigma},t)\delta(d_X(\vec{\sigma},2\tau)=d).
\end{equation}

Then $x_0(p,q)$ is the loss probability with the phase diagram shown in~\autoref{fig:phase}. Our goal now is to understand the mechanism of this loss by studying how the full distribution of distances evolves in time.

Generalizing~\autoref{eq:x0-flow}, 
we obtain
\begin{align}
x_d(\tau + 1) = f_{d,q}(\sum_{i=0}^d \tilde{x}_d(\tau) \tilde{x}_{d-i}(\tau))
\end{align}
where
\begin{equation}
\tilde{x}_d(\tau) = f_{d,q}(x_d(\tau) [1 - x_d(\tau)- 2 \sum_{i=0}^{d-1} x_i(\tau)]).
\end{equation}
\autoref{fig:distr-dist} shows the distribution of $x_d$ within the coding phase, at $q=0.03$ and $x_0(0)=p=x_s(q)$. By this choice of initial condition, $x_0$ is independent of $\tau$, while the remaining weight of the distribution shifts to larger $d$ with increasing depth. The distribution has a power-law tail to small $d$, and the exponent of this power law is independent of $\tau$. {With increasing $q$, the distribution shifts toward smaller $d$ and the exponent of the power law decreases.}

As one way of characterizing this distribution, let's consider its first moment. Conditioned on survival, the average distance after an even number of layers is\footnote{$x_d(2\tau)$ is identically zero for $d>2^\tau$, since $d_X(\tau)=2^\tau$ in the absence of errors.}
\begin{equation}\label{eq:ave-dist}
d(p,q,\tau) = \frac{\sum_{i=0}^{2^\tau} d x_d(\tau)}{1 - x_0(\tau)}.
\end{equation}

\begin{figure*}[hbtp]
\subfloat[]{\includegraphics[width=0.34\linewidth]{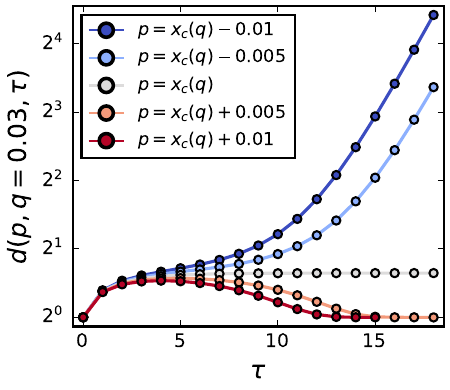}\label{fig:dist-means}}
\subfloat[]{
\includegraphics[width=0.36\linewidth]{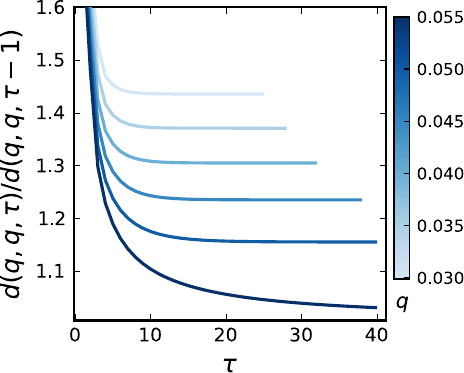}
\label{fig:dist-ratio}}
\subfloat[]{\includegraphics[width=0.28\linewidth]{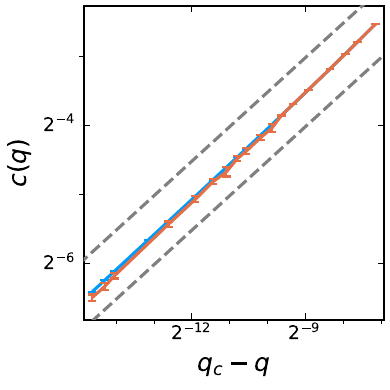}\label{fig:dist-scaling}}
\caption{(a) Average conditional code distance conditioned on survival (\autoref{eq:ave-dist}), at $q=0.03$, tuning $p$ through the critical point. (b) Ratio between the conditional code distance in consecutive time steps, at $p=q=0.03,0.035,0.04,0.045,0.05,0.055$ from light to dark. (c) Rate of exponential growth of $d(q,q,\tau)$ as a function of error rate $q$ [\autoref{eq:cq}] (blue data points), as well as the exponential growth of the quantity $d_{peak}(q,q,\tau)$ defined in~\autoref{eq:dpeak} of the Supplement~\cite{supp-ref} (orange). Gray dashed lines show $\sim (q_c-q)^{1/2}$.} 
\end{figure*}

There are three different trends of $d(p,q,\tau)$, depending on $p$ and $q$. Along the critical curve $(q < q_c, p=x_c(q))$ and in the noncoding phase, $d(p,q,\tau)$ saturates at $O(1)$ (\autoref{fig:dist-means})\footnote{Unconditioned on survival, i.e. only taking the numerator in~\autoref{eq:ave-dist}, the average distance of course converges to 0 in the noncoding phase.}. In the coding phase, $(q < q_c, p < x_c(q))$, the conditional distance grows exponentially (\autoref{fig:dist-ratio}:
\begin{equation}\label{eq:cq}
d(p,q,\tau) \sim \exp[c(q) \tau].
\end{equation}
As $q\nearrow q_c$, the rate of growth vanishes with the same exponent as the correlation length (\autoref{eq:nu}): $c(q) \sim \sqrt{q_c - q}$ (\autoref{fig:dist-scaling}).

The third trend is at $p=q=q_c$. Along $p=q$, running the circuit forward in time, the distance $d(\vec{\sigma},t)$ for a given realization of errors is in some ways analogous to a random walk with an absorbing wall at $d=0$: the walker can move left or right depending on which locations are heralded in that time step, and once the logical is fully erased, it cannot be recovered (hence $d=0$ is absorbing). However, there are some striking differences which make this analogy imperfect. 
In an unbiased random walk with an absorbing boundary~\cite{Fisher1984}, the survival probability decays as $\sim t^{-1/2}$ and the average distance from the wall, conditioned on survival, grows as $t^{1/2}$. In contrast, in the Bell tree model at $q=q_c$, the survival probability converges to a finite value $1-x_0^*$ (cf.~\autoref{eq:wc}) with the deviations decaying as $\sim 1/\tau$, and the average conditional distance grows as $d(q_c,q_c,\tau) \sim \tau^{1.2}$~\cite{supp-ref}. These discrepancies are partly related to how we define "time": in a single layer, the walker can move by up to $d(t)$ steps, so that at late times, the probability distribution of $d(\tau+1)$ given that $d(\tau)=d^*\gg 1$ becomes approximately Gaussian.

\section{Discussion}\label{sect:conclude}
In this work, we have investigated statistical mechanics models associated with quantum circuits on trees. Leveraging the tractability of tree tensor networks, we have identified classes of gates that produce high-performing quantum codes with transitions under i.i.d. surface and bulk errors. The tree structure induces a recursion relation for the distribution of logical probabilities, making the maximum likelihood decoding problem the quantum analog to spin glass models on trees~\cite{Mezard2006}.  When the errors are heralded, the recursion relation can be evaluated exactly, enabling analytic calculation of the phase diagram and many features of the code, such as the conditional code distance, as a function of bulk error rate $q$ and boundary rate $p$.

\gs{These statistical mechanics models are interesting in their own right, as models of spin glasses with a subtler phase diagram than the well-studied Bethe lattice Ising model. Notably, the quantumness of the code forces the bulk transition to be discontinuous, but with a divergent correlation length associated with the merging and annihilation of a pair of fixed points.}

\gs{From a quantum error correction perspective, it is also noteworthy that code concatenation with just two-qubit gates can produce high-performing codes, robust to errors both at the end and in the bulk of the encoding circuit. Past theoretical studies have mainly focused on constructions where the inner code itself can correct at least one error, necessitating $b \geq 5$~\cite{Poulin2006}. The encoding circuit for a $b>2$ inner code contains several two-qubit gates, each of which incurs some errors. Thus, our simple bulk error model, consisting of i.i.d. errors on each branch but no errors ``internal'' to the nodes, is a reasonable first approximation to the noise processes in actual binary tree circuits, but becomes less realistic for larger branching number.} 

\gs{Real quantum devices have errors not captured in our models, such as correlated multi-qubit Pauli errors and coherent noise. Nevertheless, the toy models analyzed in this work can guide understanding of coding phases in these devices. For example, while errors can never be perfectly heralded in practice, ``erasure conversion'' protocols such as those implemented on neutral atom platforms ~\cite{Wu2022,Sahay2023,Scholl2023,Ma2023} can convert all but a small fraction of errors to erasures. These protocols lead to significantly higher thresholds, and the corresponding toy model on trees --- that of i.i.d., partially heralded errors --- is amenable to both analytical (low-temperature expansions around the perfectly heralded fixed point) and numerical (population dynamics) techniques.}

\gs{In the main text, we focused on two concatenated codes: (1) a non-CSS self-concatenated code with optimal distance scaling, and (2) the Bell tree, a CSS code obtained by alternately concatenating bit flip and phase flip repetition codes. In the analysis of bulk errors, we chose to study the latter, with independent bit and phase flips, so that the spin glass phase is described by a pair of independent distributions $Q(m_x), Q(m_z)$. Under more general Pauli channels, the ``magnetization'' becomes a 3-component quantity less conducive to visualization, but in principle can be analyzed. {As an intermediate step, one can consider the ``qutrit Bell tree'' with independent errors, which has a 2-component magnetization.} A more formidable task is analyzing the recursion relations and resultant phases in non-stabilizer concatenated codes, i.e., trees composed of non-Clifford gates, in the spirit of recent studies of measurement-induced phase transitions in tree circuits~\cite{Nahum2021,Feng2023,Ferte2024haar,Feng2024}.}



Finally, all of these techniques start at the boundary and proceed inwards towards the root, whereas the quantum circuit runs in the opposite direction. We leave to future inquiry the question of how to understand the phase diagram from the latter point of view, evolving forward in time. Since the syndrome is only measured at the end, logical information is necessarily degraded with time (once a logical is lost, it cannot be recovered), so even in the coding phase, the failure probability at the fixed point is nonzero. In a forthcoming work, we drive the fixed-point failure probability to zero by introducing syndrome checks into the bulk, at the cost of losing this simple recursive structure~\cite{Sommers2025}.

\begin{acknowledgments} We acknowledge helpful conversations with Sarang Gopalakrishnan, Charles (ChunJun) Cao, Brad Lackey, Iman Marvian, Shiv Akshar Yadavalli, Vedika Khemani, Benedikt Placke, and Tibor Rakovszky. We also thank Siddhant Midha for comments on an earlier version of the manuscript. This research was supported in part by NSF QLCI grant OMA-2120757, including an Institute for Robust Quantum Simulation (RQS) seed grant. Numerical work was completed using computational resources managed and supported by Princeton Research Computing, a consortium of groups including the Princeton Institute for Computational Science and Engineering (PICSciE) and the Office of Information Technology's High Performance Computing Center and Visualization Laboratory at Princeton University.
\end{acknowledgments}

%% file: R-tensor.tikz
\begin{tikzpicture}
	\begin{pgfonlayer}{nodelayer}
		\node [style=U] (0) at (0, 0) {};
		\node [style=none] (1) at (-1, 1) {};
		\node [style=none] (2) at (1, 1) {};
		\node [style=none] (3) at (-1, -1) {};
		\node [style=none] (4) at (1, -1) {};
		\node [style=none] (5) at (-1.2, 1.2) {$\scriptstyle{\alpha}$};
		\node [style=none] (6) at (1.2, 1.2) {$\scriptstyle{\beta}$};
		\node [style=none] (7) at (1.2, -1.2) {$\scriptstyle{j}$};
		\node [style=none] (8) at (-1.2, -1.2) {$\scriptstyle{i}$};
	\end{pgfonlayer}
	\begin{pgfonlayer}{edgelayer}
		\draw [style=gray] (1.center) to (4.center);
		\draw [style=gray] (3.center) to (2.center);
	\end{pgfonlayer}
\end{tikzpicture}

%% file: stabilizer.tikz
\begin{tikzpicture}
	\begin{pgfonlayer}{nodelayer}
		\node [style=stabilizer] (0) at (0, -0.5) {$\alpha$};
		\node [style=none] (1) at (0, 1) {};
	\end{pgfonlayer}
	\begin{pgfonlayer}{edgelayer}
		\draw [style=gray] (1.center) to (0);
	\end{pgfonlayer}
\end{tikzpicture}

%% file: physical.tikz
\begin{tikzpicture}
	\begin{pgfonlayer}{nodelayer}
		\node [style=none] (0) at (0, -1) {};
		\node [style=physical] (1) at (0, 0.75) {};
	\end{pgfonlayer}
	\begin{pgfonlayer}{edgelayer}
		\draw [style=gray] (1) to (0.center);
	\end{pgfonlayer}
\end{tikzpicture}

%% file: error.tikz
\begin{tikzpicture}
	\begin{pgfonlayer}{nodelayer}
		\node [style=none] (0) at (0, 1) {};
		\node [style=none] (1) at (0, -1) {};
		\node [style=error] (2) at (0, 0) {};
		\node [style=none] (3) at (0, 0) {$\alpha$};
	\end{pgfonlayer}
	\begin{pgfonlayer}{edgelayer}
		\draw [style=gray] (0.center) to (1.center);
	\end{pgfonlayer}
\end{tikzpicture}

%% file: tree-enumerator.tikz
\begin{tikzpicture}
	\begin{pgfonlayer}{nodelayer}
		\node [style=U] (0) at (-3.5, 0) {};
		\node [style=U] (1) at (3.5, 0) {};
		\node [style=U] (2) at (0, -3) {};
		\node [style=U] (3) at (-5.5, 2.5) {};
		\node [style=U] (4) at (-1.5, 2.5) {};
		\node [style=U] (5) at (1.5, 2.5) {};
		\node [style=U] (6) at (5.5, 2.5) {};
		\node [style=stabilizer] (8) at (4.5, -0.75) {};
		\node [style=error] (9) at (-6, 3.5) {};
		\node [style=error] (10) at (-5, 3.5) {};
		\node [style=error] (11) at (-2, 3.5) {};
		\node [style=error] (12) at (-1, 3.5) {};
		\node [style=error] (13) at (1, 3.5) {};
		\node [style=error] (14) at (2, 3.5) {};
		\node [style=error] (15) at (5, 3.5) {};
		\node [style=error] (16) at (6, 3.5) {};
		\node [style=none] (17) at (4.5, -0.75) {$\scriptstyle{\alpha}$};
		\node [style=stabilizer] (18) at (-2.5, -0.75) {};
		\node [style=none] (19) at (-2.5, -0.75) {$\scriptstyle{\alpha}$};
		\node [style=stabilizer] (20) at (-4.5, 1.75) {};
		\node [style=stabilizer] (21) at (-0.5, 1.75) {};
		\node [style=stabilizer] (22) at (2.5, 1.75) {};
		\node [style=stabilizer] (23) at (6.5, 1.75) {};
		\node [style=physical] (24) at (-6.5, 4.5) {};
		\node [style=physical] (25) at (-4.5, 4.5) {};
		\node [style=physical] (26) at (-2.5, 4.5) {};
		\node [style=physical] (27) at (-0.5, 4.5) {};
		\node [style=physical] (28) at (0.5, 4.5) {};
		\node [style=physical] (29) at (2.5, 4.5) {};
		\node [style=physical] (30) at (4.5, 4.5) {};
		\node [style=physical] (31) at (6.5, 4.5) {};
		\node [style=none] (32) at (-4.5, 1.75) {$\scriptstyle{\alpha}$};
		\node [style=none] (33) at (-0.5, 1.75) {$\scriptstyle{\alpha}$};
		\node [style=none] (34) at (2.5, 1.75) {$\scriptstyle{\alpha}$};
		\node [style=none] (35) at (6.5, 1.75) {$\scriptstyle{\alpha}$};
		\node [style=none] (36) at (-1, -3.75) {};
		\node [style=stabilizer] (37) at (1, -3.75) {};
		\node [style=none] (38) at (1, -3.75) {$\scriptstyle{\alpha}$};
		\node [style=none] (39) at (-6, 3.5) {$\scriptstyle{0}$};
		\node [style=none] (40) at (-5, 3.5) {$\scriptstyle{0}$};
		\node [style=none] (41) at (-2, 3.5) {$\scriptstyle{0}$};
		\node [style=none] (42) at (-1, 3.5) {$\scriptstyle{0}$};
		\node [style=none] (43) at (1, 3.5) {$\scriptstyle{0}$};
		\node [style=none] (44) at (2, 3.5) {$\scriptstyle{0}$};
		\node [style=none] (45) at (5, 3.5) {$\scriptstyle{0}$};
		\node [style=none] (46) at (6, 3.5) {$\scriptstyle{0}$};
	\end{pgfonlayer}
	\begin{pgfonlayer}{edgelayer}
		\draw [style=gray, bend right, looseness=0.75] (0) to (2);
		\draw [style=gray, bend right, looseness=0.75] (2) to (1);
		\draw [style=gray, bend right=15, looseness=0.75] (5) to (1);
		\draw [style=gray, bend right=15, looseness=0.75] (1) to (6);
		\draw [style=gray, bend right=15, looseness=0.75] (3) to (0);
		\draw [style=gray, bend right=15, looseness=0.75] (0) to (4);
		\draw [style=gray] (1) to (8);
		\draw [style=gray] (0) to (19.center);
		\draw [style=gray] (3) to (20);
		\draw [style=gray] (4) to (21);
		\draw [style=gray] (5) to (22);
		\draw [style=gray] (6) to (23);
		\draw [style=gray] (24) to (3);
		\draw [style=gray] (25) to (3);
		\draw [style=gray] (26) to (4);
		\draw [style=gray] (4) to (27);
		\draw [style=gray] (5) to (28);
		\draw [style=gray] (29) to (5);
		\draw [style=gray] (30) to (6);
		\draw [style=gray] (6) to (31);
		\draw [style=thick] (2) to (36.center);
		\draw [style=gray] (2) to (38.center);
	\end{pgfonlayer}
\end{tikzpicture}

%% file: error-init.tikz
\begin{tikzpicture}
	\begin{pgfonlayer}{nodelayer}
		\node [style=physical] (0) at (0, 1) {};
		\node [style=none] (1) at (0, -1) {};
		\node [style=error] (2) at (0, 0) {};
		\node [style=none] (3) at (0, 0) {$\alpha$};
	\end{pgfonlayer}
	\begin{pgfonlayer}{edgelayer}
		\draw [style=gray] (0) to (1.center);
	\end{pgfonlayer}
\end{tikzpicture}

%% file: recursion-bulk.tikz
\begin{tikzpicture}
	\begin{pgfonlayer}{nodelayer}
		\node [style=none] (0) at (-1.25, 1.25) {};
		\node [style=none] (1) at (1.25, 1.25) {};
		\node [style=U] (4) at (0, 0) {};
		\node [style=bigU] (5) at (0, 0) {};
		\node [style=bulkerror] (6) at (0, -1.5) {};
		\node [style=none] (9) at (-1.5, 1.5) {$\bm{\eta}_1$};
		\node [style=none] (10) at (1.5, 1.5) {$\bm{\eta}_2$};
		\node [style=none] (11) at (0, 0) {$\tilde{R}$};
		\node [style=none] (12) at (0, -2.675) {$i$};
		\node [style=none] (13) at (0, -2.25) {};
		\node [style=none] (14) at (0, -1.5) {$\alpha$};
	\end{pgfonlayer}
	\begin{pgfonlayer}{edgelayer}
		\draw (5) to (6);
		\draw (0.center) to (5);
		\draw (5) to (1.center);
		\draw (6) to (13.center);
	\end{pgfonlayer}
\end{tikzpicture}

%% file: recursion.tikz
\begin{tikzpicture}
	\begin{pgfonlayer}{nodelayer}
		\node [style=none] (0) at (-1.25, 1.25) {};
		\node [style=none] (1) at (1.25, 1.25) {};
		\node [style=U] (4) at (0, 0) {};
		\node [style=bigU] (5) at (0, 0) {};
		\node [style=none] (6) at (0, -1.5) {};
		\node [style=none] (9) at (-1.5, 1.5) {$\bm{\eta}_1$};
		\node [style=none] (10) at (1.5, 1.5) {$\bm{\eta}_2$};
		\node [style=none] (11) at (0, 0) {$\tilde{R}$};
		\node [style=none] (12) at (0, -1.925) {$i$};
	\end{pgfonlayer}
	\begin{pgfonlayer}{edgelayer}
		\draw (5) to (6.center);
		\draw (0.center) to (5);
		\draw (5) to (1.center);
	\end{pgfonlayer}
\end{tikzpicture}

%% file: bibliography.bbl
%

%% file: supplement.tex
\title{Supplemental Information: Dynamically generated concatenated codes and their phase diagrams}
\maketitle
\onecolumngrid

The Supplement is organized as follows:
\begin{itemize}
    \item \autoref{app:bell} discusses several equivalent representations of the Bell tree.
    \item \autoref{app:weight-enum} reviews the weight enumerator formalism and uses it to derive recursion relations for the code distance. Different classes of operator spreading are also explored.
    \item \autoref{app:unheralded-methods} elaborates on methods for probing phase diagrams under unheralded noise.
 \item \autoref{app:unheralded-results} presents numerical results for a range of codes under surface depolarizing noise.
\item \autoref{app:unheralded-bell-bulk} provides additional details on the phase diagram of the Bell tree under unheralded bulk errors (\autoref{sect:bell-unheralded} of the main text), including a model with postselection.
    \item \autoref{app:surface-erasures} elaborates upon the analytical techniques used to study heralded errors and demonstrates them for a range of tree codes. We also present a model of deterministically placed surface errors.
    \item \autoref{app:bulk} provides more detail on the phase diagram of the Bell tree under heralded errors in the bulk (\autoref{sect:bell} of the main text).
\end{itemize}

\section{Representations of the Bell tree}\label{app:bell}
The Bell tree can be written in several equivalent ways which will be used throughout this Supplement. 

The first version alternates between two layers of encoding (the ``copy'' and ``delocalizer'' nodes in the main text). In odd layers, the stabilizer inputs are $Z$, and the unitary gate $U_o=\CNOT$. In even layers, the stabilizer inputs are $X$, and the unitary gate is 
\begin{equation}
U_e = \mathrm{NOTC} \equiv \ket{00}\bra{00} + \ket{10}\bra{10} + \ket{01}\bra{11}+ \ket{11}\bra{01}.
\end{equation}

\begin{figure}[hbtp]
\subfloat[]{
\centering
\input{tree-H.tikz}\label{fig:zx}}
\subfloat[]{
\includegraphics[width=0.4\linewidth]{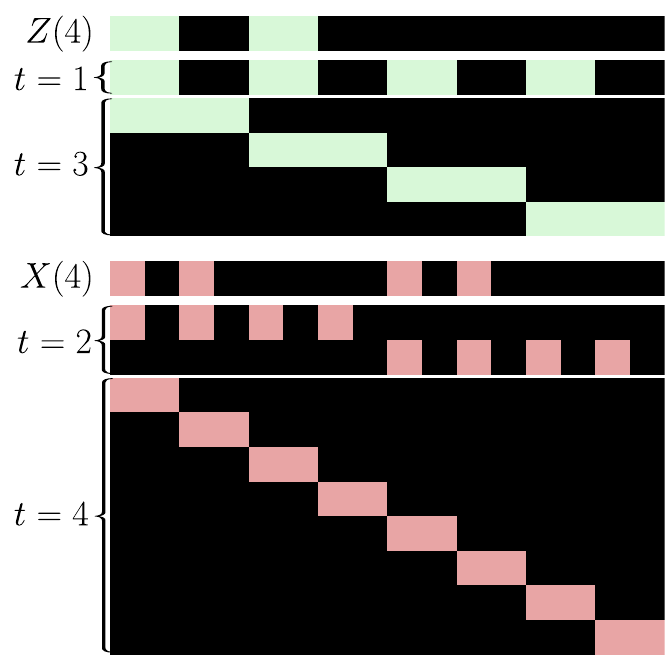}\label{fig:tableau}}
\caption{(a) The Bell tree as a ZX diagram, obtained by fusing stabilizer inputs into adjacent spiders~\cite{VandeWetering2020}. (b) Visual depiction of X- and Z-type stabilizer generators and logical operators in a depth 4 Bell tree. \gs{$Z(4)$ and $X(4)$ are logical operators, while the rows labeled $t=n, n=1,...,4$ are the stabilizer generators associated with inputs in layer $n$ of the tree.}}
\end{figure}

This circuit can be elegantly expressed as a phase-free ZX diagram~\cite{VandeWetering2020}, shown in~\autoref{fig:zx}: it is a binary tree alternating between three-legged "Z spiders" (green):
\begin{equation}
\input{green-spider.tikz} = \ket{00}\bra{0} + \ket{11}\bra{1}
\end{equation}
and three-legged "X spiders" (red):
\begin{equation}
\input{red-spider.tikz} = \ket{++}\bra{+} + \ket{--}\bra{-}.
\end{equation}

In this representation, the weight of an $X$ operator fed into the root node (thick line in~\autoref{fig:zx}) doubles in odd layers (since $U_o (XI) U_o^\dag = XX$) and stays the same in even layers (since $U_e (XI) U_e^\dag = XI$). We will denote this operator by $X(t)$. Conversely, the weight of a $Z(t)$ doubles in even layers ($U_e (ZI) U_e^\dag = ZZ$) and stays the same in odd layers ($U_o (ZI) U_o^\dag = ZI$).\footnote{These alternating layers of growth provide intuition behind the code distance~\autoref{eq:d2}, but we emphasize that the code distance is the minimum weight over \textit{all} logical representatives, (of which there are $2^{2^t-1}$ for each logical class). The Bell tree is a special example where the bare logical operators are also minimum-weight representatives.}~\autoref{fig:tableau} shows the stabilizer and logical generators at $t=4$.

The CNOT-NOTC representation lends itself to a classical version of the tensor $\tilde{R}$, with bond dimension $2$ rather than $4$. If the only possible errors are bit flips, we need not keep track of the $Z$ and $Y$ stabilizers, so to each CNOT (copy) node we can associate the tensor
\begin{equation}\label{eq:copy-R}
\tilde{R}_{\mathrm{copy}} = e_I^{II} + e_X^{XX}
\end{equation}
and to each NOTC (delocalizer) node, the tensor
\begin{equation}\label{eq:deloc-R}
\tilde{R}_{\mathrm{deloc}} = e_I^{II} + e_I^{XX} + e_X^{XI} + e_X^{IX}.
\end{equation}
For phase flip errors, the roles of the CNOT and NOTC nodes are swapped. From~\autoref{eq:copy-R} and~\autoref{eq:deloc-R}, combined with the definitions of $f_0(\{\bm{\eta}\})$ (\autoref{eq:f-eta}) and $m$ (\autoref{eq:mag}), we recover the recursion relations for the magnetization,~\autoref{eq:copy-f-m} and~\autoref{eq:deloc-f-m}, respectively. We also recover the single-parameter flow equations~\autoref{eq:flow} for heralded bit and phase flips.

A second way of representing the Bell tree, which puts it in the form of~\autoref{fig:tree} with identical gates and stabilizer inputs at each node, is to define $U=(H\otimes H) \CNOT$ and take each stabilizer input to be $Z$, yielding the $\tilde{R}$ tensor stated in~\autoref{eq:tildeR-bell}. Note that in this formulation, $X(t)$ alternates between all $Z$'s and all $X$'s. Pushing the Hadamards through the CNOT gate and canceling them in pairs of every two layers, we recover the CNOT-NOTC representation. 

Adding a Hadamard gate to the root of the tree, we obtain a third representation, matching the definition in Ref.~\cite{Yadavalli2023}, with $Z$ stabilizer inputs and gate $U = \CNOT (H \otimes \mathbbm{1})$. This representation gives meaning to the name "Bell tree": $U$ transforms a product state in the $Z$ basis into a Bell state, e.g., $U \ket{00}= (\ket{00} + \ket{11})/\sqrt{2}$.

\section{Tensor Enumerators for Code Distance}\label{app:weight-enum}
In this section we elaborate on the method for evaluating the distance of a tree code and derive recursion relations for three classes of high-performing codes composed of identical nodes. We also present results on tree codes with non-identical gates and discuss different classes of operator spreading.

\subsection{Identical nodes}
We claimed in the main text that the fastest-growing code distance in a tree code where each node is identical is $d(t) \propto 1.521^t$ [\autoref{eq:best-distance}]. To derive this scaling, we contract the tensor enumerator of a representative of this class [\autoref{eq:best-tensor}] with the vector $(w,x,z,y)$ on each of the leaves to obtain the complete vector enumerator of the local two-qubit code: 
\begin{equation}\label{eq:best-A}
\vec{A}_1^{(L)}(w,x,z,y) = (w^2+y^2, z^2 + x^2, 2wy, 2xz).
\end{equation}
The recursion relation~\autoref{eq:recursion-A} yields
\begin{equation}
    \mathbf{A}^{(L)}_{t+1}(w,x,z,y) = (I_t^2+Y_t^2) e_I + (Z_t^2 + X_t^2) e_X + 2 Y_t I_t e_Z + 2 X_t Z_t e_Y,
\end{equation}
which implies
\begin{equation}
d_X(t+1) = 2 \min(d_X(t), d_Z(t)), \qquad d_Z(t+1) = d_Y(t), \qquad d_Y(t+1) = d_X(t) + d_Z(t).
\end{equation}
From the initial conditions $d_X(1) = 2, d_Z(1) = 1, d_Y(1) = 2$, we find that $d_Z(t) < d_X(t), d_Y(t)$ for all $t$, and thus
\begin{equation}
d(t) = d_Z(t) = d(t-2) + 2 d(t-3)
\end{equation}
as stated in~\autoref{eq:best-distance}. $d(t)$ is sequence A159288 in Ref.~\cite{oeis}, whose asymptotic exponential growth $d(t) \propto a^t$ is obtained as the real solution to the polynomial equation
\begin{equation}\label{eq:poly-root}
a^3 - a - 2 = 0 \rightarrow a = 1.521...
\end{equation}

Turning to codes with distance $d(t) = 2^{\lfloor t/2 \rfloor}$ [\autoref{eq:d2}], first consider the Bell tree, expressed in the identical-gate form with $U=(H\otimes H) \CNOT$. From the associated tensor enumerator~\autoref{eq:tildeR-bell}, contracted with two copies of $(w,x,z,y)$, we obtain
\begin{equation}\label{eq:complete-enum-ex}
\mathbf{A}^{(L)}_1(w,x,z,y) = (w^2+x^2) e_I + (z^2 + y^2) e_X + 2xw e_Z + 2zy e_Y,
\end{equation}
which implies
\begin{equation}
d_X(t+1) = 2 \min(d_Y(t), d_Z(t)), \qquad d_Z(t+1) = d_X(t), \qquad d_Y(t+1) = d_Y(t) + d_Z(t).
\end{equation}
From the initial conditions $d_X(1)=2,d_Z(1)=1,d_Y(1)=2$, we find that  $d_Y(t)>d_Z(t)$ for all $t$, and thus 
\begin{equation}
d(t) = d_Z(t) = 2d(t-2),
\end{equation}
as stated in~\autoref{eq:d2}. Altogether,
\begin{equation}
d_X(t) = 2^{\lceil t/2 \rceil}, \quad d_Z(t) = 2^{\lfloor t/2 \rfloor}, \quad d_Y(t) = d_X(t) + d_Z(t).
\end{equation}

For completeness, we note that there is a second class with distance scaling~\autoref{eq:d2}, which is inequivalent to the Bell tree. A representative circuit in this class has the level-1 vector enumerator
\begin{equation}
    \mathbf{A}^{(L)}_1(w,x,z,y) = (xy + w^2) e_I + (xy + z^2) e_X + (x+y)w e_Z + (x+y)z e_Y
\end{equation}
and the associated distances
\begin{equation}
d_X(t) = 2^{\lceil t/2 \rceil}, \quad d_Z(t) = 2^{\lfloor t/2 \rfloor}, \quad d_Y(t) = d_X(t-1) + d_Z(t-1).
\end{equation}
This class of codes has inferior performance to the $d(t)\propto 1.521^t$ class, and unlike the Bell tree class, does not contain CSS codes. Thus, we have not conducted a thorough analysis of its phase diagrams.

\subsection{Non-identical nodes}\label{app:non-identical}

Can we do any better than $d(t) \propto 1.521^t$ by choosing different isometries at different nodes? To answer this question, recall that the weight enumerator of a binary tree code composed of not-necessarily-identical nodes is simply the output of the tensor network contraction in~\autoref{eq:tree-enum}.  

The computational complexity for contracting this tensor network is $O(\log(n))$ when all the nodes are identical, or $O(n)$ in the general case, where $n$ is the number of physical qubits~\cite{Cao2024expansion}. It should be noted, however, that this complexity only accounts for the matrix multiplications and does not account for the growing cost of multiplying and storing polynomials as the degree increases. Fortunately, if we are interested only in the distance and not in the full weight enumerator polynomial, we can avoid the symbolic representation of the polynomials altogether, and just construct a tensor network for $d_0 = 0, d_1 = d_X, d_2 = d_Z, d_3 = d_Y$. 

To wit, for a given three-legged tensor $\tilde{R}$ and ``logical index'' $j$, there are two pairs of ``physical indices'' $(\alpha_j, \beta_j), (\alpha'_j, \beta'_j)$ for which $\tilde{R}^{\alpha\beta}_j$ is nonzero. Now suppose we have prepared two stabilizer codes (say, two depth $t$ trees) each encoding one logical qubit, with vector enumerators $\vec{A}^{(L)}_{1}(u), \vec{A}^{(L)}_{2}(u)$. Then, if we contract the logical legs of codes 1 and 2 with the left and right outgoing legs of $\tilde{R}$, respectively,
\begin{equation}\label{eq:join}
    \input{join-tensor.tikz},
\end{equation}
the resulting code (e.g., a depth-$t+1$ tree) has
\begin{equation}\label{eq:distance-update}
d_j = \min(d_{\alpha_j,1} + d_{\beta_j,2}, d_{\alpha'_j,1} + d_{\beta'_j,2}).
\end{equation}

More formally, and to automate the process of computing $d(t)$ for arbitrary, not necessarily balanced, tree geometries, consider the tensor network with the replacements: $(w,x,z,y)\rightarrow (0,1,1,1)$ on the leaves, and $0\rightarrow\infty, 1\rightarrow 0$ in each $\tilde{R}$ tensor. Then if we contract this tensor network using the tropical algebra, wherein each "multiply" operation is $+$, and each "add" operation is $\min$~\cite{maclagan2021}, the output vector is $(d_0(t), d_1(t), d_2(t), d_3(t))$.

To find the optimal distance depth $t+1$ tree, we can search through all possible $\tilde{R}$ and all possible left and right depth $t$ subtrees, joining them via~\autoref{eq:join} and~\autoref{eq:distance-update}. This exhaustive search would quickly get out of hand, but for a simple observation: if there is a pair of depth $t$ trees $A$ and $B$ such that $d_{j,A} \leq d_{j,B}$ for all $j$, then for a given $\tilde{R}$ and tree $C$, the code distance of the depth $t+1$ tree obtained by contracting $A,C$ with $\tilde{R}$ is no larger than the distance of the tree obtained from $B,C$. Thus, we can remove all such trees $A$ from the pool of possible depth $t$ trees. 

\begin{figure}[t]
\subfloat[]{\scalebox{0.7}{\input{optimal-tree.tikz}}\label{fig:optimal-tree}}
\subfloat[]{\includegraphics[width=0.3\linewidth]{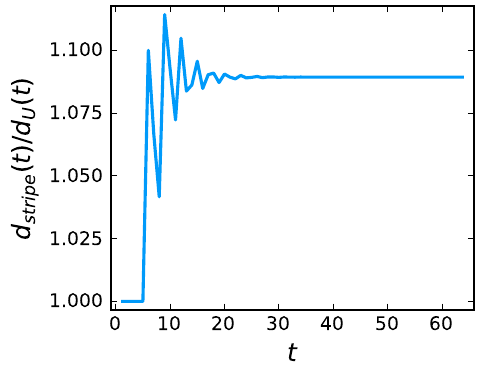}}
\caption{(a) Optimal-distance binary tree: $V = (\mathbbm{1} \otimes R_{(1,1,1)}[-\pi/3]) \iSWAP$ in the third layer, $U = (\mathbbm{1} \otimes R_X[\pi/2]) \iSWAP$ in all other layers, with $X$ stabilizer inputs on all gates. (b) Ratio of optimal distance (striped tree in (a)) to optimal self-concatenated distance (tree with $U$ at every node).\label{fig:stripe}}
\end{figure}

Applying this optimization up to depth $t=14$, we find that the optimal gate sequence (\autoref{fig:optimal-tree}) is in fact quite close to the self-concatenated code with the node defined in~\autoref{eq:best-tensor}: feed in an $X$ stabilizer input to each gate, and apply the gate $U = (\mathbbm{1} \otimes R_X[\pi/2]) \iSWAP$ at every node, \textit{except} in the layer $t=3$, where it is replaced by $V = (\mathbbm{1} \otimes R_{(1,1,1)}[-\pi/3]) \iSWAP$. The single layer of $V$ nodes gives a slight boost to the distance, which persists at all depths with a constant factor $\approx 1.08$ improvement (\autoref{fig:stripe}), but since it is only a "boundary effect" the code distance still grows asymptotically as $1.521^t$.

At depths $t>14$, we have verified explicitly that this striped tree achieves the highest distance among all trees with only two different node types, and conjecture that this is also optimal across any number of unique nodes. Some intuition about why this is the best we can do is offered in the next section.

\begin{figure}
\includegraphics[width=0.4\linewidth]{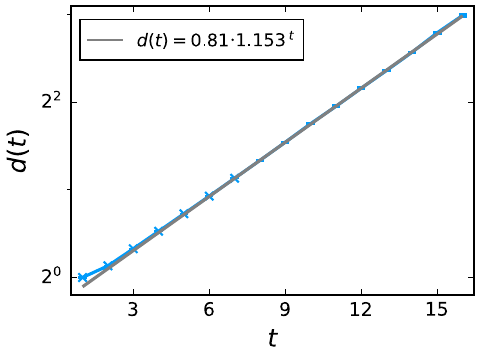}
\caption{\label{fig:random-dist} Average code distance of a tree code prepared by a binary tree circuit with random two-qubit Clifford gates. Each data point is averaged over 3000 samples. Data points marked with x's ($t=1$ to $t=7$) are excluded from the exponential fit.  
}
\end{figure}

We also examine the code distance of binary tree codes in which each gate is a random Clifford gate. Random Clifford circuits are the workhorse behind many theoretical developments in quantum error correction~\cite{Brown2012,Brown13}, and on a brickwork geometry, random Clifford circuits applied to product initial states generate high-threshold finite-rate codes in logarithmic depth~\cite{Gullans21}. But on the tree geometry, random gates do much worse than the nonrandom constructions we have found. On average, the distance is exponential in $t$, but with a much smaller base ($\approx 1.153$) (\autoref{fig:random-dist}).

\subsection{Transfer matrix for operator spreading}
The growth of the code distance is upper bounded by the growth of a "bare" logical operator $X(t)$, $Y(t)$, or $Z(t)$, i.e., the operator spreading of an $X$, $Y$, or $Z$ fed into the root node, while feeding in the identity on all the stabilizer inputs. This time-evolved operator is one choice of logical representative; multiplying it by some element of the stabilizer group can, in general, reduce its weight, which is why it is only an upper bound on $d_X, d_Y, d_Z$.

For a given gate $U$, we analyze the operator spreading on a tree with $U$ applied at every node by constructing a 3x3 transfer matrix $\mathbf{T}$, whose $(ij)$ entry is the number of $P_i$'s in $U (P_j I) U^\dag$. (Here $i, j$ run from $1$ to $3$.) The transfer matrix is a repackaging of the tensor $R_{i0}$\footnote{$\mathbf{T}$ is a permutation of the matrix $D_{xyz}$ in Prop. 3 of Ref.~\cite{Yadavalli2023}.}:
\begin{equation}
\mathbf{T}_{ij} = 2 R^{ii}_{j0} + \sum_{k\neq i} \left(R^{ik}_{j0} + R^{ki}_{j0}\right).
\end{equation}

The absolute value of the largest eigenvalue $\lambda_1$ quantifies the exponential growth of the operators. If $U$ is not an entangling gate (e.g. the product of single-site unitaries), all eigenvalues lie on the unit circle, as no growth is possible. In other cases (e.g., $U=\iSWAP$), $|\lambda_1|=1$ and is degenerate, and $\mathbf{T}$ is defective, allowing for linear growth. Otherwise, if all of the bare logicals grow exponentially, the corresponding gate becomes a candidate for making a "good code", \textit{with the appropriate choice of stabilizer input}. 

There are two classes of gates---where gates in the same class are related by a change of basis, and have the same transfer matrix eigenvalues---which, for the appropriate choice of stabilizer input, encode $d(t) \propto 1.52^t$ codes. All such codes are characterized by the same vector enumerator (\autoref{eq:best-A}) up to a permutation of $(x,z,y)$ and $e_X,e_Z,e_Y$, and thus perform identically under unbiased noise.

In one class, $\mathbf{T}$ has the eigenvalues $(-1,1,2)$. The leading eigenvector is associated with a \textit{bare} logical operator growing in weight as $2^t$, whereas $d_X,d_Z,d_Y$ are all constrained to grow as $1.52^t$. For this class of gates, only one choice of stabilizer input yields the optimal growth of code distance; the other two choices result in $d(t)=1$ for all $t$.

The second class of gates, which includes $U = (\mathbbm{1} \otimes R_X[\pi/2]) \iSWAP$, produce transfer matrices whose eigenvalues are the roots of the same cubic equation (\autoref{eq:poly-root}) obtained from the distance recursion relation. In fact, the bare logical operators $X(t), Y(t), Z(t)$ are also the lowest weight logical representatives, for the appropriate choice of stabilizer input. In other words, members of this class saturate the inequality between bare operator spreading and code distance, whereas gates with $\lambda_1>1.521...$ have a large gap between bare logical weight and code distance. Curiously, $1.521$ falls just short the leading eigenvalue of the \textit{average} transfer matrix, $1.6$, which is the rate of operator spreading in a random Clifford, or Haar-random, tree.

For any two-qubit Clifford gate, there is always at least one choice of stabilizer input which results in $d(t)=1$ for all $t$. This immediately implies that turning some of the stabilizer inputs into gauge qubits to make a subsystem code destroys the exponential growth of code distance.

\section{Tensor Enumerators for Decoding}\label{app:unheralded-methods}
In the main text we outlined a numerical method for evolving the distribution $Q(\bm{\eta})$ of normalized coset enumerators. In this section, we connect the tensor network language to the spacetime code, derive the recursion relation~\autoref{eq:recursion-i}, and present alternative numerical approaches. 
\subsection{Relation to the spacetime code}\label{app:spacetime}
The problem of decoding errors that appear in the bulk of a circuit can be phrased in terms of the spacetime code of Refs.~\cite{Bacon2017,Gottesman2022,Delfosse2023}.
For our purposes, the spacetime code is a subsystem code in which each possible error location corresponds to a qubit. To each gate, one assigns a set of gauge qubits---for the two-qubit gates in the tree, these gauge generators are precisely the stabilizer generators of the encoding state in~\autoref{eq:gate-enum}. 

Assign each spacetime qubit a label, $j$, and suppose the probability of applying the Pauli $P_{\alpha}$ at this location is $q_\alpha(j)$. Now suppose that the circuit suffers a fault $E$, which is a Pauli operator defined on the spacetime qubits. The ``effect'' of a fault, denoted eff$(E)$ in Ref.~\cite{Delfosse2023}, is the $N$-qubit Pauli string obtained by propagating the fault $E$ to the leaves of the tree. The decoder reads out the syndrome $s$ of eff$(E)$, and seeks to determine the most likely correction operator to apply to the leaves.

The logical class probabilities can be computed in a similar fashion to the case of end-of-circuit errors, but the four components of the coset enumerator [\autoref{eq:coset-enumerator}] now have a slightly different interpretation. For a canonical fault $E_s$, where the Pauli applied at spacetime location $j$ is $P_{\alpha_j}$, contract the tensor network with $F^{\alpha_j}$ at location $j$, evaluated at $\vec{f} = \vec{q}(j)$. For example, if the error rate is $\vec{p}$ on the leaves (yellow coloring) and $\vec{q}$ on the links (cyan coloring), then the tensor network for a depth 3 tree is\footnote{Note that we can always choose $E_s$ to have support only on the leaves, i.e. $\alpha_9=\alpha_{10}=...=\alpha_{15}=0$ in~\autoref{eq:coset-enumerator-bulk}.}
\begin{equation}\label{eq:coset-enumerator-bulk}
    \vec{A}^{(L, E_s)}_t = \scalebox{1.1}{\input{tree-enumerator-bulk.tikz}},
\end{equation}
where, since the model of choice has no errors on the stabilizer inputs, we have suppressed those legs.

The $e_L$ component of~\autoref{eq:coset-enumerator-bulk} enumerates all faults whose effect is stabilizer-equivalent to $\mathrm{eff}(E_s)\overline{L}$. In terms of the spacetime code, these faults are operators that are equivalent, under multiplication by spacetime stabilizer generators \textit{or} gauge generators, to $E_s \cdot \mathrm{spackle}(L)$, where $\mathrm{spackle}(L)$ is the spacetime operator whose support on the spacetime qubits at depth $t$ is the time-evolved logical operator $L(t)$~\cite{Bacon2017}. 
\subsection{Derivation of Eq.~\ref{eq:recursion-i}}
The distribution of logical class probabilities, $Q^{(t)}(\bm{\eta})$, is a distribution over all choices of $\{\alpha\}$ in~\autoref{eq:coset-enumerator-bulk}, where different fault patterns with the same syndrome give rise to the same $\bm{\eta}$ up to a logical permutation. This ``logical permutation''---i.e., a choice of the canonical fault $E_s$---manifests as a symmetry in~\autoref{eq:distr}: $Q^{(t)}(\bm{\eta}) = Q^{(t)}(\bm{\eta}')$ for any pair of probability vectors $\bm{\eta}, \bm{\eta}'$ where
\begin{equation}\label{eq:symmetry}
\exists \, j \, \mathrm{s.t.} \, \eta'_i = \eta_{i \pauliplus j} \, \forall i.
\end{equation}
This simplifies the recursion equation for $Q^{(t)}$ in the presence of bulk errors, because if the required symmetry is present at depth $t$,  we can choose $\alpha=0$ in~\autoref{eq:recursion} rather than taking a sum over $\alpha$. Plugging in the expression for $f_0(\bm{\eta})$ [\autoref{eq:f-eta}] and taking branching number $b=2$ for simplicity yields
\begin{equation}\label{eq:distr-2}
Q^{(t+1)}(\bm{\eta}) \propto \int \prod_{i=0}^3 \delta\left(\eta^i - \sum_{j_1,j_2,k} \eta_1^{j_1} \eta_2^{j_2} \tilde{R}^{j_1 j_2}_k q_{k \pauliplus i} / z(\bm{\eta}_1, \bm{\eta}_2)\right) z(\bm{\eta}_1, \bm{\eta}_2) dQ^{(t)}(\bm{\eta}_1) dQ^{(t)}(\bm{\eta}_2).
\end{equation}
Now consider the conditional distribution $Q^{(t+1)}_\alpha$. Plugging in the definition~\autoref{eq:distr-i}, we find
\begin{align}
    \tilde{Q}^{(t+1)}_\alpha(\bm{\eta}) &\propto \int \eta^\alpha \prod_{i=0}^3 \delta\left(\eta^i - \sum_{j_1,j_2,k} \eta_1^{j_1} \eta_2^{j_2} \tilde{R}^{j_1 j_2}_k q_{k \pauliplus i} / z(\bm{\eta}_1, \bm{\eta}_2)\right) z(\bm{\eta}_1, \bm{\eta}_2) dQ^{(t)}(\bm{\eta}_1) dQ^{(t)}(\bm{\eta}_2) \notag \\
    &= \int \sum_{j_1,j_2,k} \eta_1^{j_1} \eta_2^{j_2} \tilde{R}^{j_1 j_2}_k q_{k \pauliplus \alpha} \, \delta(\bm{\eta} - f_0(\bm{\eta})) dQ^{(t)}(\bm{\eta}_1) dQ^{(t)}(\bm{\eta}_2) \notag \\	
    & \propto \int \sum_{j_1,j_2,k} \tilde{R}^{j_1 j_2}_k q_{k \pauliplus \alpha} \delta(\bm{\eta} - f_0(\bm{\eta})) d\tilde{Q}_{j_1}^{(t)}(\bm{\eta}_1) d\tilde{Q}_{j_2}^{(t)}(\bm{\eta}_2) = \mathrm{\autoref{eq:recursion-i}}.
\end{align}

The factor of $z(\bm{\eta})$ in~\autoref{eq:distr-2}, which fortuitously cancels in the conditional distribution, can be traced back to the weighting of syndrome classes in~\autoref{eq:distr}. A ``quenched average'' ---treating each syndrome as equally likely---would not have this factor, making it simpler to simulate the unconditional distribution $Q^{(t)}$ directly. As noted in the main text, however, the quenched average is unnatural from the perspective of spin glasses, as the ``quenched average boundary condition'' neglects the correlations associated with exponentially many pure Gibbs states~\cite{Mezard2001,Mezard2006}.\footnote{The Ising model is an exceptional case where the computation using the quenched boundary yields the same transition temperature $T_{SG}$~\cite{Chayes1986}, but this robustness to the details of the boundary distribution does not hold in general.} From the error correction perspective, the necessity of $z(\bm{\eta})$ is even clearer, since in the absence of postselection, the average over syndromes must account for their relative likelihood of being observed.

\subsection{Numerical methods}
In the population dynamics method outlined in~\autoref{sect:population}, applied to a $b$-ary tree, each element of the population at depth $t$ is obtained by resampling $b$ elements from the populations at depth $t-1$. One danger of this method is that the resampling introduces correlations between elements of the population. In particular, close to a transition, random fluctuations away from the true distribution early in recursion can become amplified under resampling. \gs{In the vicinity of an unstable fixed point at $p_c$ with correlation length $\xi_c$, we have to resample of order $\xi_c$ times before the  system flows substantially away from this point. With each resampling, the effective $p$ is shifted randomly by an amount of order $M^{-1/2}$. From this heuristic argument, the population dynamics is unstable in an interval $\Delta p \propto (\xi_c/M)^{1/2}$ around $p_c$.} 

In practice, to check for instability at each error rate, we performed $N$ independent runs, with $N$ ranging from 10 to 20. Quantities were averaged within each run, yielding a set of $N$ estimators. In each figure, the solid curves and error bars/ribbons are, respectively, the average and standard error across the $N$ estimators. Instabilities in the method can also lead to systematic errors with population size $M$, so we verified that our results did not depend on $M$ for a sequence of large enough $M$. To generate~\autoref{fig:tree-opt}, we took $M$ ranging from $10^5$ to $2.8 \times 10^6$, using larger population sizes closer to the threshold.

\subsubsection{Continuous-time dynamics}
A single time step of the population dynamics method in the main text consists of $dM$ iterations of~\autoref{step1}-\ref{step2}, where $d=4$ in the general case and $d=2$ for independent bit/phase flips. In a ``continuous-time'' variation on this method, we keep steps 1-3 the same but modify~\autoref{step4} to 
\begin{enumerate}[label={\arabic*$'$.}]
\setcounter{enumi}{3}
\item Randomly remove one element of $S_i^{(t)}$ and replace it with $f(\{\bm{\eta}\})$ to produce $S_i^{(t+1)}$.
\end{enumerate}

This change of ``timing'' does not affect the fixed point distribution, but it does modify the transient. In particular, the unstable plateau is easier to detect in the continuous-time version, so we use this version to identify the unstable fixed points in~\autoref{fig:fixed-points} by initializing at $p$ close to $p_c(q)$.\footnote{To account for the alternating layers of gates in the CNOT-NOTC Bell tree, we first group together subtrees of depth 2 into one node,  obtaining a tree with identical nodes and branching number $b=4$.}

\subsubsection{Direct sampling}
An alternative method for sampling from $Q(\bm{\eta})$ is to directly sample error patterns and evaluate the corresponding coset enumerators. That is, we draw $M$ independent samples as follows:
\begin{enumerate}
\item At each possible error location $i$, set $E_i = I, X, Y, Z$ with probability $1-p,p_x,p_y,p_z$.
\item Evaluate the coset enumerator $\coset{E}$ by contracting the tree tensor network, inserting $F^{E_i}$ at each location $i$.
\item For sample $j$, let $\bm{\eta}_j = \coset{E}/|\coset{E}|_1$.
\end{enumerate}

This ``direct sampling'' method has a few advantages compared to the population dynamics method. From a conceptual standpoint, the connection to syndrome-based decoding is more clear. If the code suffers an error $E$, the decoder succeeds if $E$ belongs to the most likely logical class with that syndrome---i.e., if $I(E\Pi) > X(E\Pi), Y(E\Pi), Z(E\Pi)$. Direct sampling is also more numerically stable than population dynamics close to the threshold. Samples are independent by definition, providing an unbiased estimator of the failure probability arbitrarily close to $p_c$.

On the other hand, we use population dynamics for most of our numerics because it allows us to reach much larger depths. Direct sampling requires a full tensor network contraction for each error pattern and each depth, becoming computationally prohibitive beyond $t\approx 10$. While such shallow depths are sometimes sufficient for probing transitions under surface errors, where the convergence to the coding fixed point is doubly exponential in $t$, larger depths are essential for analyzing models with bulk noise, where the convergence is only a simple exponential.

\section{Additional numerics on depolarizing errors}\label{app:unheralded-results}
\subsection{Intermediate phase in the Bell tree}\label{app:unheralded-bell}
\begin{figure}[t]
\includegraphics[width=\linewidth]{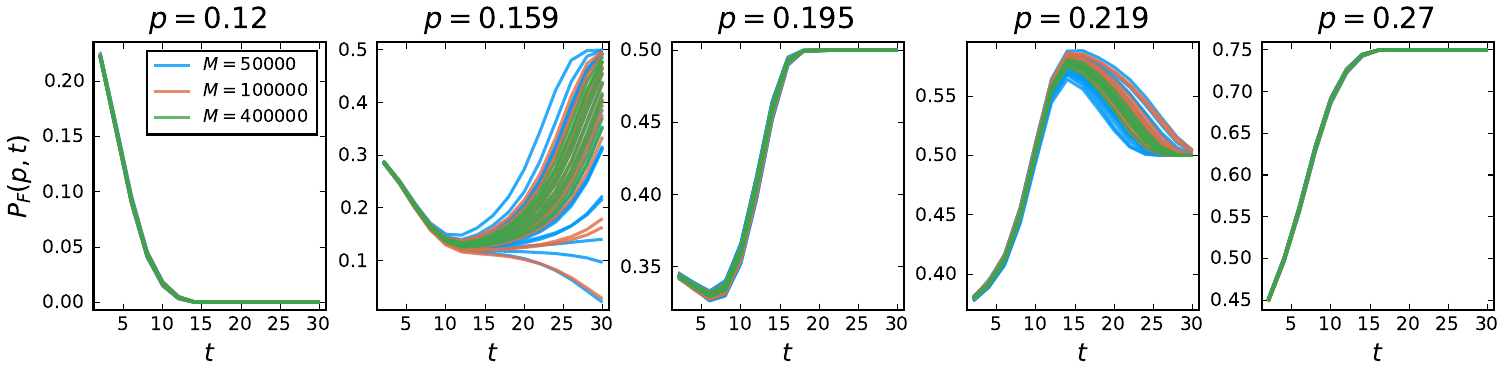}
\caption{Logical failure probability of the optimal decoder for the Bell tree subject to depolarizing noise at rate $p$ on the leaves, as a function of even depth $t$. In each panel, separate curves show independent runs (averaged within a run according to~\autoref{eq:pfail-sample}), with 20 runs at $M=5 \times 10^4$ (blue) and $M=10^5$ (orange), and 10 runs at $M=4\times 10^5$ (green).\label{fig:bell-panels}}
\end{figure}

One example where the behavior at shallow depths is deceiving is the Bell tree subject to depolarizing noise on the leaves. As shown in~\autoref{fig:bell-panels}, at small $p$ (within the coding phase) and large $p$ (within the noncoding phase), $P_F(p,t)$ is monotonically decreasing or increasing, respectively, as a function of even $t\geq 2$. For noise rates in the interval $[0.158, 0.22]$, however, $P_F(p,t)$ has an extremum at intermediate $t$, so the fixed point cannot be inferred just from the trend at low depths. Indeed, $[0.158,0.22]$ is the approximate extent of the intermediate phase where the average failure probability at the fixed point is $P_F(p,t)=1/2$, meaning that only a classical bit survives. The three phases---coding, classical, and noncoding---manifest as three plateaus in~\autoref{fig:bell-depol}, which shows the failure probability as a function of $p$. Note that the intermediate phase is roughly symmetric about the hashing bound, which is also the error rate where shallow depths have an approximate crossing.

\autoref{fig:bell-panels} also showcases the instability of the population dynamics method that arises near transitions. At $p=0.159$, in the vicinity of the lower threshold, different independent runs begin to diverge after reaching an unstable plateau between $t=10$ and $t=16$. At the smaller population sizes, some runs approach the classical coding fixed point $P_F=1/2$, while others approach the coding fixed point ($P_F=0$). Large variance across runs is also visible at $p=0.219$, near the higher threshold. Values of $(p,t)$ where the population dynamics method is unstable are therefore excluded from~\autoref{fig:bell-depol}.

\subsection{Comparison of codes}\label{app:subthreshold}
\begin{figure}[hbtp]
\centering
\subfloat[]{
\includegraphics[width=0.4\linewidth]{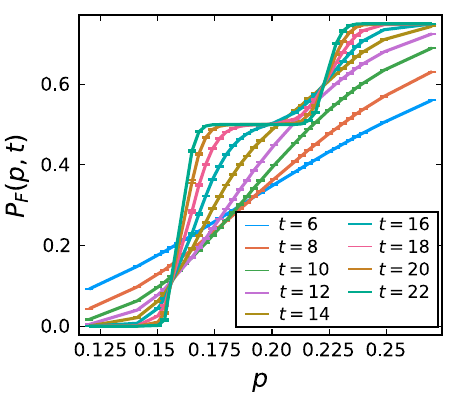}\label{fig:bell-depol}
}
\subfloat[]{
\includegraphics[width=0.45\linewidth]{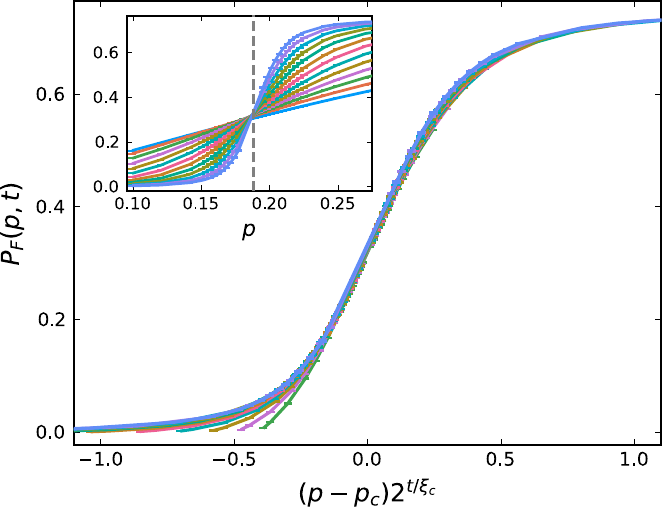}\label{fig:random-depol}}
\caption{Failure probability of (a) depth-$t$ Bell tree codes and (b) random binary tree codes, subject to depolarizing noise on the leaves. The main panel of (b) shows a scaling collapse with $p_c=0.188$, $\xi_c=7.1$, and $t$ ranging from 8 to 28. The inset shows the unscaled data with $t$ ranging from 4 to 28.}
\end{figure}
In~\autoref{app:weight-enum} of this Supplement, we presented two examples of tree codes in which not all gates are identical: binary trees composed of random gates, and a modification of the $d(t)\propto 1.52^t$ self-concatenated in which the third layer of the tree is populated with different gates (\autoref{fig:stripe}a). A natural coding question, then, is how these various codes compare for a fixed system size (tree depth) and error model.

\autoref{fig:random-depol} shows the logical failure probability for the random tree code at varying depths. While the threshold $p_c\approx 0.188$ is similar to that of the optimal-distance self-concatenated code, the transition is significantly broader, $\xi_c \approx 7.1$, and the suppression of the logical failure probability below threshold is much weaker.

\begin{figure}[t]
\centering
\subfloat[]{
\includegraphics[width=0.4\linewidth]{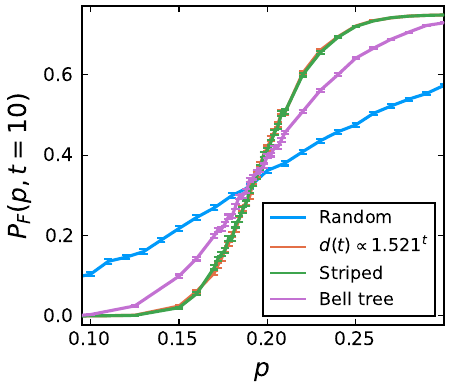}
\label{fig:comp-t10}
}
\subfloat[]{
\includegraphics[width=0.5\linewidth]{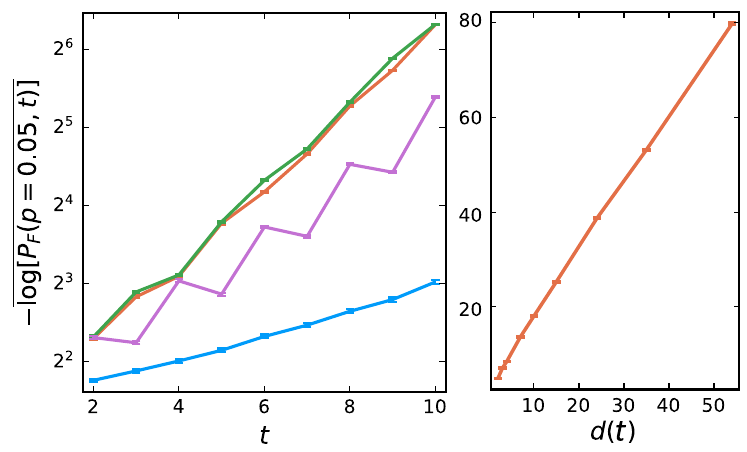}\label{fig:subthreshold}}
\caption{(a) Comparison of the logical failure probability as a function of $p$, the rate of depolarizing surface errors, applied to different tree codes of depth $t=10$. ``Striped'' (green) refers to the code generated by the tree in~\autoref{fig:stripe}(a), whose distance is a constant factor larger than the best self-concatenated code (orange). (b) Subthreshold scaling of the logical failure probability at $p=0.05$, as a function of depth $t$. The rightmost panel shows $-\log[P_F(p,t)]$ for the self-concatenated $d(t)\propto 1.52^t$ code vs. the distance $d(t)$.}
\end{figure}

In~\autoref{fig:comp-t10}, the logical failure probability of the four codes---random, self-concatenated $d(t)\propto 1.52^t$, ``striped'' optimal-distance tree, and the Bell tree---at depth $t=10$ is compared. As expected, the curve is sharper for the codes with higher distance.\footnote{Note that at this relatively shallow depth, the intermediate phase in the Bell tree is not discernible.}

As noted above, in the coding phase of models with bulk errors, the convergence to the coding fixed point is a simple exponential, with correlation time $\xi_s(q)$. When errors occur only on the surface, however, the correlation time is zero, and the decay of the failure probability becomes doubly exponential in $t$, as shown in the left panel of~\autoref{fig:subthreshold}. To wit, well below the threshold, the failure probability is expected to decay exponentially with the code distance
\begin{equation}\label{eq:subthreshold}
P_F(p,t) \propto e^{-c d(t)}.
\end{equation}
The right panel of~\autoref{fig:subthreshold}, which shows the linear trend in $-\log[P_F(p=0.05,t)]$ vs. $d(t)$ for the optimal-distance self-concatenated code, is consistent with this scaling.

\section{Unheralded errors in the bulk of the Bell tree}\label{app:unheralded-bell-bulk}
In this section we return to the error model of independent bit and phase flips in the Bell tree, with surface and bulk rates $p$ and $q$ respectively.
\subsection{Correlation length in the coding phase}\label{app:lambda}
\begin{figure}[hbtp]
\includegraphics[width=0.4\linewidth]{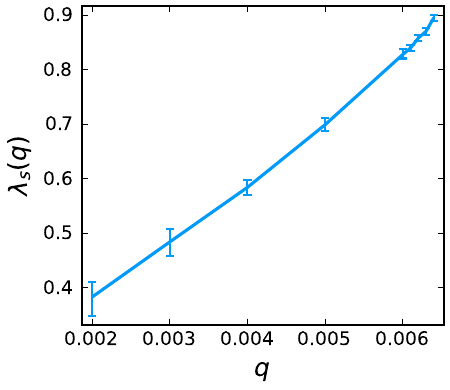}
\caption{Base of the exponential convergence toward the coding fixed point, $\lambda_s(q)$, in the coding phase of the Bell tree with unheralded noise.\label{fig:fits-unheralded}}
\end{figure}

At a fixed $q<q_c$, the logical X and Z failure probabilities converge to the respective fixed points $P_X^s(q)$, $P_Z^s(q)$, in the coding phase [\autoref{fig:fixed-points-unheralded}]. The correlation time $\xi_s(q)$ is a bulk property which does not depend on $p$ as long as $p<p_{cx}(q), p_{cz}(q)$ respectively. To obtain precise estimates, we perform fits of the form 
\begin{equation}
P_X(p,q,\tau) = P_X^s(q) + c(p,q) \lambda_{fit}(p,q)^\tau
\end{equation}
for a range of $p$ within the coding phase, and likewise for $P_Z$. The parameter $c(p,q)$ is negative for small $p$ (convergence to the fixed point from below) and positive for large $p$. The fit parameter $\lambda_{fit}(p,q)$ displays a weak upward drift with $p$, but remains within a fairly small interval for fixed $q$.~\autoref{fig:fits-unheralded} shows the trend in $\lambda_s(q)$, where at each $q$ we averaged over the fits from $P_X$ and $P_Z$ at several $p$. The increasing trend with $q$ is consistent with the expected behavior, $\lambda_s(q)\rightarrow 1$ as $q\rightarrow q_c$.

The error bars in \autoref{fig:fits-unheralded} are the standard error across $\lambda_{fit}(p,q)$ for different $p$. This only captures the statistical uncertainty, but at small $q$, where the rapid exponential convergence restricts the fit to a very small interval, the inferred $\lambda_s(q)$ also appears to be a systematic overestimate. {More accurate fits might be obtained by fitting observables such as the coherent information, which smoothly decays over a longer interval.}

\subsection{Postselection and ferromagnetic thresholds}\label{app:ferro}
The models considered in the main text have quenched disorder, in the pattern of heralded locations and/or the syndrome that is measured from the leaves. In the latter case, we can remove disorder by restricting to a single syndrome class, e.g., postselecting on observing the trivial syndrome. We can then take the ``canonical error'' $E_s$ to be the identity string, so the coset enumerator~\autoref{eq:coset-enumerator} reduces to the standard vector enumerator.\footnote{If there is a nonzero rate of bulk errors, this will be the vector enumerator of the spacetime subsystem code:~\autoref{eq:coset-enumerator-bulk} evaluated at $\alpha_j=0$ for all $j$.} Correspondingly, the associated stat mech Hamiltonian $H_{E_s}$~\cite{Chubb2021} has no quenched disorder.

We can then read off the postselected failure probability from~\autoref{eq:pfail-s}, as
\begin{equation}\label{eq:pfail-trivial}
P(\mathrm{failure \, | trivial \, syndrome}) = 1 - \frac{\max(\vec{A}_t^{(L)})}{B_t} = 1 - \frac{I_t}{B_t}
\end{equation}
where the second equality holds at low error rates since the logical identity class contains the leading-order contribution (ground states).

In practice, postselecting on the all-zero syndrome is unrealistic, since the probability of observing this syndrome vanishes with increasing system size. But this is a perfectly fine statistical mechanics model, and the absence of disorder makes it analytically tractable even when the errors are unheralded. For example, using a MacWilliams identity, which relates the A and B type enumerators, Ref.~\cite{Cao2024expansion} proves that under depolarizing errors, \textit{any} stabilizer code (encoding any number of logical qubits) has an ``error detection threshold'' of $p=1/2$, at which the failure probability~\autoref{eq:pfail-trivial} is independent of system size.

On trees, the trivial syndrome has an intuitive meaning as choosing ``unfrustrated'' boundary conditions, so that $z(\bm{\eta})$ is maximized at every step of the recursion. Consider first the classical repetition code. There the only outcomes which yield a trivial syndrome are the all-up or all-down bit strings: these are simply the fully polarized boundary conditions used to probe the paramagnet - ferromagnet transition in the Bethe lattice Ising model~\cite{Chayes1986,Baxter2007}. This transition occurs at a higher temperature than the spin glass transition at the reconstruction threshold: in the language of error correction, below $T_{FM}$, the trivial syndrome becomes decodable, but only below $T_{SG}$ does a \textit{typical} syndrome become decodable. Thus, the (in)stability of the paramagnetic fixed point discussed in~\autoref{sect:ising} must be defined with respect to a certain class of perturbations on the surface: stability towards the ferromagnet refers to perturbations in $\langle m \rangle$, while stability towards the spin glass refers to perturbations with $\langle m \rangle = 0, \langle |m| \rangle \neq 0$~\cite{Chayes1986}.

\begin{figure}[t]
\subfloat[]{
\includegraphics[width=0.45\linewidth]{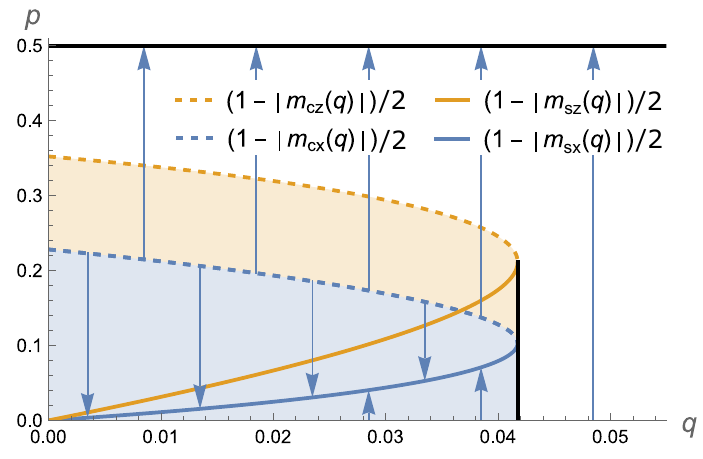}\label{fig:bell-ferro}}\hfill
\subfloat[]{
\includegraphics[width=0.45\linewidth]{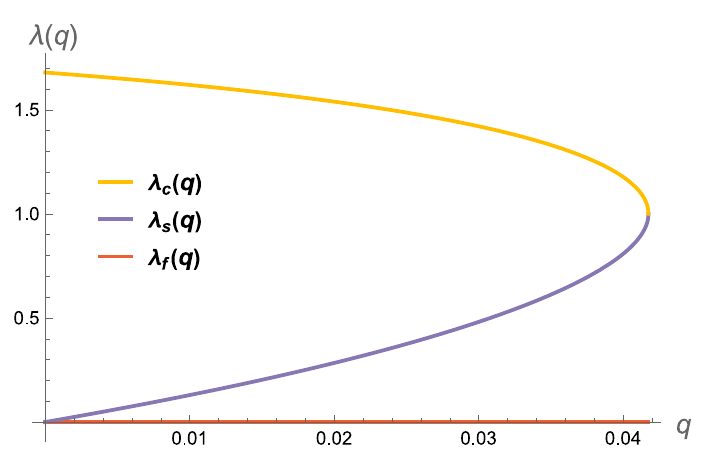}\label{fig:lambda-ferro}}
\caption{(a) Phase diagram and fixed points of the Bell tree with heralded bit and phase flips at rate $p$ on the surface and rate $q$ in the bulk, postselecting on the all zero syndrome. As in~\autoref{fig:phase-unheralded} and~\autoref{fig:phase} in the main text, the blue shaded region is the coding phase, the orange shaded region is where only classical information survives, and the unshaded region is the noncoding phase. Dashed and solid curves are unstable and stable fixed points of~\autoref{eq:ferro-m}. (b) Eigenvalues of the two-step flow equations~\autoref{eq:ferro-m} at the critical point (yellow curve), coding fixed point (purple curve), and noncoding fixed point (red line). (Compare~\autoref{fig:lambda}.)}
\end{figure}

Now consider our representative CSS model: the Bell tree with independent bit/phase flips. Plugging $m_1 = m_2$ into~\autoref{eq:copy-f-m} and~\autoref{eq:deloc-f-m}, the two-step flow equations for the logical X and Z ``magnetizations'' become
\begin{equation}\label{eq:ferro-m}
    m_x(\tau+1) = \frac{2m_x(\tau)^2 (1-2q)^2}{1 + m_x(\tau)^4 (1-2q)^2}, \qquad m_z(\tau+1) = \frac{4 m_z(\tau)^2 (1 - 2 q)^3}{(1 + m_z(\tau)^2)^2}.
\end{equation}
The fixed points of~\autoref{eq:ferro-m} are plotted in~\autoref{fig:bell-ferro}. Note that since the flow has been reduced to one variable, rather than a full distribution, the unstable fixed points once again coincide with the phase boundaries, unlike when we average over syndromes [\autoref{fig:phase-unheralded} and \autoref{fig:fixed-points}]. 

The phase diagram and critical behavior of this model is nearly identical to that of the heralded model studied in the main text, aside from minor quantitative details. For example, in the absence of bulk errors, $\lambda_c = 1.67857...$, corresponding to $\xi_c=2.6765$ in the sense of~\autoref{eq:scaling}. While slightly narrower than the (unpostselected) transition under heralded surface errors ($\xi_c=3.27056...$), this scaling is still broader than the $\xi_c = 2$ we might expect from the typical variations in the number of errors. 

Turning on bulk errors, the coding phase survives up to $q=q_{FM}$, where
\begin{equation}
27(1-2q_{FM})^6 -16 = 0 \rightarrow q_{FM} = 0.04175...
\end{equation}
At $q=q_{FM}$, the coding and critical fixed points merge into one marginal fixed point (\autoref{fig:lambda-ferro}). The correlation length in the coding phase has the same divergence as in the heralded model, $\log \lambda_s(q) \sim (q_{FM}-q)^{-1/2}$. 

Comparing to the unheralded error model without postselection, as one would expect, $q_{FM}$ is significantly larger than $q_c\approx 0.0066$, as are the surface thresholds $p_{cx}(q), p_{cz}(q)$ for $q<q_{FM}$. Thus, the spin glass phase occupies only a small subregion of the ``postselected ferromagnet'' phase. Within the spin glass phase, the ferromagnet defines the global minimum in the energy landscape, but the configuration space also contains an exponential number of local minima, whose structure was plotted in~\autoref{fig:mag-distr}. Having now studied the ferromagnetic transition, we can now precisely define the quantity $m^{max}(q)$ which appears in that figure:
\begin{equation}
m^{max}_{x,z}(q) = |m_{x,z}^s|(q).
\end{equation}
Again, in contrast to the Ising model/repetition code, the noncoding (paramagnetic) fixed point remains stable for all $q$: the coding phase (whether a ferromagnet or spin glass) can only be accessed at a sufficiently small $p$. 
\subsection{Magnetization distribution}
\begin{figure}[hbtp]
\includegraphics[width=0.6\linewidth]{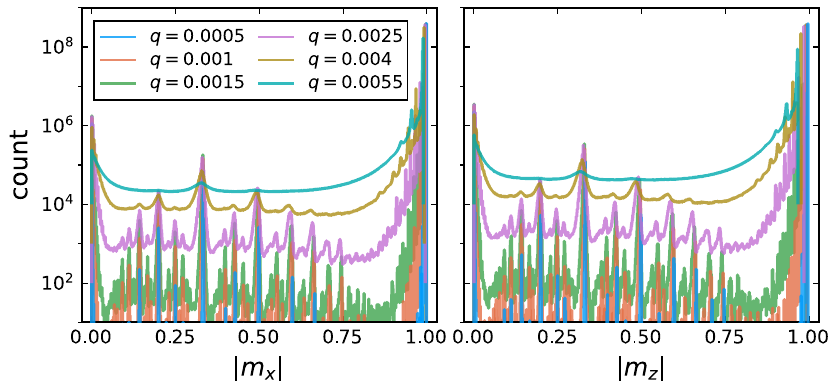}
\caption{\label{fig:mag-distr-full} Histogram of $|m_x|$ (left) and $|m_z|$ (right). Total count at each $q$ is $4\times 10^8$, aggregated from 100 independent runs with $M=4\times 10^6$, 2 populations each, $\tau=11,...,15$. Data are binned into bins of width 0.001.}
\end{figure}
Deep within the coding phase, \autoref{fig:mag-distr-full} shows the distributions of magnetizations $|m_x|$ and $|m_z|$ for the same values of $q$ as in the main text (\autoref{fig:mag-distr}), but now zoomed out to display the full range of $m$ and up to higher $q$. In addition to the series of peaks near $|m|=1$, there are strong contributions near $|m|=0$, $|m|=1/3$, as well as many other weaker peaks. 

To see where these peaks come from, consider keeping only the leading order in $z=q/(1-q)$ in the update rules~\autoref{eq:copy-f-m} and~\autoref{eq:deloc-f-m}. The recursion terminates at a finite number of delta function peaks, so that the leading-order distributions at even depths are
\begin{subequations}
\begin{align}
Q^s(|m_x|) &= (1 - 30z)\delta(|m_x|-(1-2z)) + 24z \delta(|m_x|-(1-8z)) + 6z \delta(|m_x|)\label{eq:mx} \\
Q^s(|m_z|) &= (1 - 60z)\delta(|m_z|-(1-6z)) + 48z \delta(|m_z|-(1-12z)) + 12z \delta(|m_z|) \label{eq:mz}.
\end{align}
\end{subequations}
The copy rule (\autoref{eq:copy-f-m}) transforms $Q^s(|m_z|)$ to $Q^s(|m_x|)$, while the delocalizer rule (\autoref{eq:deloc-f-m}) transforms $Q^s(|m_x|)$ to $Q^s(|m_z|)$. The peaks with the largest weight coincide to leading order with the ``ferromagnetic'' fixed points shown in~\autoref{fig:bell-ferro}:
\begin{equation}
    m^{max}_x = 1 - 2z - O(z^2), \quad m^{max}_z = 1 - 6z - O(z^2).
\end{equation}

Expanding to second order, the recursion does not converge with a finite number of terms. However, several notable features appear after just a few iterations. The peak at $m^{max}$ splits into $m^{max}(q)-18nz^2$ where $n$ is a non-negative integer. Additional peaks with weight $O(z^2)$ appear at $m^{max}(q) - 12 z -O(z^2)$ and at $1/3$.

\section{Heralded surface errors}\label{app:surface-erasures}
In this section, we delve deeper into the phase diagrams under heralded surface errors for the representative non-random tree in the main text (\autoref{fig:surface-e}), as well as the Bell tree and random binary trees. On the Bell tree, we also construct a model without disorder for which the transition is narrowed to a width of one error.

\subsection{General theory}
One way to interpret the heralded error channel (\autoref{eq:heralded-channel}) is as follows: with probability $p_a$, an eavesdropper (E) is given full access to the physical qubit, while with probability $p_x,p_y,p_z$, E is only allowed to measure in the $X$, $Y$, or $Z$ basis respectively (and thus can dephase the qubit in that basis). From this perspective, $x(t)=1$ means that, given the eavesdropping pattern on the leaves of a depth $t$ tree, E can recover $X_L$ (but not $Y_L$ or $Z_L$). The tree structure induces a flow of the initial vector $\pi(0) = (n=1-|\mathbf{p}|, x=p_x,z=p_z,y=p_y,a=p_a)$ backwards from the leaves. By generalizing beyond Ref.~\cite{Ferte2024}, which considered surface erasures ($\pi_0 = (1-p,0,0,0,p)$), we can access parts of the phase diagram that ordinary erasures would never flow to.

In the main text, we plotted the failure probability of an optimal decoder given this noise model [\autoref{eq:pfail-erasure}]. A related quantifier of the ability to decode is the average mutual information $I(R:E)$ between the initial reference $R$ and the eavesdropper/environment $E$:
\begin{equation}\label{eq:mutual}
I(R:E) = S(\rho_R) + S(\rho_E) - S(\rho_{RE}) = x + y + z + 2 a = 1 + a - n.
\end{equation}
The mutual information is a natural quantity to work with because it is odd under reflection $n \leftrightarrow a$, which is a symmetry of the gates.

From the decoding perspective, each bit of mutual information lost to the environment degrades the recovery probability by a factor of $1/2$: an optimal decoder can perfectly recover the information fed into the root if $I(R:E)=0$; can perfectly recover only a classical bit (but must randomly guess half of the qubit) if $I(R:E)=1$; and makes a fully random guess if $I(R:E)=0$. This leads to~\autoref{eq:pfail-erasure} for the average failure probability.

\subsection{Optimal-distance self-concatenated code}
From the tensor enumerator of our representative self-concatenated $d(t)\propto 1.52^t$ code [\autoref{eq:best-tensor}], we obtain the flow equation:
\begin{align}\label{eq:best-flow}
n' &= n(n + 2x + 2z), \quad
x' = x^2 + z^2, \quad 
z' = y(2-y)+2na \notag \\
y' &= 2xz, \quad 
a' = a(a + 2x + 2z).
\end{align}

In the main text, we identified three fixed points of~\autoref{eq:best-flow}, reached from ordinary erasures at rate $p$: stable fixed points at $n=1$ and $a=1$, and a critical point at $p_c=1/2$. The full phase diagram also includes several multicritical points and length-2 limit cycles which are fixed points of  the two-step flow equations. This richer phase diagram is accessed by applying biased heralded errors.

Since $d(t)\propto 1.521^t$ is the most favorable distance scaling of all binary tree codes, it is worth comparing the transition to more traditional concatenated codes, which correspond to trees with higher branching number. As an example, take the local code to be the five-qubit (perfect) code, the smallest code that can correct any single-qubit error ($d=3$)~\cite{ECZoo}. At concatenation depth $t$, the resulting code has $5^t$ qubits and distance $3^t$. Tuning the erasure rate $p$ on the leaves takes the system through an unstable fixed point at $(n,a)=(1/2,1/2)$, i.e. the initial condition $p=1/2$ is itself a fixed point. This point is multicritical, with a two-fold degenerate leading eigenvalue of $\lambda=15/8$. This eigenvalue corresponds to $\xi_c = 1/\log_5(15/8) = 2.560...$, meaning that as a function of the number of qubits $N$, the transition is slightly sharper than in the binary tree codes.

\subsection{Random gates}\label{app:herald-random}
Next we consider the phase diagram under heralded errors of random binary tree codes, i.e., the ensemble of binary tree codes in which each node is a random two-qubit Clifford gate, with average distance scaling shown in~\autoref{fig:random-dist}. The flow equations are:
\begin{align}
n'= n(6-n-4a)/5, \quad a'=a(6-a-4n)/5, \quad x'=z'=y'=(5+a(a-6)+n(n-6)+8na)/15.
\end{align}
The random ensemble has a permutation symmetry of $x, y,$ and $z$, so the fixed points lie in the two-dimensional subspace of unbroken permutation symmetry, parameterized by $(n,a)$.\footnote{This should be contrasted with the random ensemble used in~\cite{Ferte2024}, in which each gate is a CNOT dressed by random \textit{single-qubit} gates: then the $\mathcal{S}_3$ symmetry is broken to a $\mathbb{Z}_2$ symmetry between $x$ and $y$, although the flows projected onto $(n,a)$ are qualitatively similar to ours.}
\begin{figure}[t]
\centering
\includegraphics[width=0.4\linewidth]{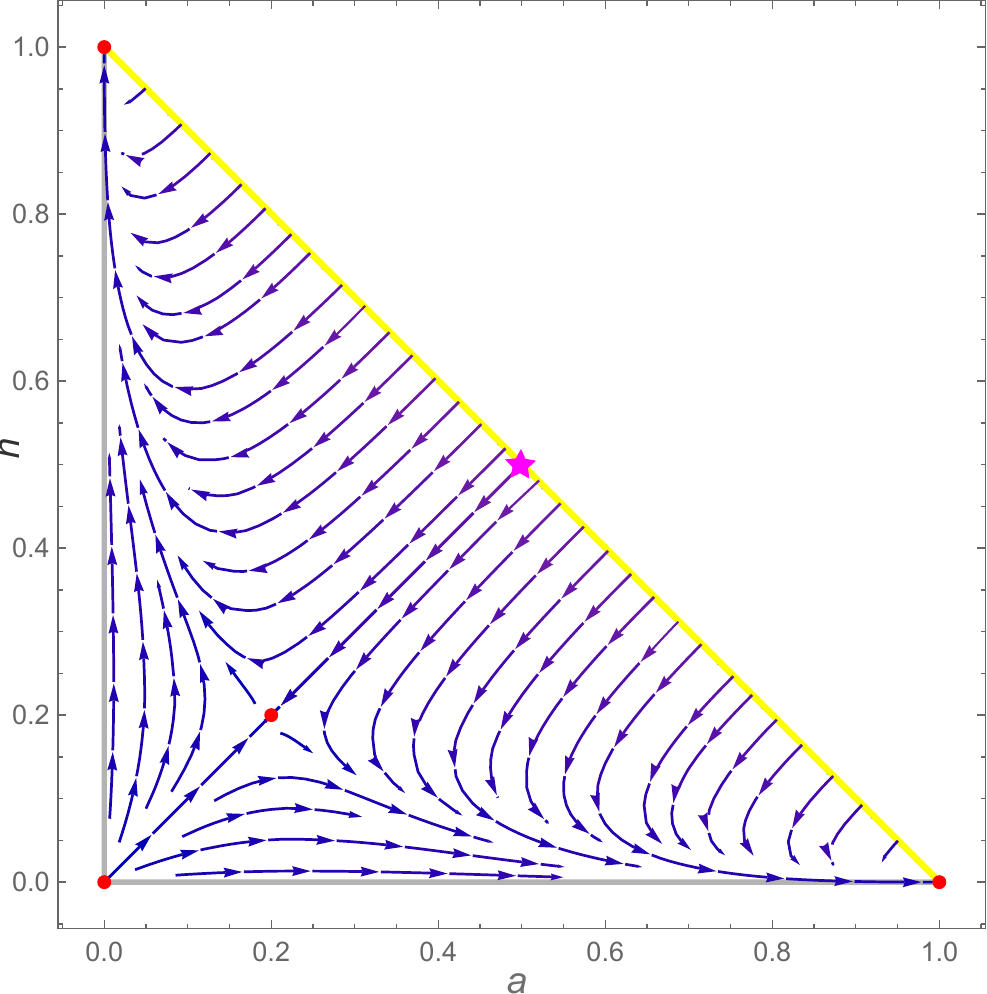}
\caption{Flow within the $(n,a)$ subspace ($x=y=z=(1-n-a)/3$) for binary tree codes with random gates, subject to heralded surface errors.\label{fig:random-flow}}
\end{figure}
As shown in~\autoref{fig:random-flow}, there are four fixed points. Under i.i.d. erasures at rate $p$, i.e. tuning the initial condition along the yellow line in the figure, the system flows erasure rate $p$ is tuned along the yellow line, the system flows to the coding phase fixed point ($n=1$) for $p<p_c=1/2$; the critical point at $(n,a)=(1/5,1/5)$ for $p=p_c$; and the noncoding phase fixed point ($a=1$) for $p>p_c$. The fourth fixed point, at $(n,a)=(0,0)$, is multicritical, with two relevant eigenvalues, and is accessed from a heralded error channel of the form $\mathcal{E}_{p_x,p_z,p_y,0}$ where $p_x+p_z+p_y=1$. 

As expected, the relevant eigenvector of $(n,a)=(1/5,1/5)$ is an odd perturbation, $\vec{v}_1 = (1,-1)/\sqrt{2}$, and we can read off the scaling behavior of the failure probability in the vicinity of the critical point from its eigenvalue $\lambda=28/25$:
\begin{equation}
    \xi_c = 1/\log_2(28/25) = 6.1126...
\end{equation}
Meanwhile, the irrelevant eigenvector is an even perturbation, $\vec{v}_1 = (1,1)/\sqrt{2}$, controlling the flow along the line $n=a$, along which $I(R:E)=1$.

Note that $\xi_c$ is significantly larger than in models with non-random gates. The relative broadness of the transition reflects the inferiority of the random-gate codes.

\subsection{Bell tree}\label{app:erasure-bell}
Since the Bell tree generates a CSS code, it is useful first to consider the single-parameter flows of the logical X and logical Z erasure probabilities, separately. As in the main text, let $x_0$ denote the probability of an undetectable logical X error, regardless of whether a Z logical is also lost. Under a CNOT gate with $Z$ stabilizer inputs, $x_0$ evolves as:
\begin{equation}\label{eq:cnotX}
x_0' = h(x_0) \equiv x_0^2
\end{equation}
while under a NOTC gate with $X$ stabilizer inputs, $x_0$ evolves as:
\begin{equation}\label{eq:notcX}
1-x_0' = h(1-x_0)
\end{equation}
The inverse occurs for logical Z loss probability, $z_0$. 

Thus, we obtain the two-step flow equations in~\autoref{eq:flow}, which at $q=0$ simplify to:
\begin{subequations}
\begin{align}\label{eq:bell-css-flow}
x_0(\tau+1) &= h(1-h(1-x_0(\tau))) = (x_0(\tau)(2-x_0(\tau)))^2 \\
z_0(\tau+1)&=1-h(1-h(z_0(\tau))) = z_0(\tau)^2(2-z_0(\tau)^2)  .
\end{align}
\end{subequations}
Note that $1-z_0(\tau)$ obeys the same flow equation as $x_0(\tau)$. This is just a manifestation of the cleaning lemma specialized to CSS codes~\cite{Bravyi2009,Kalachev2022}: if a given region contains $N_x$ X logicals, then its complement contains $k-N_x$ Z logicals, where in this case $k=1$. 
 
The flows of $x_0, z_0$ define a two-dimensional subspace of the four-dimensional space $\mathcal{S}$ of $\pi$ in which $X_L$ and $Z_L$ are lost independently. Mathematically, this subspace is defined by
\begin{equation}\label{eq:x0z0}
\pi = ((1-x_0)(1-z_0), x_0(1-z_0), z_0(1-x_0), 0, x_0 z_0).
\end{equation}
so that the average mutual information with the environment (\autoref{eq:mutual}) evaluates to:
\begin{equation}\label{eq:mutual-css}
I(R:E) = x_0 + z_0.
\end{equation}
In the absence of bulk errors, each of the flows $x_0, z_0$ has three fixed points, as obtained in the main text. Tuning the rate of heralded X and Z errors independently, we therefore obtain 9 fixed points in the $(x_0, z_0)$ subspace, one at each pairing of $(x_s(0), x_c(0), x_f) = (0,p_c,1)$ and $(z_s(0), z_c(0), z_f) = (0, 1-p_c, 1)$, where $p_c=(3-\sqrt{5})/2$.\footnote{Interestingly, $(1-p_c)/p_c$ is the golden ratio.} 

A richer phase diagram is obtained if we take initial conditions in the full four-dimensional parameter space. Taking the Bell tree representation in which each gate is $(H\otimes H) \CNOT$, the one-step flow equations are:
\begin{align}\label{eq:bell-flow}
n' &= n(n + 2z + 2y), \quad
x' = z^2 + y^2, \quad 
z' = 2na+(2-x)x \notag \\
y' &= 2zy, \quad 
a' = a(a + 2z + 2y).
\end{align}

\begin{figure}[hbtp]
\includegraphics[width=\linewidth]{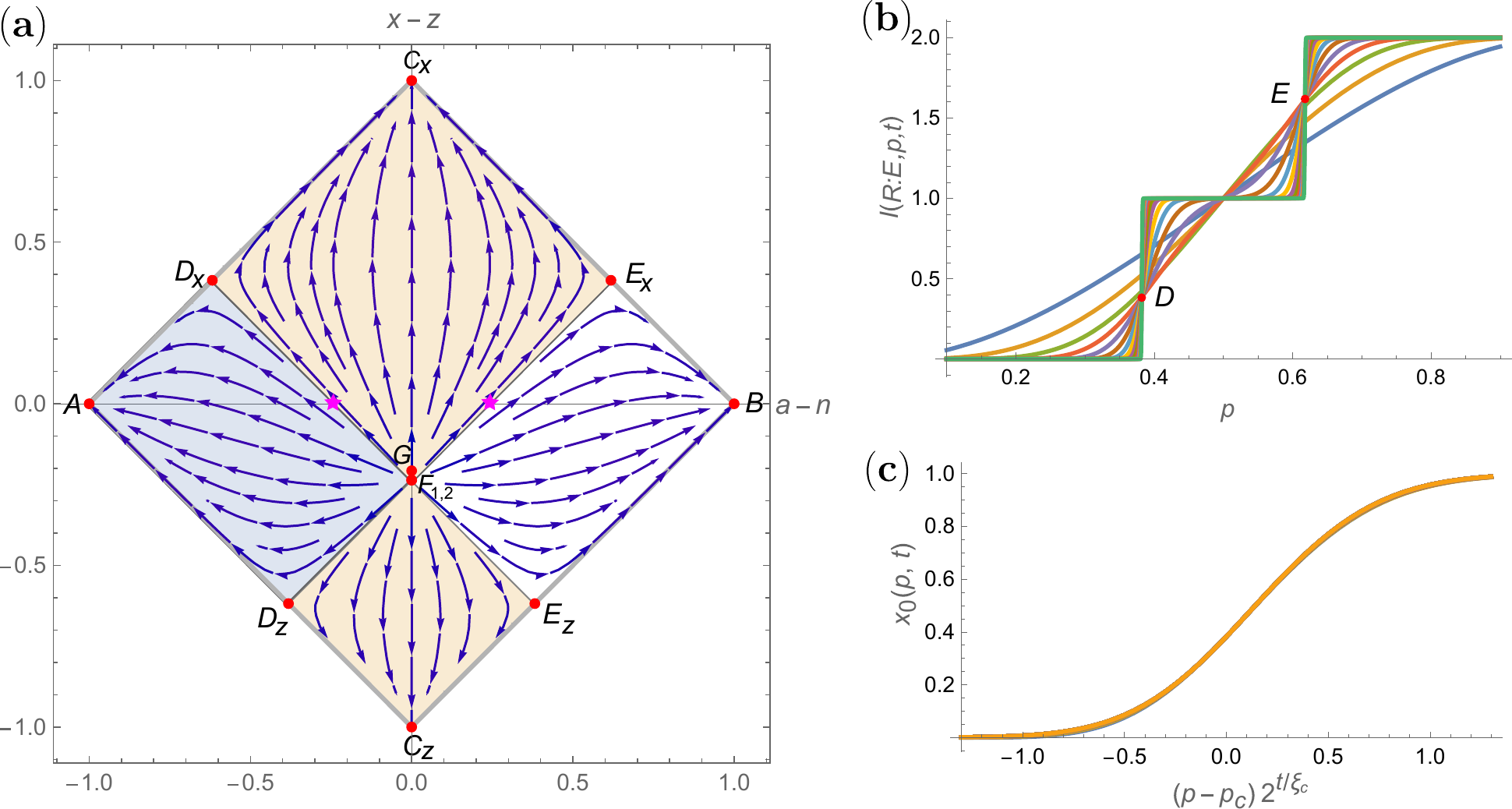}
\caption{(a) Phase diagram and two-step flows projected onto the axes $a-n$, $x-z$. The points A-F are described in the text; those with subscripts $x,z$ map to each other under one step of the flow. Streamlines show the direction of the projected flow for $y=0$. Blue, orange, and unshaded regions indicate the quantum coding, classical coding, and noncoding phases, respectively, as in~\autoref{fig:phase} of the main text. (b) Mutual information in the Bell tree under heralded noise with $x(0)=z(0),y(0)=0, a(0)-n(0)=2p-1$, at even depths 2-30. (c) Scaling collapse of $x_0(p,t)$, for even depths 6-120, with $\xi_c=3.27056...$ and $p_c=(3-\sqrt{5})/2$.  \label{fig:intermediate}}
\end{figure}
To visualize the flows, in~\autoref{fig:intermediate}a, we project onto the axes $(a-n, x-z)$. Owing to the $\mathbbm{Z}_2$ symmetry of the flow equations under $n\leftrightarrow a$, the flow diagram is symmetric under reflection about the vertical axis. Distinct points within $\mathcal{S}$ are mapped to distinct points under this projection: $(x_0, z_0) \rightarrow (a-n,x-z)=(x_0+z_0-1, x_0-z_0)$. The streamlines in the plot show the projected two-step flows within the subspace $y=0$, which is preserved under~\autoref{eq:bell-flow}. Within that subspace, the projected flows are uniquely determined by the projected coordinates, i.e. $(a(t+2)-n(t+2)) - (n(t) - a(t))$ and $(x(t+2)-z(t+2))-(x(t)-z(t))$ are functions of $(a(t)-n(t), x(t)-z(t))$ alone. 

The only stable fixed points under two steps of~\autoref{eq:bell-flow}, at the four corners of the phase diagram in~\autoref{fig:intermediate}a, in fact belong to $\mathcal{S}$: Point A, at $(a-n,x-z)=(-1,0)$, corresponds to the quantum coding phase at $(x_0=0,z_0=0)$; point $B=(a-n,x-z)=(1,0)$, corresponds to the noncoding phase at $(x_0=1,z_0=1)$; and a pair of "classical bit" fixed points at $C_x=(0,1)$ and $C_z=(0,-1)$, corresponding to $(x_0=1,z_0=0)$ and $(x_0=0,z_0=1)$ respectively, cycle between each other under one step. 

The remaining (unstable) fixed points fall into two groups. The first group contains pairs of critical points belonging to the same length-2 limit cycle of the one-step flow, and again belong to $\mathcal{S}$: $(x_0, z_0) = (p_c,0), (1-p_c,0)$ ($D_x$ and $D_z$ in~\autoref{fig:intermediate}a) and its mirror image, $E_x, E_z = (x_0,z_0) = (1,1-p_c), (p_c,1)$. The two-step flow matrix at each of these points has one relevant eigenvalue, $\lambda_c=6-2\sqrt{5}$, which corresponds to $\xi_c=1/\log_4(\lambda_c)=3.27056...$ at each of these transitions. This scaling can be inferred from the independent flows of $x_0$ and $z_0$ (cf~\autoref{eq:lambda}): $\lambda_c(0) = g'_0(x)|_{x=p_c}$. 

The second group contains multicritical points which are invariant under one step of the flow. One of these, $F_1$, belongs to $\mathcal{S}$, with $(x_0=p_c,z_0=1-p_c)$ (point F in~\autoref{fig:intermediate}c). Since the one-parameter flows $x_0$ and $z_0$ are both critical at this point, it has two relevant eigenvalues, both equal to $\lambda_c$. Meanwhile, $(1,-1,-1,1)$ is an eigenvector with $\lambda < 1$, and corresponds to a perturbation in the direction orthogonal to the projected space. That is, $F_1$ is an attractor within the one-dimensional subspace $\mathcal{F} = \{\pi | y=0, n=a, x-z=2p_c-1\}$, which projects onto the single point $(a-n=0,x-z=2p_c-1)$ in the figure. $\mathcal{F}$ is invariant under the flow dynamics, and within it, all points flow to $F_1$. $\mathcal{F}$ also contains a completely unstable fixed point, $F_2 =\tilde{\pi}=(0,p_c,1-p_c,0)$, which has four relevant eigenvalues, all $\lambda_c$. Finally, there is one fixed point with $y\neq 0$: the point $\tilde{\pi}=(0,1-1/\sqrt{2}, 1/2, 0)$, which projects onto $G=(0, -(\sqrt{2}-1)/2)$, and has three relevant eigenvalues.

~\autoref{fig:intermediate}b shows the mutual information at different depths for a standard one-parameter path through the phase diagram: initial conditions at $y=0, a-n=2p-1,x-z=0$, which map onto the horizontal axis of~\autoref{fig:intermediate}a. Such initial conditions include heralded X and Z errors applied independently at an equal rate $p$, as in the main text ($(x_0, z_0)=(p,p)$), as well as ordinary erasures at rate $p$ ($(n,a)=(1-p,p)$). Moving left to right along the horizontal axis, the stars at $\pm(1-2p_c)$, which flow toward the critical points $D$ and $E$ respectively, mark the boundaries of the intermediate phase in which only a classical bit survives. To see the correlation length $\xi_c=3.27056...$ more clearly, in~\autoref{fig:intermediate}c we show a scaling collapse of $x_0(p,t)$ for the case of independent X and Z errors. To circumvent the intermediate phase, one could instead take the rate of heralded X and Z errors to be different, e.g. to take the system through the multicritical point $F_1$. 

\subsection{Balanced erasures}
Heralded errors have a pattern of where the errors happened, and this pattern can be random (as in the main text and the analysis above) or (in models) deterministic. The choice of model can modify the critical behavior. For example, when i.i.d. erasure errors are applied to random stabilizer codes, the transition is rounded by $\sqrt{N}$ fluctuations in the number of errors, but fixing the fraction of qubits that are erased narrows this transition to a width of $O(1)$ error~\cite{Gullans21}. Thus, removing some disorder (in that case, disorder in the total number of erasures) uncovers a first-order transition.

Here, we remove disorder in a more drastic fashion, not only fixing the number of erasures, but also restricting their placement to be ``maximally balanced'' on the leaves of a tree. The construction is recursive. Consider assigning $M(t)$ erasures to the leaves of a tree of depth $t$. This tree contains two depth $t-1$ subtrees, to which we assign $\lfloor M(t)/2 \rfloor$, $\lceil M(t)/2 \rceil$ erasures. The recursion step is iterated until we reach the leaves, where some leaves are assigned one erasure and others are left untouched. 

On the Bell tree, we can separately track the variables $x_0$, $z_0$. The copy update rule~\autoref{eq:cnotX} becomes
\begin{equation}\label{eq:cnotX-balanced}
    x_0'(M) = x_0(\lfloor M/2 \rfloor) x_0(\lceil M/2 \rceil)
\end{equation}
while the delocalizer update rule~\autoref{eq:notcX} becomes
\begin{equation}
    1-x_0'(M) = (1 -x_0(\lfloor M/2 \rfloor)) (1-x_0(\lceil M/2 \rceil)).
\end{equation}
For a given $M(t)$, the above protocol defines an ensemble of maximally balanced trees. A straightforward proof by induction shows that every tree in this ensemble has the same logical outcome: either every erasure pattern results in no logical error ($x_0=0$), or every pattern results in an undetectable logical error ($x_0=1$). Moreover, a maximally balanced pattern of $M$ errors is contained within some maximally balanced pattern of $M+1$ errors, so $x_0(M)$ is monotonic in $M$. This results in a first-order transition of width 1 erasure, i.e. $\xi_c=1$.

With $x_0, z_0$ only taking values $0, 1$, the two-step flow equations for a depth $t=2\tau$ tree become
\begin{equation}\label{eq:bell-flow-balanced}
x_0(M, \tau+1) = x_0(\lceil \lfloor M/2 \rfloor /2 \rceil, \tau), \qquad z_0(M,\tau+1) = z_0(\lfloor \lceil M/2 \rceil /2 \rfloor, \tau).
\end{equation}
Let $M^*_{x,z}(\tau)$ denote the threshold at which $x_0(M,\tau)$ or $z_0(M,\tau)$ jumps from 0 to 1.~\autoref{eq:bell-flow-balanced} implies
\begin{equation}
M^*_x(\tau + 1) = 4 M^*_x(\tau) - 2, \qquad M^*_z(\tau + 1) = 4 M^*_z(\tau) - 1
\end{equation}
which, imposing the initial conditions $M^*_x(0) = M^*_z(0) = 1$, have the solutions
\begin{equation}\label{eq:threshold-M}
M^*_x(\tau) = (2^{2\tau}+2)/3, \qquad M^*_x(\tau) = (2^{2\tau+1}+1)/3.
\end{equation} 
Noting that a depth $2\tau$ tree contains $2^{2\tau}$ erasures, the X and Z thresholds in~\autoref{eq:threshold-M} correspond, asymptotically, to erasure fractions of $1/3$ and $2/3$, respectively. Thus, changing the error model in this way not only narrows the transition, but it also shifts it to a different erasure fraction (from $(3-\sqrt{5})/2$ and $(\sqrt{5}-1)/2$, respectively). This shift in the critical erasure fraction is possible because the leaves are a nonzero fraction of the system.

A natural next question is whether we can analogously sharpen the transition under bulk errors, by introducing errors on the links in a deterministic, balanced fashion. We have yet to find such a construction; in all the deterministic models of bulk erasures that we have tried, the failure probability as a function of erasure fraction is not smooth. We do not know whether this failure to sharpen the transition is just because we have not found a good way of reducing the randomness, or whether the rounding of the transition is a more "intrinsic" feature that cannot be reduced by choosing a "good" pattern of the errors.

\section{Heralded errors in the bulk of the Bell tree}\label{app:bulk}
In this section, we elaborate on the critical behavior in the Bell tree with heralded bulk errors. As in the main text, let $x_d(q,\tau)$ denote the probability that the conditional X distance of a depth $2\tau$ tree is $d$. 

\subsection{Survival/loss}
For $q\leq q_c$, the flow equations for $x_0(q,\tau)$ and $z_0(q,\tau)$ [\autoref{eq:flow}] each have three fixed points, as shown in~\autoref{fig:phase} of the main text. Here $q_c$ is the lone real root of the polynomial
\begin{equation}
f(q) = 32q^3 - 96q^2 + 96q - 5.
\end{equation}

The intermediate phase in which only a classical bit survives arises due to the alternating patterns of gates in the Bell tree [\autoref{fig:zx}]. In fact, defining $x_0(q,\tau+1/2)$ and $z_0(q,\tau+1/2)$ with integer $\tau$ to be the loss probabilities in odd depth trees, we find:
\begin{align}
x_0^*(q) &\equiv \lim_{\tau\rightarrow\infty} x_0(q,\tau) = \lim_{\tau\rightarrow\infty} z_0(q,\tau+1/2) \notag \\
z_0^*(q) &\equiv \lim_{\tau\rightarrow\infty} z_0(q,\tau) = \lim_{\tau\rightarrow\infty} x_0(q,\tau+1/2),
\end{align}
where the outcome of the limit---coding fixed point $x_s$, critical point $x_c$, or noncoding fixed point $x_f=1$---depends on the initial condition (surface heralding rate). Thus, for a fixed $q$, the threshold in $p$ for a logical $Z$ error to become undetectable is higher at even depths, but lower at odd depths. 

\begin{figure}[t]
\subfloat[]{
\includegraphics[width=0.45\linewidth]{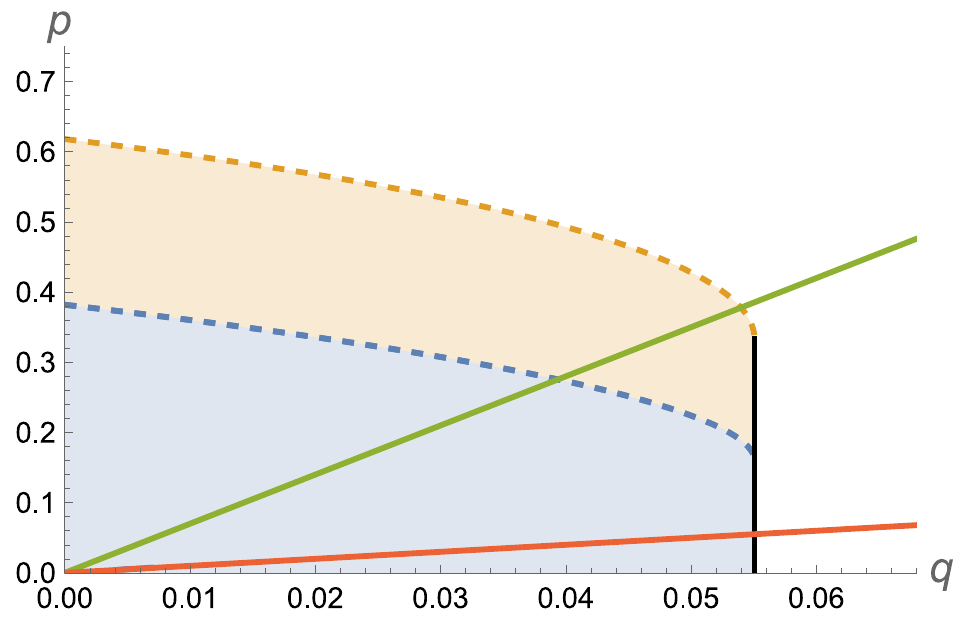}
\label{fig:fixed-points}}\hfill
\subfloat[]{
\includegraphics[width=0.45\linewidth]{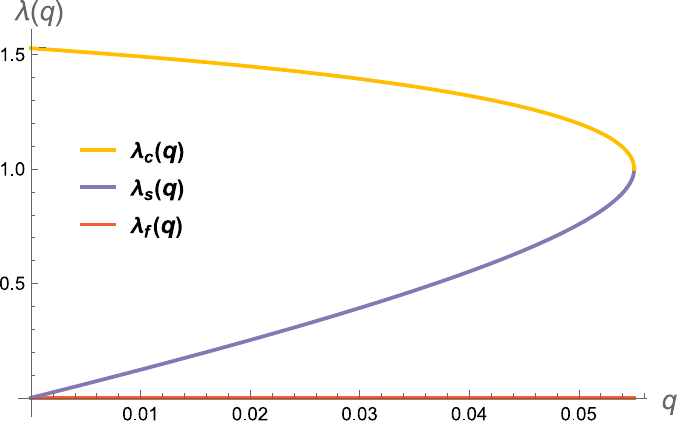}\label{fig:lambda}}
\\
\subfloat[]{
\includegraphics[width=0.45\linewidth]{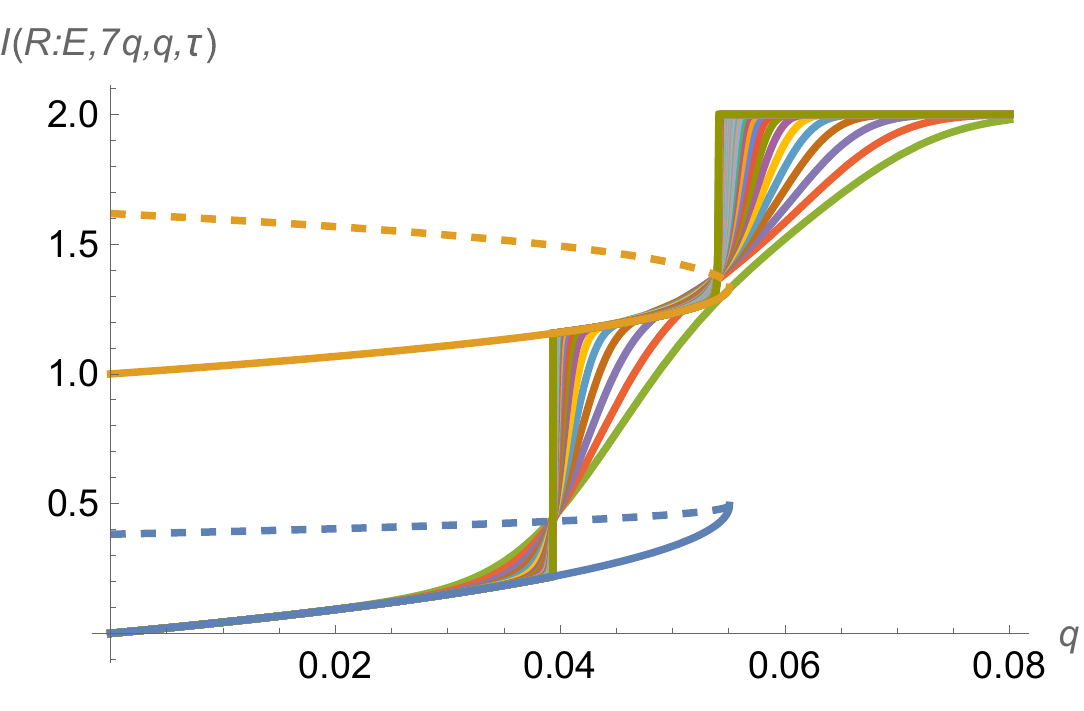}
\label{fig:7q}}\hfill
\subfloat[]{
\includegraphics[width=0.45\linewidth]{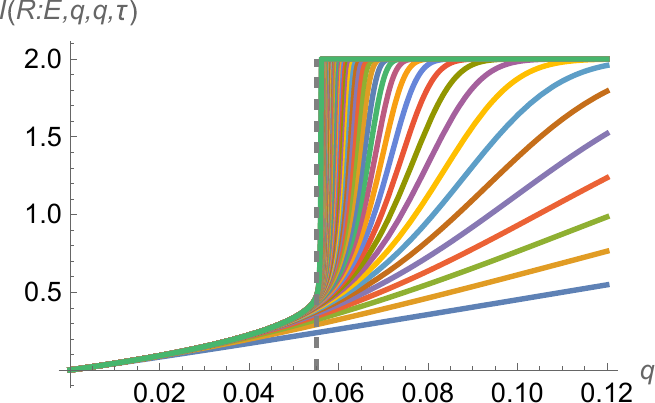}
\label{fig:q}}
\\
\caption{(a) Phase diagram of the Bell tree under heralded bit and phase flips at rate $p$ on the leaves and rate $q$ on the links. The green and red lines define the cuts through parameter space, $p=7q$ and $p=q$, respectively. (b) $\lambda(q)$ evaluated at each of the three fixed points [\autoref{eq:lambda-v2}]. (c) Mutual information as a function of $q$ along the line $p=7q$, for $\tau=5,...,40$. The blue solid curve, blue dashed curve, orange solid curve, and orange dashed curve are $x_s(q) + z_s(q)$, $x_c(q) + z_s(q)$, $1+z_s(q)$, and $1+z_c(q)$, respectively. (d) Mutual information as a function of $q$ along the line $p=q$, for $\tau=1,...,60$. Gray dashed line is at $q=q_c=0.05505...$.}
\end{figure}

In~\autoref{fig:fixed-points} we show again the fixed points, along with two cuts through the phase diagram: a green cut along the line $p=7q$ and a red cut along the line $p=q$, corresponding to the physical setting is where the bulk and surface error rates are the same.

Linearizing around a given fixed point, let $\lambda(q)$ denote the base of the exponential convergence toward a stable fixed point $(\lambda<1$), or growth away from an unstable fixed point $(\lambda>1)$, i.e.:
\begin{equation}\label{eq:lambda-v2}
    x_0(q,\tau) - x_0^*(q) \propto \lambda(q)^\tau, \quad \lambda(q) = g'_q(x)|_{x=x^*(q)}
\end{equation}
(cf.~\autoref{eq:lambda}).\footnote{As to be expected, for a given $q$, we obtain the same eigenvalues at the coding phase fixed points ($\lambda_s(q)$ at $x_s(q) < z_s(q)$), noncoding fixed points ($\lambda_f(q)=0$ at $z_f(q)=x_f(q)=1$), and critical points ($\lambda_c(q)$ at $x_c(q)<z_c(q)$).}

Along the cut $p=7q$, the $X$ and $Z$ thresholds occur at the intersection of the green line with the dashed phase boundaries, at
\begin{equation}
    q_x = x_c(q_x)/7 \rightarrow q_x = 0.0394, \quad 
    q_z=z_c(q_z)/7 \rightarrow q_z = 0.0539
\end{equation}
separated by an intermediate ``classical'' phase. Thus, as shown in~\autoref{fig:7q}, as $\tau\rightarrow\infty$, the mutual information $I(R:E,7q,q,\tau)$ approaches $x_s(q) + z_s(q)$ for $q<q_x$; $x_c(q) + z_s(q)$ at $q=q_x$; $x_f + z_s(q) = 1+z_s(q)$ for $q_x < q < q_z$; $x_f + z_c(q) = 1 + z_c(q)$ at $q=q_z$; and $x_f + z_f = 2$ for $q>q_z$.  
The scaling behavior in the vicinity of both transitions is mediated by the relevant eigenvalue $\lambda_c(q_{x,z})$. Comparing to~\autoref{eq:scaling},
\begin{equation}
    \xi_c(q) = 1/\log_4(\lambda_c(q))
\end{equation}
so as $\lambda_c(q)$ decreases with increasing $q$, the correlation length increases, resulting in a broader transition. For example, $\xi_c(q_x)=4.913,\xi_c(q_z)=14.983$ are both larger than the critical correlation length under surface errors, $\xi_c(0)=3.271$. As $q\rightarrow q_c$, $\xi_c(q)$ diverges: physically, increasing $q$ dilutes the ``bulk coupling'', so that a perturbation to the surface ``field'' $p$ drives the system more slowly away from the unstable fixed point.

Meanwhile, the cut $p=q$ circumvents the critical points, and the intermediate phase, entirely, since $q<x_c(q),z_c(q)$ for all $q<q_c$. Therefore, the threshold transition now occurs at $q=q_c$, where the coding fixed point and critical fixed point merge and become marginal. This transition is first-order, since $I(R,E;q)$ jumps from $x_s(q_c) + z_s(q_c) = (4-2^{4/3})/3$ to $x_f+z_f=2$. However, it still is characterized by a diverging length scale---the correlation length $\xi_s(q)$---whose divergence, as $(q_c-q)^{-1/2}$, yields the critical exponent $\nu=1/2$ [\autoref{eq:nu}].

In the main text, we asserted that the exponential convergence toward the coding fixed point for $p=q<q_c$ is replaced with an algebraic decay of the form $w(q,\tau) \sim 1/\tau$ (\autoref{eq:wc}) at $q=q_c$. This comes from the fact that
\begin{equation}
\frac{dw}{d\tau} \approx w(\tau+1,q_c)-w(\tau,q_c) = -c w(\tau,q_c)^2 + O(w^3),
\end{equation}
so to leading order, $w(\tau,q_c) \propto 1/\tau$. 

\subsection{Conditional distance}
\begin{figure}[hbtp]
\subfloat[]{
\includegraphics[width=0.3\linewidth]{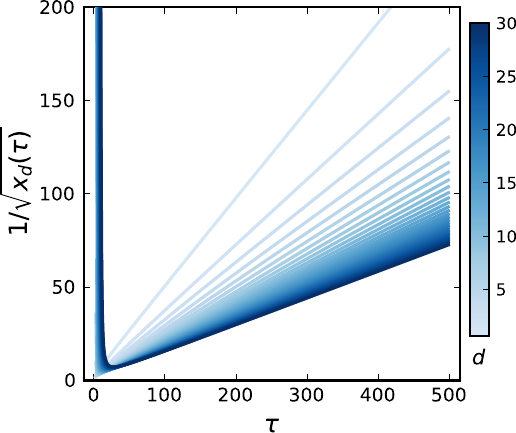}\label{fig:distr-qc}}
\hfill
\subfloat[]{
\includegraphics[width=0.29\linewidth]{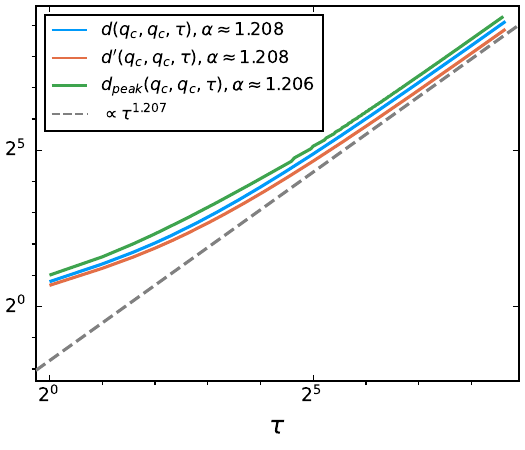}\label{fig:d-qc}}
\subfloat[]{
\includegraphics[width=0.37\linewidth]{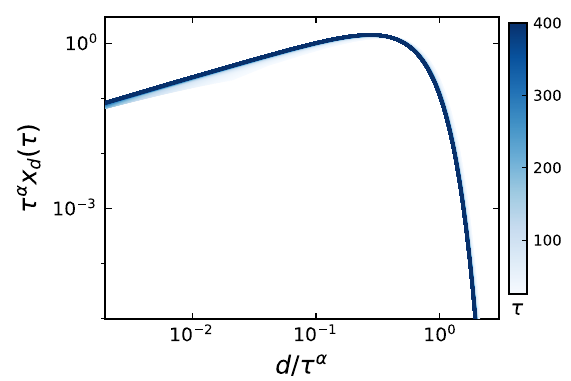}\label{fig:distr-log}}
\caption{Scaling behavior of the distribution of conditional code distance in the Bell tree subject to heralded bit and phase flips at $p=q=q_c$. (a) $1/\sqrt{x_d(\tau)}$ vs. $\tau$, for $d=1$ (lightest color) up to $d=30$ (darkest). (b) $d(q_c,q_c,\tau)$ [\autoref{eq:ave-dist}], $d'(q_c,q_c,\tau)$ [\autoref{eq:dprime}], and $d_{peak}(q_c,q_c,\tau)$ [\autoref{eq:dpeak}]. Gray dashed curve is $\propto \tau^{1.207}$. (c) Rescaled distribution with $\alpha = 1.207$, $\tau\in[25,400]$.}
\end{figure}

At $q=q_c$, plugging $x_0(\tau)=x^* + c/\tau$ into the flow equation for $x_1$ yields, to leading order, $x_1(\tau) \sim 1/\tau^2$. This is unsurprising, since a code with $d=1$ is just one well-placed heralded error away from being lost, hence $x_1$ should be proportional to the loss rate $dw/d\tau$. In fact, for all $d\geq 1$, $x_d(\tau)$ decays as $\sim 1/\tau^2$ at late times, after reaching a peak at intermediate times (\autoref{fig:distr-qc}). 

In the main text, we noted that the averaged conditional distance, $d(p,q,\tau)$, grows linearly in time at $p=q=q_c$. A related trend appears in the quantity $d_{peak}(\tau)$, defined as:
\begin{equation}\label{eq:dpeak}
d_{peak}(p,q, \tau) = \max \{d: \mathrm{argmax}_s [x_d(s)] \leq \tau \},
\end{equation}
that is, the largest $d$ such that $x_d$ reaches a maximum before time step $\tau$. Like $d(q,q,\tau)$, $d_{peak}(q,q,\tau)$ grows exponentially with $\tau$ for $q<q_c$, with a correlation time that diverges as $(q-q_c)^{-1/2}$, and grows as a power law at $q=q_c$. Both quantities, along with the code distance $d'$ not conditioned on survival,
\begin{equation}\label{eq:dprime}
d'(p,q,\tau) = \sum_{i=0}^{2^{\tau}} d x_d(\tau)
\end{equation}
are plotted in~\autoref{fig:d-qc} for $p=q=q_c$. Three-parameter fits to the function $c_1 + c_2 \tau^{\alpha}$ for $\tau \in [150,400]$ yield $\alpha \approx 1.2$, where $c_1$ is small. The rescaled distribution is shown in~\autoref{fig:distr-log}.

\subsection{Random walk interpretation}
In the main text, we commented on the analogy between code distance $d_X(\sigma,t)$ conditioned on an error realization $\sigma$ and the position in a random walk with an absorbing wall. Whereas the recursion relations employed throughout this work describe the flow \textit{backwards} from the leaves to the root, for this analogy it is more natural to consider the evolution \textit{forward} in time, from the root to the leaves. 

To evaluate the bias in the random walk, we examine the quantity $\Pi(d,d^*,\tau)$, the probability distribution of $d_X(\sigma,2\tau)$ over erasure patterns $\sigma$ such that $d(\sigma, 2(\tau-1))=d^*$. 

\begin{figure}
\subfloat[]{
\includegraphics[width=0.33\linewidth]{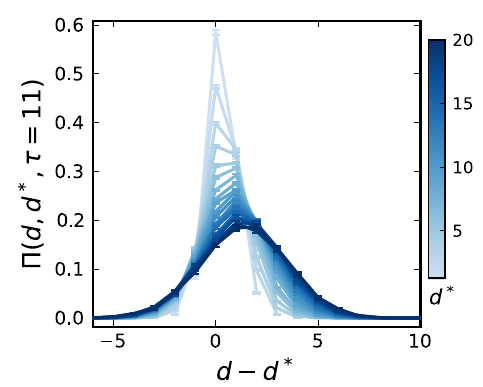}\label{fig:walk-distr}
}
\subfloat[]{\includegraphics[width=0.61\linewidth]{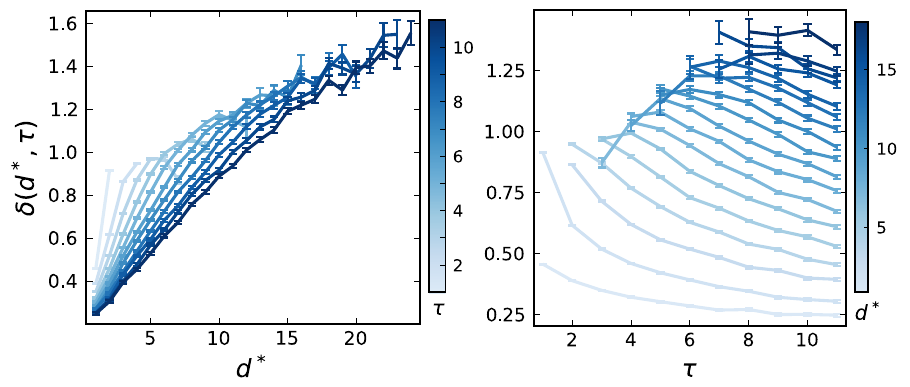}\label{fig:bias}}
\caption{Behavior of the conditional code distance at $p=q=q_c$. (a) Distribution of the conditional distance at $\tau+1$, given that the distance at $\tau=11$ is $d^*$. Curves from light to dark are $d^*=1,...,20$. Error bars are $\sqrt{\Pi(d,d^*,\tau)(1-\Pi(d,d^*,\tau))}/N(d^*)$ where $N(d^*)$ is the number of samples with $d_X(\sigma,2\tau)=d^*$. (b) Bias (\autoref{eq:bias}) as a function of $d^*$ (left) and $\tau$ (right). In the left panel, darker curves correspond to larger $\tau$, while in the right panel, darker curves correspond to larger $d^*$. Only data points with $N(d^*)>1000$ are included.}
\end{figure}

~\autoref{fig:walk-distr} shows the distribution for $\tau=11$ and $d^*$ ranging from 1 to 20, obtained by sampling 480000 error patterns on depth $T=24$ trees. As $d^*$ increases, the distribution broadens and approaches a Gaussian, since each time step contains many opportunities for $d$ to change. At sufficiently large $d^*$ and $\tau$, the distribution is roughly independent of $\tau$, and is always peaked at $d>d^*$; that is, the bias
\begin{equation}\label{eq:bias}
    \delta(d^*,\tau) = \sum_d \Pi(d; d^*,\tau) d - d^*
\end{equation}
is positive. 

\autoref{fig:bias} shows $\delta(d^*,\tau)$ as a function of $d^*$ (left) and $\tau$ (right). At fixed $\tau$, $\delta(d^*,\tau)$ increases with $d^*$ before reaching a peak. At a fixed $d^*$, $\delta(d^*,\tau)$ decreases at large $\tau$, but appears to be asymptoting toward a positive constant $\delta_{\infty}(d^*)$.

Moving away from $q_c$ adds or suppresses the effective branching of the tree, which naively adds a bias
$\sim (q_c-q)d$  to this ``walker.'' Indeed, $\delta_\infty(d^*, q) = m(q) d^* + b(q)$, where $m(q)$ grows linearly in $q_c - q$. {The linear trend in the bias should be contrasted with the exponent $\nu=1/2$ governing the transition, suggesting that the square root singularity is an emergent phenomenon not tied directly to the bias in the random walk.}

\subsection{Interface pinning}\label{app:mutual}
The interpretation of the coding threshold as an interface pinning transition is related to concepts from entanglement membrane theory, which maps the entanglement entropy across a timelike or spacelike cut in a local circuit onto the free energy of an interface separating the subsystem of interest from its complement~\cite{Nahum2017op,Nahum2017entanglement,Jonay2018,Zhou2019,Zhou2020,Bao2020,Jian2020,Li2021,Li2023,Sang2023}. The precise statistical mechanics model for this interface depends on the quantity being computed; in random circuits, for example, the membrane associated with the average purity is an Ising domain wall, whereas the replica limit involved in calculating the average Renyi entropy gives rise to a directed polymer in a random environment. Other characteristics of the model depend on the nature of the dynamics: random vs. nonrandom, Clifford vs. non-Clifford, etc~\cite{Sommers2024}.

Introducing measurements or noise into the circuit or along its boundary can cause the membrane to become pinned to randomness in the bulk, or bound to a surface~\cite{Gullans21,Li2023,Lovas2023,Sommers2024}. This gives a physical meaning to the transition between a phase at low noise rate, where logical information fed into the circuit survives to long times, and the phase at high noise rate where the information is ``lost.'' The survival of the logical information is quantified by the mutual information $I(R:E)$, which consists of three entropic terms (cf.~\autoref{eq:mutual}): $S_R$, the entropy of the reference, is the number of logical qubits fed into the circuit; $S_E$, the entropy of the environment; and $S_{RE}$, the entropy of the joint reference + environment. Each term is associated with a minimal membrane, $M_R$, $M_E$, and $M_{RE}$. The mutual information is zero when $M_{RE} = M_{R} \cup M_E$, i.e. the membrane for $S_{RE}$ is just the disjoint union of the membranes for $S_R$ and $S_E$.

In the main text, we referred to a single interface, which lives along the time slice where the logical information is lost. This interface can be thought of as the extra part of the membrane that needs to be inserted to $M_{R} \cup M_E$ to make $M_{RE}$. In the noncoding phase, the interface can be placed anywhere in the bulk at late enough times, whereas in the coding phase, it has some probability to be pinned near the root, or is absent altogether. In Clifford circuits, entanglement membranes behave like interfaces in classical models at zero temperature~\cite{Sommers2024}, so the effective model for the interface in question should also be at $T=0$.

%% file: green-spider.tikz
\begin{tikzpicture}
	\begin{pgfonlayer}{nodelayer}
		\node [style=none] (2) at (-1, 1) {};
		\node [style=none] (3) at (1, 1) {};
		\node [style=none] (10) at (0, -1.25) {};
		\node [style=z spider] (14) at (0, 0) {};
	\end{pgfonlayer}
	\begin{pgfonlayer}{edgelayer}
		\draw (14) to (10.center);
		\draw (14) to (2.center);
		\draw (14) to (3.center);
	\end{pgfonlayer}
\end{tikzpicture}

%% file: red-spider.tikz
\begin{tikzpicture}
	\begin{pgfonlayer}{nodelayer}
		\node [style=none] (2) at (-1, 1) {};
		\node [style=none] (3) at (1, 1) {};
		\node [style=none] (10) at (0, -1.25) {};
		\node [style=x spider] (14) at (0, 0) {};
	\end{pgfonlayer}
	\begin{pgfonlayer}{edgelayer}
		\draw (14) to (10.center);
		\draw (14) to (2.center);
		\draw (14) to (3.center);
	\end{pgfonlayer}
\end{tikzpicture}

%% file: tree-H.tikz
\begin{tikzpicture}
	\begin{pgfonlayer}{nodelayer}
		\node [style=z spider] (0) at (0, 0) {};
		\node [style=x spider] (1) at (-3.5, 2) {};
		\node [style=x spider] (2) at (3.5, 2) {};
		\node [style=z spider] (3) at (-1.75, 4) {};
		\node [style=z spider] (4) at (-5.25, 4) {};
		\node [style=z spider] (5) at (1.75, 4) {};
		\node [style=z spider] (6) at (5.25, 4) {};
		\node [style=x spider] (7) at (-6.25, 6) {};
		\node [style=x spider] (8) at (-4.25, 6) {};
		\node [style=x spider] (9) at (-2.75, 6) {};
		\node [style=x spider] (10) at (-0.75, 6) {};
		\node [style=x spider] (11) at (0.75, 6) {};
		\node [style=x spider] (12) at (2.75, 6) {};
		\node [style=x spider] (13) at (4.25, 6) {};
		\node [style=x spider] (14) at (6.25, 6) {};
		\node [style=none] (29) at (0, -1) {};
		\node [style=none] (30) at (0, -1) {};
		\node [style=none] (31) at (-6.75, 7) {};
		\node [style=none] (32) at (-5.75, 7) {};
		\node [style=none] (33) at (-4.75, 7) {};
		\node [style=none] (34) at (-3.75, 7) {};
		\node [style=none] (35) at (-3.25, 7) {};
		\node [style=none] (36) at (-2.25, 7) {};
		\node [style=none] (37) at (-0.25, 7) {};
		\node [style=none] (38) at (-1.25, 7) {};
		\node [style=none] (39) at (1.25, 7) {};
		\node [style=none] (40) at (0.25, 7) {};
		\node [style=none] (41) at (2.25, 7) {};
		\node [style=none] (42) at (3.25, 7) {};
		\node [style=none] (43) at (4.75, 7) {};
		\node [style=none] (44) at (3.75, 7) {};
		\node [style=none] (45) at (5.75, 7) {};
		\node [style=none] (46) at (6.75, 7) {};
	\end{pgfonlayer}
	\begin{pgfonlayer}{edgelayer}
		\draw (1) to (0);
		\draw (0) to (2);
		\draw (1) to (4);
		\draw (1) to (3);
		\draw (7) to (4);
		\draw (4) to (8);
		\draw (9) to (3);
		\draw (3) to (10);
		\draw (11) to (5);
		\draw (12) to (5);
		\draw (5) to (2);
		\draw (2) to (6);
		\draw (13) to (6);
		\draw (6) to (14);
		\draw [style=thick] (30.center) to (0);
		\draw (7) to (31.center);
		\draw (32.center) to (7);
		\draw (33.center) to (8);
		\draw (8) to (34.center);
		\draw (35.center) to (9);
		\draw (9) to (36.center);
		\draw (38.center) to (10);
		\draw (37.center) to (10);
		\draw (40.center) to (11);
		\draw (39.center) to (11);
		\draw (41.center) to (12);
		\draw (42.center) to (12);
		\draw (44.center) to (13);
		\draw (43.center) to (13);
		\draw (45.center) to (14);
		\draw (46.center) to (14);
	\end{pgfonlayer}
\end{tikzpicture}

%% file: join-tensor.tikz
\begin{tikzpicture}
	\begin{pgfonlayer}{nodelayer}
		\node [style=bigU] (0) at (-2, 0.5) {};
		\node [style=bigU] (1) at (2, 0.5) {};
		\node [style=none] (2) at (-2, 0.5) {$\scriptstyle{\mathbf{A}^{(L)}_1}$};
		\node [style=none] (3) at (2, 0.5) {$\scriptstyle{\mathbf{A}^{(L)}_2}$};
		\node [style=U] (4) at (0, -1.25) {};
		\node [style=none] (5) at (0, -1.25) {$\scriptstyle{\tilde{R}}$};
		\node [style=none] (6) at (0, -2.5) {};
	\end{pgfonlayer}
	\begin{pgfonlayer}{edgelayer}
		\draw (3.center) to (5.center);
		\draw (5.center) to (6.center);
		\draw (0) to (5.center);
	\end{pgfonlayer}
\end{tikzpicture}

%% file: optimal-tree.tikz
\begin{tikzpicture}
	\begin{pgfonlayer}{nodelayer}
		\node [style=U] (0) at (-10.25, 0) {};
		\node [style=U] (1) at (-3.75, 0) {};
		\node [style=U] (2) at (-7, -3) {};
		\node [style=U] (3) at (-12, 2.5) {};
		\node [style=U] (4) at (-8.5, 2.5) {};
		\node [style=U] (5) at (-5.5, 2.5) {};
		\node [style=U] (6) at (-2, 2.5) {};
		\node [style=stabilizer] (8) at (-2.75, -0.75) {};
		\node [style=none] (17) at (-2.75, -0.75) {$\scriptstyle{X}$};
		\node [style=stabilizer] (18) at (-9.25, -0.75) {};
		\node [style=none] (19) at (-9.25, -0.75) {$\scriptstyle{X}$};
		\node [style=stabilizer] (20) at (-11, 1.75) {};
		\node [style=stabilizer] (21) at (-7.5, 1.75) {};
		\node [style=stabilizer] (22) at (-4.5, 1.75) {};
		\node [style=stabilizer] (23) at (-1, 1.75) {};
		\node [style=physical] (24) at (-12.75, 4.25) {};
		\node [style=physical] (25) at (-11.25, 4.25) {};
		\node [style=physical] (26) at (-9.25, 4.25) {};
		\node [style=physical] (27) at (-7.75, 4.25) {};
		\node [style=physical] (28) at (-6.25, 4.25) {};
		\node [style=physical] (29) at (-4.75, 4.25) {};
		\node [style=physical] (30) at (-2.75, 4.25) {};
		\node [style=physical] (31) at (-1.25, 4.25) {};
		\node [style=none] (32) at (-11, 1.75) {$\scriptstyle{X}$};
		\node [style=none] (33) at (-7.5, 1.75) {$\scriptstyle{X}$};
		\node [style=none] (34) at (-4.5, 1.75) {$\scriptstyle{X}$};
		\node [style=none] (35) at (-1, 1.75) {$\scriptstyle{X}$};
		\node [style=stabilizer] (37) at (-5.25, -3.75) {};
		\node [style=none] (38) at (-5.25, -3.75) {$\scriptstyle{X}$};
		\node [style=U] (93) at (0, -6.25) {};
		\node [style=none] (94) at (-1, -7.55) {};
		\node [style=stabilizer] (95) at (1, -7.5) {};
		\node [style=none] (96) at (1, -7.5) {$\scriptstyle{X}$};
		\node [style=none] (97) at (0, -6.25) {$U$};
		\node [style=none] (99) at (-7, -3) {$U$};
		\node [style=none] (102) at (-2, 2.5) {$U$};
		\node [style=none] (103) at (-8.5, 2.5) {$U$};
		\node [style=none] (106) at (-5.5, 2.5) {$U$};
		\node [style=none] (107) at (-12, 2.5) {$U$};
		\node [style=none] (108) at (-10.25, 0) {$V$};
		\node [style=none] (109) at (-3.75, 0) {$V$};
		\node [style=U] (110) at (3.25, 0) {};
		\node [style=U] (111) at (9.75, 0) {};
		\node [style=U] (112) at (6.5, -3) {};
		\node [style=U] (113) at (1.5, 2.5) {};
		\node [style=U] (114) at (5, 2.5) {};
		\node [style=U] (115) at (8, 2.5) {};
		\node [style=U] (116) at (11.5, 2.5) {};
		\node [style=stabilizer] (117) at (10.75, -0.75) {};
		\node [style=none] (118) at (10.75, -0.75) {$\scriptstyle{X}$};
		\node [style=stabilizer] (119) at (4.25, -0.75) {};
		\node [style=none] (120) at (4.25, -0.75) {$\scriptstyle{X}$};
		\node [style=stabilizer] (121) at (2.5, 1.75) {};
		\node [style=stabilizer] (122) at (6, 1.75) {};
		\node [style=stabilizer] (123) at (9, 1.75) {};
		\node [style=stabilizer] (124) at (12.5, 1.75) {};
		\node [style=physical] (125) at (0.75, 4.25) {};
		\node [style=physical] (126) at (2.25, 4.25) {};
		\node [style=physical] (127) at (4.25, 4.25) {};
		\node [style=physical] (128) at (5.75, 4.25) {};
		\node [style=physical] (129) at (7.25, 4.25) {};
		\node [style=physical] (130) at (8.75, 4.25) {};
		\node [style=physical] (131) at (10.75, 4.25) {};
		\node [style=physical] (132) at (12.25, 4.25) {};
		\node [style=none] (133) at (2.5, 1.75) {$\scriptstyle{X}$};
		\node [style=none] (134) at (6, 1.75) {$\scriptstyle{X}$};
		\node [style=none] (135) at (9, 1.75) {$\scriptstyle{X}$};
		\node [style=none] (136) at (12.5, 1.75) {$\scriptstyle{X}$};
		\node [style=stabilizer] (137) at (8.25, -3.75) {};
		\node [style=none] (138) at (8.25, -3.75) {$\scriptstyle{X}$};
		\node [style=none] (139) at (6.5, -3) {$U$};
		\node [style=none] (140) at (11.5, 2.5) {$U$};
		\node [style=none] (141) at (5, 2.5) {$U$};
		\node [style=none] (142) at (8, 2.5) {$U$};
		\node [style=none] (143) at (1.5, 2.5) {$U$};
		\node [style=none] (144) at (3.25, 0) {$V$};
		\node [style=none] (145) at (9.75, 0) {$V$};
	\end{pgfonlayer}
	\begin{pgfonlayer}{edgelayer}
		\draw [style=gray, bend right=15, looseness=0.75] (0) to (2);
		\draw [style=gray, bend right=15, looseness=0.75] (2) to (1);
		\draw [style=gray, bend right=15, looseness=0.75] (5) to (1);
		\draw [style=gray, bend right=15, looseness=0.75] (1) to (6);
		\draw [style=gray, bend right=15, looseness=0.75] (3) to (0);
		\draw [style=gray, bend right=15, looseness=0.75] (0) to (4);
		\draw [style=gray] (1) to (8);
		\draw [style=gray] (0) to (19.center);
		\draw [style=gray] (3) to (20);
		\draw [style=gray] (4) to (21);
		\draw [style=gray] (5) to (22);
		\draw [style=gray] (6) to (23);
		\draw [style=gray] (24) to (3);
		\draw [style=gray] (25) to (3);
		\draw [style=gray] (26) to (4);
		\draw [style=gray] (4) to (27);
		\draw [style=gray] (5) to (28);
		\draw [style=gray] (29) to (5);
		\draw [style=gray] (30) to (6);
		\draw [style=gray] (6) to (31);
		\draw [style=gray] (2) to (38.center);
		\draw [style=gray, bend left=15] (93) to (2);
		\draw [style=thick] (93) to (94.center);
		\draw [style=gray] (93) to (95);
		\draw [style=gray, bend right=15, looseness=0.75] (110) to (112);
		\draw [style=gray, bend right=15, looseness=0.75] (112) to (111);
		\draw [style=gray, bend right=15, looseness=0.75] (115) to (111);
		\draw [style=gray, bend right=15, looseness=0.75] (111) to (116);
		\draw [style=gray, bend right=15, looseness=0.75] (113) to (110);
		\draw [style=gray, bend right=15, looseness=0.75] (110) to (114);
		\draw [style=gray] (111) to (117);
		\draw [style=gray] (110) to (120.center);
		\draw [style=gray] (113) to (121);
		\draw [style=gray] (114) to (122);
		\draw [style=gray] (115) to (123);
		\draw [style=gray] (116) to (124);
		\draw [style=gray] (125) to (113);
		\draw [style=gray] (126) to (113);
		\draw [style=gray] (127) to (114);
		\draw [style=gray] (114) to (128);
		\draw [style=gray] (115) to (129);
		\draw [style=gray] (130) to (115);
		\draw [style=gray] (131) to (116);
		\draw [style=gray] (116) to (132);
		\draw [style=gray] (112) to (138.center);
		\draw [style=gray, bend right=15] (97.center) to (139.center);
	\end{pgfonlayer}
\end{tikzpicture}

%% file: tree-enumerator-bulk.tikz
\begin{tikzpicture}
	\begin{pgfonlayer}{nodelayer}
		\node [style=U] (0) at (-3.5, 0) {};
		\node [style=U] (1) at (3.5, 0) {};
		\node [style=U] (2) at (0, -2.5) {};
		\node [style=U] (3) at (-5.5, 2.5) {};
		\node [style=U] (4) at (-1.5, 2.5) {};
		\node [style=U] (5) at (1.5, 2.5) {};
		\node [style=U] (6) at (5.5, 2.5) {};
		\node [style=error] (9) at (-6, 3.5) {};
		\node [style=error] (10) at (-5, 3.5) {};
		\node [style=error] (11) at (-2, 3.5) {};
		\node [style=error] (12) at (-1, 3.5) {};
		\node [style=error] (13) at (1, 3.5) {};
		\node [style=error] (14) at (2, 3.5) {};
		\node [style=error] (15) at (5, 3.5) {};
		\node [style=error] (16) at (6, 3.5) {};
		\node [style=physical] (24) at (-6.5, 4.5) {};
		\node [style=physical] (25) at (-4.5, 4.5) {};
		\node [style=physical] (26) at (-2.5, 4.5) {};
		\node [style=physical] (27) at (-0.5, 4.5) {};
		\node [style=physical] (28) at (0.5, 4.5) {};
		\node [style=physical] (29) at (2.5, 4.5) {};
		\node [style=physical] (30) at (4.5, 4.5) {};
		\node [style=physical] (31) at (6.5, 4.5) {};
		\node [style=none] (36) at (0, -4.5) {};
		\node [style=none] (39) at (-6, 3.5) {$\scriptstyle{\alpha_1}$};
		\node [style=bulkerror] (47) at (-4.5, 1.25) {};
		\node [style=bulkerror] (48) at (-2.5, 1.25) {};
		\node [style=bulkerror] (49) at (2.5, 1.25) {};
		\node [style=bulkerror] (50) at (4.5, 1.25) {};
		\node [style=bulkerror] (51) at (-1.75, -1.25) {};
		\node [style=bulkerror] (52) at (1.75, -1.25) {};
		\node [style=bulkerror] (53) at (0, -3.75) {};
		\node [style=none] (54) at (-5, 3.5) {$\scriptstyle{\alpha_2}$};
		\node [style=none] (55) at (-2, 3.5) {$\scriptstyle{\alpha_3}$};
		\node [style=none] (56) at (-1, 3.5) {$\scriptstyle{\alpha_4}$};
		\node [style=none] (57) at (1, 3.5) {$\scriptstyle{\alpha_5}$};
		\node [style=none] (58) at (2, 3.5) {$\scriptstyle{\alpha_6}$};
		\node [style=none] (59) at (5, 3.5) {$\scriptstyle{\alpha_7}$};
		\node [style=none] (60) at (6, 3.5) {$\scriptstyle{\alpha_8}$};
		\node [style=none] (61) at (-4.5, 1.25) {$\scriptstyle{\alpha_9}$};
		\node [style=none] (62) at (-2.5, 1.25) {$\scriptstyle{\alpha_{\scalebox{0.4}{10}}}$};
		\node [style=none] (63) at (2.5, 1.25) {$\scriptstyle{\alpha_{\scalebox{0.4}{11}}}$};
		\node [style=none] (64) at (4.5, 1.25) {$\scriptstyle{\alpha_{\scalebox{0.4}{12}}}$};
		\node [style=none] (65) at (-1.75, -1.25) {$\scriptstyle{\alpha_{\scalebox{0.4}{13}}}$};
		\node [style=none] (66) at (1.75, -1.25) {$\scriptstyle{\alpha_{\scalebox{0.4}{14}}}$};
		\node [style=none] (67) at (0, -3.75) {$\scriptstyle{\alpha_{\scalebox{0.4}{15}}}$};
	\end{pgfonlayer}
	\begin{pgfonlayer}{edgelayer}
		\draw [style=gray] (0) to (2);
		\draw [style=gray] (2) to (1);
		\draw [style=gray] (5) to (1);
		\draw [style=gray] (1) to (6);
		\draw [style=gray] (3) to (0);
		\draw [style=gray] (0) to (4);
		\draw [style=gray] (24) to (3);
		\draw [style=gray] (25) to (3);
		\draw [style=gray] (26) to (4);
		\draw [style=gray] (4) to (27);
		\draw [style=gray] (5) to (28);
		\draw [style=gray] (29) to (5);
		\draw [style=gray] (30) to (6);
		\draw [style=gray] (6) to (31);
		\draw [style=thick] (2) to (36.center);
	\end{pgfonlayer}
\end{tikzpicture}